\documentclass{opendatalab}

\usepackage[utf8]{inputenc}
\usepackage[T1]{fontenc}
\usepackage{textcomp}
\usepackage{xcolor}
\usepackage[most]{tcolorbox}
\usepackage{longtable}
\usepackage{listings}
\usepackage[numbers]{natbib}
\usepackage{enumitem}
\usepackage{graphicx}
\usepackage{xspace} 
\usepackage{hyperref}
\usepackage{algorithm}
\usepackage[noend]{algpseudocode}
\usepackage{pifont}
\usepackage{array}
\usepackage{amssymb}
\usepackage{amsfonts}

\definecolor{codegreen}{rgb}{0,0.6,0}
\definecolor{codegray}{rgb}{0.5,0.5,0.5}
\definecolor{codepurple}{rgb}{0.58,0,0.82}
\definecolor{backcolour}{rgb}{0.95,0.95,0.92}
\definecolor{promptcolor}{HTML}{D1D0F2}
\definecolor{promptcolorheader}{HTML}{bdbcec}
\definecolor{comment}{RGB}{82, 145, 145}
\definecolor{keyword}{RGB}{219, 68, 150} 
\definecolor{string}{RGB}{82, 145, 145}

\newcommand{\github}{\raisebox{-1.5pt}{\includegraphics[height=1.05em]{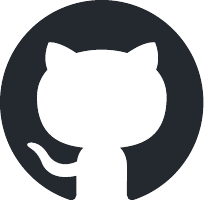}}\xspace}
\newcommand{\web}{\raisebox{-1.5pt}{\includegraphics[height=1.05em]{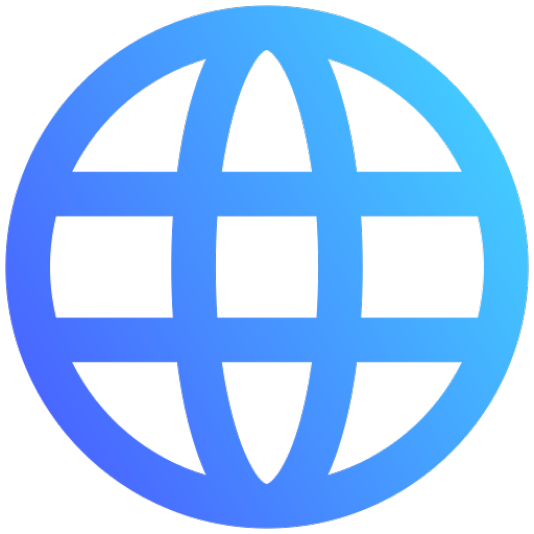}}\xspace}
\newcommand{\huggingface}{\raisebox{-1.5pt}{\includegraphics[height=1.05em]{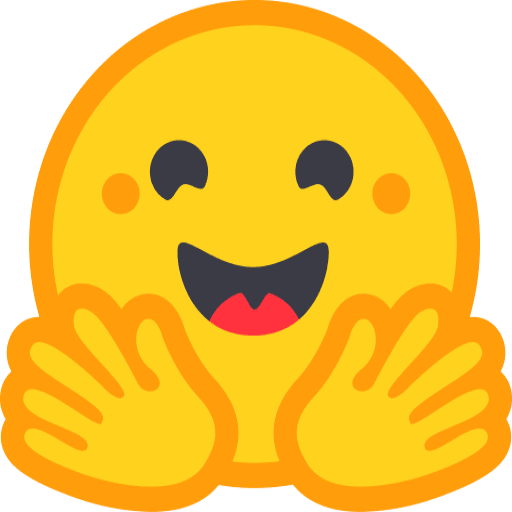}}\xspace}
\newcommand{\cmark}{\ding{51}} % 定义对号
\newcommand{\xmark}{\ding{55}} % 定义叉号
\definecolor{promptcolor}{HTML}{E3F0FA}
\definecolor{promptcolorheader}{HTML}{B5D6ED}
\definecolor{prompttitletext}{HTML}{1B3A5C}

\newtcolorbox{promptbox}[1][]{
  enhanced, breakable,
  top=0.3em,bottom=0.3em,left=0.5em,right=0.5em,
  toptitle=0.3em,bottomtitle=0.2em,boxsep=0pt,
  colframe=promptcolorheader, colback=promptcolor!50, boxrule=0.5pt,
  width=\columnwidth, 
  coltitle=prompttitletext,
  title={\footnotesize #1} 
}
\lstdefinestyle{promptstyle}{
    backgroundcolor=\color{backcolour},   
    commentstyle=\color{codegreen},
    keywordstyle=\color{magenta},
    numberstyle=\tiny\color{codegray},
    stringstyle=\color{codepurple},
    basicstyle=\ttfamily\footnotesize,
    breakatwhitespace=false,         
    breaklines=true,                 
    captionpos=b,                    
    keepspaces=true,                 
    numbers=left,                    
    numbersep=5pt,                  
    showspaces=false,                
    showstringspaces=false,
    showtabs=false,                  
    tabsize=2
}
\lstdefinestyle{codestyle}{
    language=Python,
    backgroundcolor={},   
    basicstyle=\ttfamily\small\color{black}, 
    commentstyle=\color{comment}, 
    keywordstyle=\color{keyword},
    stringstyle=\color{comment}, 
    showstringspaces=false,        
    frame=none,                  
    numbers=none,                 
    breaklines=true,           
    tabsize=4,                  
    aboveskip=0.5em,
    belowskip=0.5em
}
\title{Programming with Data: Test-Driven Data Engineering for Self-Improving LLMs from Raw Corpora}
\author[1]{Chenkai Pan$^\dagger$}
\author[2]{Xinglong Xu$^\dagger$}
\author[2]{Yuhang Xu}
\author[3]{Yujun Wu}
\author[1]{Siyuan Li}
\author[1*]{Jintao Chen}
\author[3*]{Conghui He}
\author[2*]{Jingxuan Wei}
\author[3*]{Cheng Tan}

\affiliation[1]{Zhejiang University}
\affiliation[2]{University of Chinese Academy of Sciences}
\affiliation[3]{Shanghai Artificial Intelligence Laboratory}

\abstract{
Reliably transferring specialized human knowledge from text into large language models remains a fundamental challenge in artificial intelligence. Fine-tuning on domain corpora has enabled substantial capability gains, but the process operates without feedback: when a model fails on a domain task, there is no method to diagnose what is deficient in the training data, and the only recourse is to add more data indiscriminately. Here we show that when a structured knowledge representation extracted from the source corpus serves as the shared foundation for both training data and evaluation, the complete data-engineering lifecycle maps onto the software development lifecycle in a precise and operative way: \textit{training data becomes source code specifying what the model should learn, model training becomes compilation, benchmarking becomes unit testing, and failure-driven data repair becomes debugging}. Under this correspondence, model failures decompose into concept-level gaps and reasoning-chain breaks that can be traced back to specific deficiencies in the data and repaired through targeted patches, with each repair cycle producing consistent improvements across model scales and architectures without degrading general capabilities. We formalize this principle as \textit{Programming with Data} and instantiate it across sixteen disciplines spanning the natural sciences, engineering, biomedicine, and the social sciences, releasing a structured knowledge base, benchmark suite, and training corpus as open resources. By demonstrating that the relationship between training data and model behaviour is structurally traceable and systematically repairable, this work establishes a principled foundation for the reliable engineering of human expertise into language models.
}
\metadatainfo{
  \parbox{\linewidth}{\centering
    \github~\href{https://github.com/OpenRaiser/ProDa}{\textbf{Code}} \quad
    \web~\href{https://openraiser.github.io/ProDa-webpage}{\textbf{Website}} \quad
    \huggingface~\href{https://huggingface.co/datasets/OpenRaiser/ProDalib}{\textbf{Dataset}}
  }
}

\begin{document}

\maketitle

% \tableofcontents

\begin{figure}[h]
\centering
\vspace{-6mm}
\includegraphics[width=0.96\textwidth]{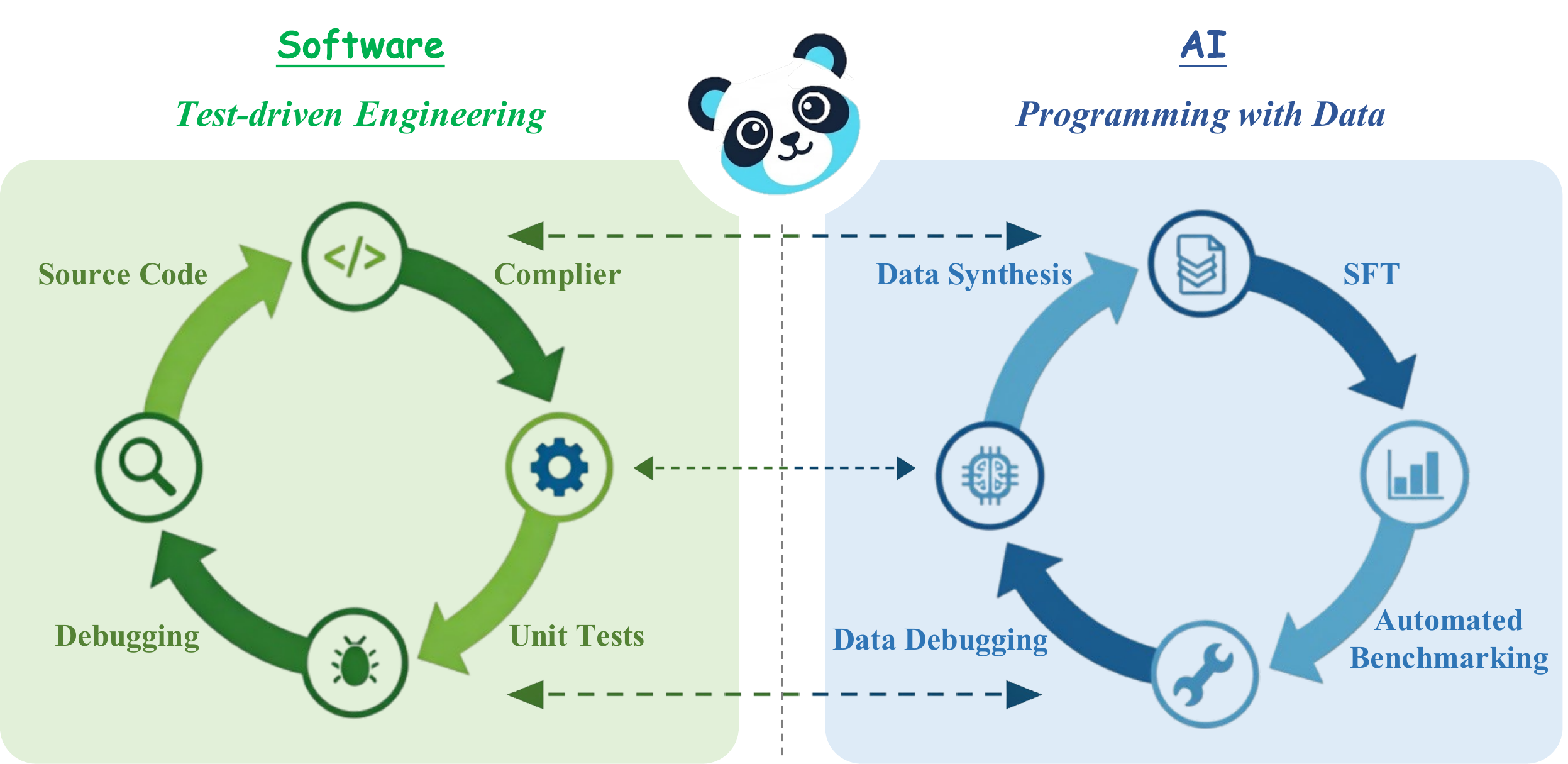}
\vspace{-2mm}
\caption{The conceptual correspondence between test-driven engineering in software and the Programming with Data for AI.
By treating data as a first-class, executable artifact, ProDa establishes a one-to-one mapping between software engineering practices and modern AI, enabling a compile–test–debug cycle that supports automated and self-evolving model improvement.
}
\vspace{-6mm}
\label{fig:teaser}
\end{figure}

\section{Introduction}
\label{sec:intro}

Reliably encoding specialized human knowledge into large language models remains a defining challenge of artificial intelligence, in large part because the vast majority of such knowledge resides in unstructured corpora, including textbooks, technical manuals, research manuscripts and clinical guidelines, whose transformation into verifiable model capabilities is the central task of data engineering~\cite{ling2025domain,dredze2024academics}. Although synthetic data generation and textbook-quality corpora have enabled substantial capability gains~\cite{honovich2023unnatural,wang2023self,mukherjee2023orca,gunasekar2023textbooks,li2023textbooks}, the prevailing workflow operates without feedback. When a fine-tuned model produces an incorrect derivation, misapplies a domain principle or hallucinates a nonexistent concept, no principled mechanism exists to trace the failure back to a specific deficiency in the training data, nor to repair it in a targeted manner~\cite{kwon2024datainf,huang2024large}. The standard remedy is undirected augmentation: indiscriminately increasing data volume or diversity~\cite{sun2025improving}, an approach that is computationally expensive, poorly interpretable and provides no structural guarantee of improvement. The entire pipeline is open-loop: evaluation can diagnose model failures, but those diagnoses carry no information about where in the data the deficiency lies and therefore cannot guide correction.

This open-loop condition has a deeper origin than is commonly recognized. The workflow that dominates domain-specific fine-tuning has been inherited, largely without modification, from the logic of pre-training. In pre-training, open-loop operation is tolerable: corpora are measured in trillions of tokens, scale itself provides a statistical form of coverage, and post-hoc evaluation on general benchmarks~\cite{hendrycks2021measuring,srivastava2023beyond} suffices to verify broad competence~\cite{liang2022holistic}. Domain-specific fine-tuning faces a fundamentally different constraint structure: source corpora are limited and often irreplaceable, domain knowledge is highly structured rather than statistically distributed, and every model failure carries diagnostic information that could, in principle, guide precise correction. Yet practitioners routinely apply the pre-training playbook to this different regime: collect domain data, synthesize instruction-tuning samples~\cite{xu2023wizardlm,taori2023alpaca,zhou2023lima}, train, evaluate on existing or ad-hoc benchmarks~\cite{huang2023c}, and when results disappoint, add more data. The benchmarks used to evaluate fine-tuned models are either borrowed from general evaluation suites or constructed independently from the training data, so that when a benchmark reveals a failure, there is no shared structure through which the failure can be localized in the data and corrected~\cite{liu2021explainaboard,kiela2021dynabench}. Evaluation diagnoses symptoms but cannot identify the pathology in the training signal. The pipeline remains open-loop because the lack of the structural prerequisite: a shared representation that links training data and evaluation at the knowledge level.
% The pipeline remains open-loop not because closing it is impossible, but because no one has supplied the structural prerequisite: a shared representation that links training data and evaluation at the knowledge level.

In developing a methodology that supplies this missing link, we recognized that the resulting lifecycle exhibits a structural correspondence with software engineering as formalized by test-driven development~\cite{beck2003test}. Before disciplined methodologies were adopted, software was produced in a similar open-loop fashion: developers wrote code, executed it and diagnosed failures after the fact, with no systematic connection between a failing test and the code responsible for the defect. The introduction of a shared specification from which both source code and test suites are derived transformed this process~\cite{brooks1995mythical,royce1987managing}: test failures became directly traceable to code defects and repairable through targeted patches, converting software construction from artisanal practice into rigorous engineering. We observe that the same structural precondition holds when training data and benchmarks are jointly derived from a common knowledge representation extracted from the source corpus: benchmark failures become traceable to identifiable gaps in the data and correctable through precise augmentation, not by surface analogy, but because the shared knowledge representation provides the same indexing structure that a shared specification provides in software.

We formalize this principle as \textbf{Programming with Data}, a paradigm that reconceptualizes the relationship between raw corpora and model capabilities. Under this paradigm, \textit{the raw corpus serves as the requirements specification} constraining what the model should know; \textit{synthesized training data serves as source code} encoding the logic the model is expected to implement; \textit{model training serves as compilation} translating human-readable data into machine-executable weights; and \textit{benchmarking serves as unit testing} verifying the compiled model against its specification. 

\begin{figure}[H]
\centering
\includegraphics[width=\textwidth]{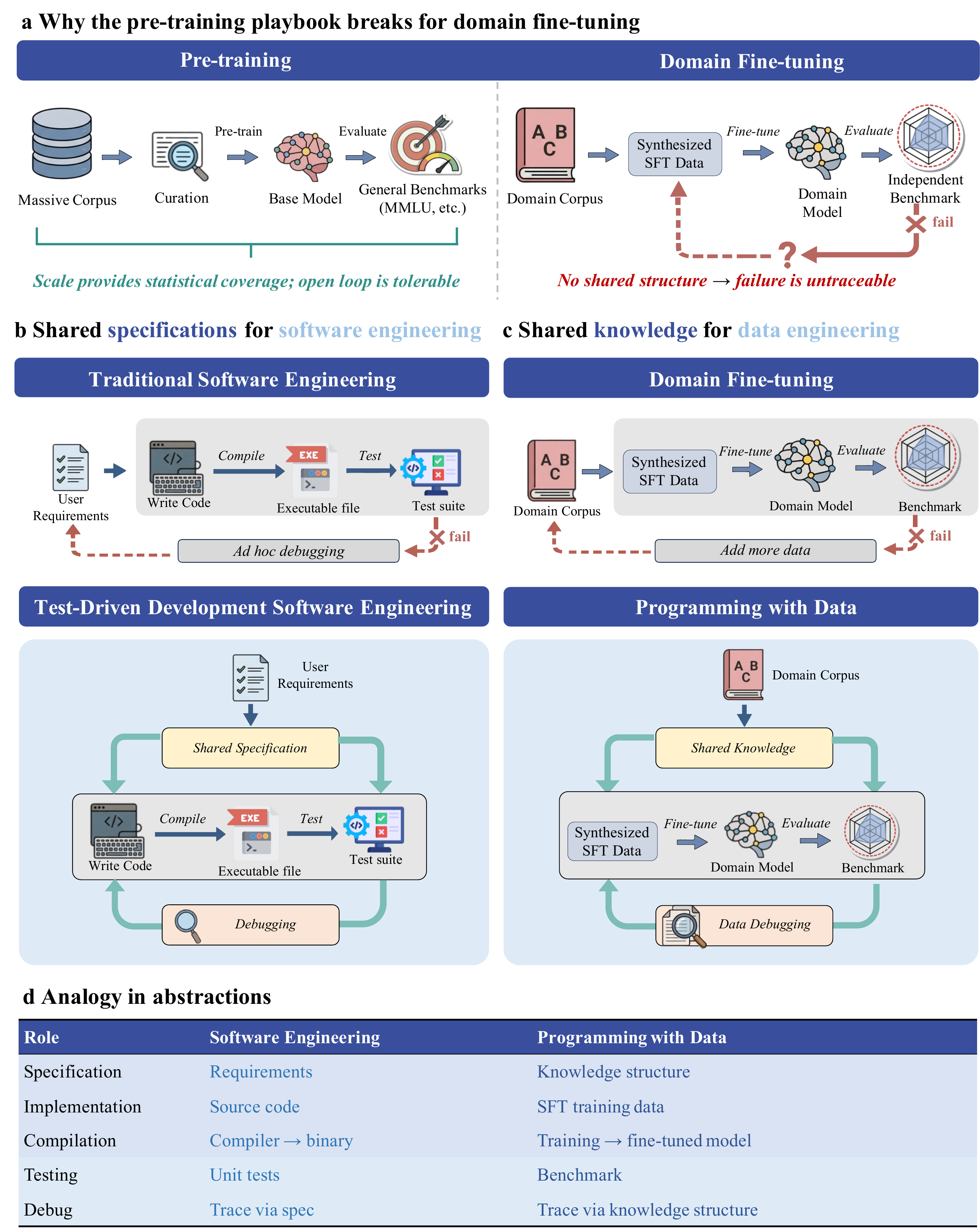}
\caption{\textbf{From open-loop data engineering to Programming with Data.}
\textbf{a}, The pre-training playbook breaks for domain fine-tuning because failures cannot be traced back to data without a shared structure.
\textbf{b}, Software engineering solved this by deriving code and tests from a shared specification.
\textbf{c}, Programming with Data applies the same principle, using a shared knowledge structure to close the loop between training data and evaluation.
\textbf{d}, Formal correspondence between the two paradigms.}
\label{fig:intro}
\end{figure}

The critical enabler is a three-level knowledge structure, comprising atomic concepts, relational triples and reasoning chains, extracted from the corpus and shared by both the training data and the benchmark. This shared structure provides the traceability that connects a test failure to a data deficit and transforms evaluation from a terminal judgement into an actionable diagnostic. To govern data quality within this paradigm, we introduce the \textbf{CORE} principle, a set of engineering standards requiring that synthesis be \textit{Contextualized} within document-level scope, \textit{Organized} into stratified knowledge layers, \textit{Rigorous} in enforcing adversarial robustness and non-overlap between training instances and evaluation items, and \textit{Evolving} through iterative refinement driven by empirical feedback.

We instantiate \textbf{Programming with Data} in the ProDa framework, which operationalizes the closed loop and the CORE standards through three tightly coupled phases. First, ProDa extracts the three-level knowledge structure from the corpus and compiles it into a benchmark before any training occurs, defining target capabilities as executable specifications. Second, it synthesizes training data from the same knowledge structure under the CORE principle, establishing baseline domain competency. Third and most critically, it treats benchmark failures as runtime errors: each failure is diagnosed as either a concept gap or a reasoning deficit, traced to the responsible nodes in the knowledge structure and repaired through a targeted data patch that is fed into the next training cycle. Because all three phases operate on the same knowledge structure, failures are not opaque signals that prompt indiscriminate data scaling but structured diagnostics that drive precise, traceable repairs.

We apply ProDa to 16 disciplines spanning natural sciences, engineering, biomedical sciences and social sciences, and release ProDaLib, an open-source resource suite containing 227k key concepts, 16k evaluation items and 160k synthesized training samples. Experiments across two model families (Llama and Qwen) and parameter scales from 3B to 32B demonstrate that each debugging iteration produces consistent improvements across every model and scale tested, with no exceptions. After a single round of data debugging, a 32B open-source model surpasses GPT-5.4, Gemini-3-flash and DeepSeek-v3.2 on the 16-discipline average, while general capabilities measured by MMLU~\cite{hendrycks2021measuring} and CEVAL~\cite{huang2023c} remain fully preserved. By establishing that the relationship between training data and model behaviour is structurally traceable and systematically repairable, this work provides a principled foundation for the reliable engineering of human expertise into language models.

\section{Programming with Data}

The Introduction established that current data engineering for LLMs is open-loop because training data and evaluation are structurally decoupled, and that closing this loop requires a shared knowledge representation linking the two. This section formalizes the correspondence that makes the closed loop principled (\S\ref{sec:correspondence}) and describes the pipeline that makes it operational (\S\ref{sec:pipeline}).

\subsection{A structural correspondence with software engineering}
\label{sec:correspondence}

The software development lifecycle proceeds through a well-understood sequence: a requirements specification defines what the system should do; source code implements that specification in a human-readable form; a compiler translates the source code into an executable binary; and a test suite verifies the binary against the specification. When a test fails, the developer traces the failure through the shared specification back to the responsible segment of source code, writes a targeted patch, and recompiles. The entire cycle is effective because the test suite and the source code are both derived from, and indexed against, the same specification. Without this shared reference, test failures would be uninterpretable signals bearing no connection to corrective action.

We observe that when a structured knowledge representation is extracted from a raw corpus and used as the common foundation for both training data and evaluation instruments, the LLM data-engineering lifecycle acquires the same structural properties. Figure~\ref{fig:intro}(d) formalizes this correspondence. The raw corpus functions as the requirements specification, defining the scope and depth of knowledge the model should acquire. \textbf{Synthesized training data functions as source code}: it encodes, in a form the training algorithm can consume, the specific concepts and relationships the model is expected to internalize. \textbf{Model training functions as compilation}, transforming human-readable data into machine-executable weights. \textbf{The fine-tuned model functions as the compiled binary}, and \textbf{benchmarks function as the test suite} that verifies whether the binary faithfully implements its specification. When the model fails a benchmark item, the failure can be traced through the shared knowledge structure to a specific deficit in the training data and repaired through a targeted data patch, corresponding directly to a bug fix in software engineering.

The critical enabler of this correspondence is the three-level knowledge structure that serves as the shared specification. We extract from each source corpus three layers of increasing complexity:

\begin{itemize}[leftmargin=2em]
\item \textbf{L1 Key Concepts}: the atomic vocabulary of a discipline—technical terms, named principles, canonical equations, and foundational definitions. Each L1 entry carries a concise canonical definition grounded in its source text.
\item \textbf{L2 Knowledge Relations}: typed pairwise associations between L1 concepts. Each L2 entry is a triple (subject, relation, object) carrying semantic content beyond co-occurrence—specialization, causal mechanism, prerequisite, contrast, and others.
\item \textbf{L3 Reasoning Chains}: multi-step inferential pathways that traverse multiple L1 concepts along L2 relations. Each chain is decomposed into discrete steps, with each step annotated by the concepts it invokes and the nature of the inferential transition.
\end{itemize}

This three-level structure is what makes the correspondence operative rather than metaphorical. When the benchmark is constructed from L3 reasoning chains and the training data from L1 concepts and L2 relations, a benchmark failure on a specific L3 chain can be decomposed: either the model lacks one of the constituent L1 concepts or L2 relations (a \textit{concept gap}), or it possesses the requisite pieces but fails to compose them in the correct inferential sequence (a \textit{reasoning deficit}). In either case, the failure points to identifiable nodes in the shared knowledge structure, and the repair targets exactly those nodes. This decomposition is the mechanism through which test failures become actionable engineering diagnostics.

\subsection{The ProDa pipeline}
\label{sec:pipeline}

ProDa instantiates \textit{Programming with Data} as an automated pipeline comprising three components (Builder, Tester, and Debugger) that operate on the shared three-level knowledge structure. To govern the quality of data synthesized within this pipeline, we adopt a set of engineering standards that we term the \textbf{CORE principle}, requiring that synthesis be \textit{Contextualized} within document-level scope, \textit{Organized} into stratified knowledge layers, \textit{Rigorous} in enforcing adversarial robustness and instance-level non-overlap between training and evaluation, and \textit{Evolving} through iterative refinement driven by empirical feedback. Each CORE standard is operationalized through a specific pipeline component.

\paragraph{The Builder: knowledge extraction and training data synthesis.}

The Builder transforms the raw corpus into the shared knowledge structure and uses it to synthesize the initial training data. Extraction proceeds top-down: L3 reasoning chains are extracted first from high-quality corpus chunks, then decomposed into L2 relational triples, and finally L1 concepts are harvested and canonicalized from L2 subjects and objects. This top-down order, rather than the conventional bottom-up sequence of named-entity recognition followed by relation extraction, guarantees that every L1 concept and L2 relation is reachable from at least one L3 chain, eliminating orphan entries that would be untestable and therefore undebuggable.

Two CORE standards govern extraction and synthesis. The \textbf{Contextualized} standard requires that each knowledge unit be grounded in the global context of its source document, not extracted from isolated fragments. When a reasoning chain spans multiple paragraphs or depends on definitions introduced earlier in the text, the extraction process must preserve these dependencies. Decontextualized extraction produces units that appear locally coherent but encode incomplete or misleading logic—analogous to source code that compiles in isolation but fails when linked into the full program. In ProDa, this requirement is operationalized through a document-level curation stage that evaluates academic depth and reasoning density before any chunking occurs, followed by a chunk-level quality scoring system that evaluates reasoning depth, prerequisite density, scenario applicability, counter-intuitive content, knowledge synthesis, and boundary integrity (Methods \S\ref{sec:methods_extraction}).

The \textbf{Organized} standard requires that knowledge be represented in a stratified, reusable structure rather than as a flat collection of question–answer pairs. The three-level structure (L1/L2/L3) satisfies this requirement directly: each level serves a distinct engineering purpose. L1 and L2 feed the synthesis of foundational training data—a mixture of open-ended questions, multiple-choice items, and true-false judgments that cover the factual and relational substrate of the domain. L3 feeds the construction of reasoning-intensive benchmarks. This separation enables independent engineering of each level and, critically, enables precise failure attribution: a benchmark failure on an L3 chain can be decomposed into deficits at the L2 or L1 level.

The Builder's output constitutes the first version of the source code: an initial training corpus grounded in the knowledge structure and ready for compilation.

\paragraph{The Tester: benchmark construction.}

The Tester constructs the benchmark from L3 reasoning chains, which encode the most demanding inferential patterns in the corpus. Each benchmark item requires the model to traverse multiple L1 concepts along L2 relations in a specific logical sequence, testing not merely whether isolated facts have been learned but whether the model can compose them into coherent reasoning. The Tester operates before any training, consistent with the test-first principle of test-driven development: the benchmark defines the criteria that the compiled model must satisfy.

The \textbf{Rigorous} standard imposes two requirements on benchmark construction. First, each item must contain adversarial distractors constructed from the same knowledge structure, so that correct responses demand discrimination between closely related concepts rather than elimination of implausible options. In ProDa, distractors are generated by perturbing L2 relations (inverting relation types, substituting semantically adjacent L1 concepts) and truncating L3 chains, producing alternatives that are locally plausible but globally inconsistent with the full reasoning pathway. Second, benchmarks and training data must maintain instance-level orthogonality: no benchmark item may be answerable by verbatim recall of a training sample. The model must generalize from the training data to the benchmark, not memorize. This orthogonality is enforced structurally: training data is synthesized from L1 and L2 entries, while benchmark items are constructed from L3 chains that require novel composition of those entries. The knowledge-level connection between the two is what enables traceability; the instance-level separation is what ensures that benchmark performance reflects genuine capability. 
% rather than memorization.

\paragraph{The Debugger: diagnosis and targeted repair.}

After the model is compiled through training and evaluated against the benchmark, the Debugger analyzes each failure. It classifies the underlying deficit into one of two categories. A \textit{concept gap} indicates that the model lacks or confuses a specific piece of domain knowledge, traceable to missing or malformed L1 and L2 entries in the training data. A \textit{reasoning deficit} indicates that the model possesses the requisite knowledge but fails to compose it correctly across multiple steps, traceable to insufficient coverage of the relevant L3 chain patterns. For each category, the Debugger applies a distinct repair strategy: concept gaps are addressed through knowledge-reinforcement samples that explicitly contrast the confused concepts with their correct definitions and boundaries, while reasoning deficits are addressed through chain-of-thought samples that scaffold the missing inferential steps with explicit intermediate reasoning.

The \textbf{Evolving} standard requires that training data not be treated as a static artifact produced once and consumed passively: it must evolve through iterative refinement driven by empirical feedback. The Debugger operationalizes this standard by treating each benchmark failure as evidence that the current data is incomplete or malformed with respect to the knowledge the item tests. The resulting patches are combined with a strategically selected subset of the original training data—chosen to cover knowledge regions that the model has already mastered, ensuring that previously acquired capabilities are preserved while new deficits are repaired. The model is then retrained on this augmented corpus, completing the loop.

\paragraph{Structural coupling.}

The three components are coupled through the shared knowledge structure, and it is this coupling that delivers the central promise of Programming with Data. When the Tester identifies a failure on a benchmark item derived from a particular L3 chain, the Debugger traces backward through the chain's constituent L2 relations and L1 concepts to identify precisely which knowledge nodes are under-represented or absent in the training data produced by the Builder. The patch generated by the Debugger targets exactly those nodes. This backward traceability—from test failure through shared structure to data deficit—is the mechanism that transforms model failures from opaque evaluation signals into actionable engineering diagnostics, and it is the property that distinguishes the Programming with Data paradigm from the open-loop workflows that currently dominate the field.

\section{Results}

\subsection{Specification: structured knowledge from raw corpora }
\label{sec:specification}

\begin{figure}[h]
\centering
\includegraphics[width=\textwidth]{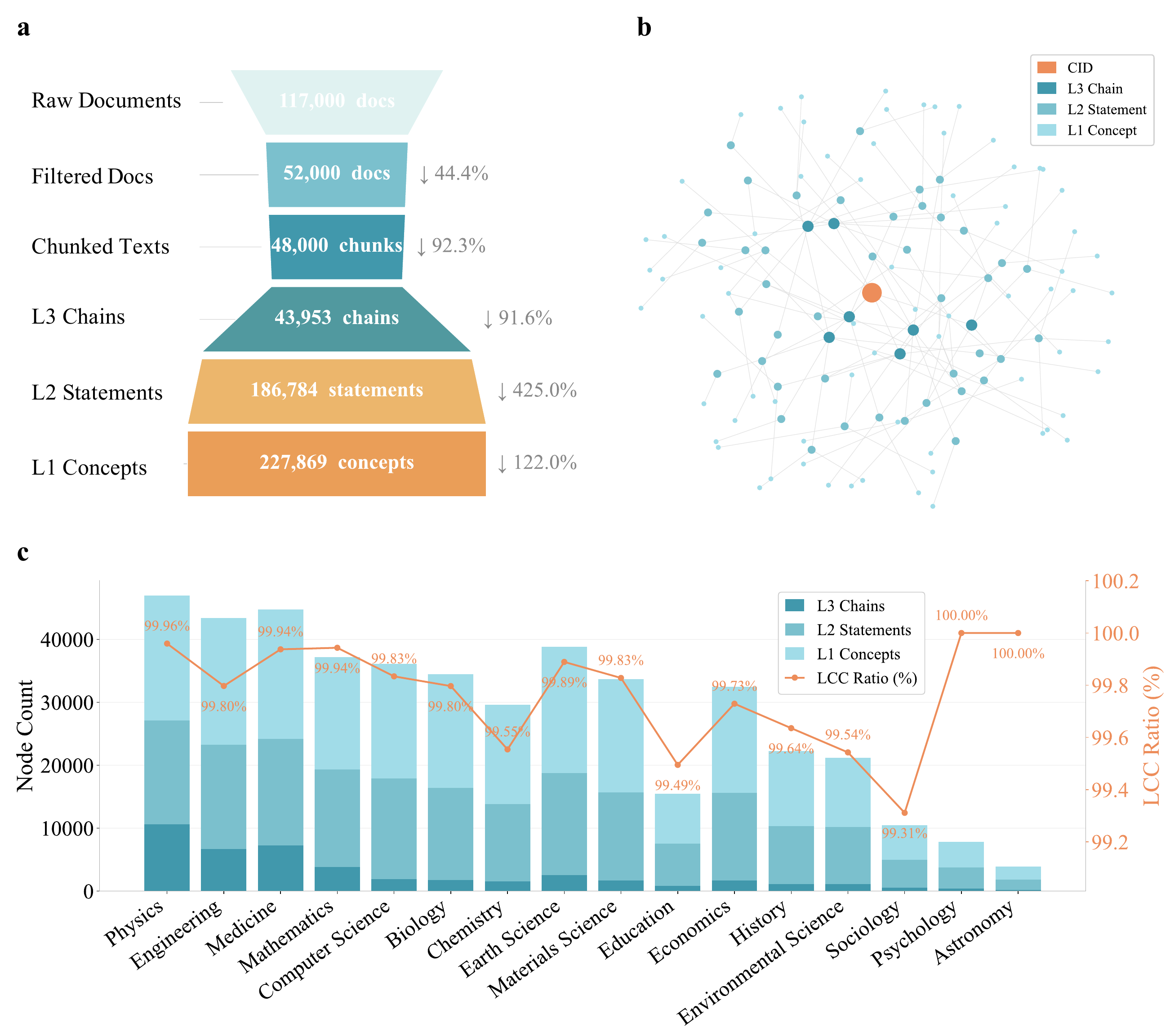}
\caption{%
\textbf{Structured knowledge extracted from 16 disciplines}. \textbf{a}, Corpus distillation pipeline. Successive filtering reduces 117,000 raw documents (~15B tokens) to 48,000 high-quality chunks, from which 43,953 L3 reasoning chains, 186,784 L2 relational statements, and 227,869 L1 atomic concepts are extracted top-down. Percentages indicate retention rates. \textbf{b}, Representative knowledge subgraph for a single corpus chunk. Node colour encodes layer membership; every lower-level node is reachable from at least one higher-level node, confirming zero orphan nodes. \textbf{c}, Per-discipline scale (stacked bars) and largest connected component ratio (orange line). All disciplines exceed 99\% connectivity.}
\label{fig:results_specification}
\end{figure}

The entire \textit{Programming with Data} pipeline rests on the premise that a structured knowledge representation can be reliably extracted from unstructured corpora at scale. We test this premise across 16 disciplines. Starting from 117,000 textbook-grade documents spanning the natural sciences, engineering, biomedicine, and social sciences, successive quality-based filtering retains 48,000 high-quality chunks comprising approximately 1.5 billion tokens, a 10:1 compression that concentrates the corpus toward reasoning-rich, conceptually dense material (Figure~\ref{fig:results_specification}a; filtering criteria in Methods \S~\ref{sec:methods_extraction}). From these chunks, top-down extraction yields 43,953 L3 reasoning chains, 186,784 L2 relational statements, and 227,869 L1 atomic concepts, totalling 458,622 knowledge nodes. Notably, the transition from L3 to L2 is expansive rather than compressive: each reasoning chain decomposes into approximately four atomic statements, reflecting the multi-step character of domain reasoning.

Figure~\ref{fig:results_specification}b illustrates the internal structure of the extracted knowledge for a representative corpus chunk. L3 reasoning chains anchor the subgraph as dark-coloured nodes; each chain decomposes into L2 relational statements, shown as medium-coloured nodes, that encode causal mechanisms, definitional relationships, and prerequisite conditions between L1 atomic concepts, rendered as light-coloured nodes. The top-down extraction order enforces a strict reachability invariant: because L2 statements are derived by decomposing L3 chains, and L1 concepts are grounded in L2 subjects and objects, every lower-level node is reachable from at least one higher-level node by construction. Across all 458,622 nodes, the orphan rate is exactly zero. This property guarantees that every concept in the training data can be tested through an L3-derived benchmark item, and that every benchmark failure can be traced back to specific L1 or L2 entries, providing the traceability on which the debugging loop depends.

The knowledge structure spans all 16 target disciplines with substantial coverage in each (Figure~\ref{fig:results_specification}c). Per-discipline node counts range from approximately 7,000 in Astronomy and Psychology to over 45,000 in Physics, Engineering, and Medicine, reflecting both the volume of available source material and the inherent conceptual density of each field. Despite this variation in scale, connectivity is uniformly high: the largest connected component encompasses at least 99.3\% of nodes in every discipline, with 11 of 16 disciplines exceeding 99.8\%. Structural differences across disciplines are also informative. Physics and Mathematics exhibit a high ratio of L3 chains to total nodes, consistent with long deductive sequences, whereas Medicine and Biology show proportionally more L1 concepts, reflecting taxonomically rich classification systems. These discipline-specific profiles suggest that the extraction pipeline adapts to the reasoning style of each field rather than imposing a uniform structure.

Together, these results establish that the ProDaLib knowledge structure provides a reliable foundation for the \textit{Programming with Data} pipeline. It is large enough to cover 16 disciplines at professional depth, connected enough that failures can be traced across knowledge layers, and structurally free of orphan nodes that would be untestable or unrepairable. We next examine whether the benchmark derived from this structure measures meaningful domain capabilities.

\subsection{Tester: validating the ProDa-16 benchmark}

\begin{figure}[h]
  \centering
  \includegraphics[width=\textwidth]{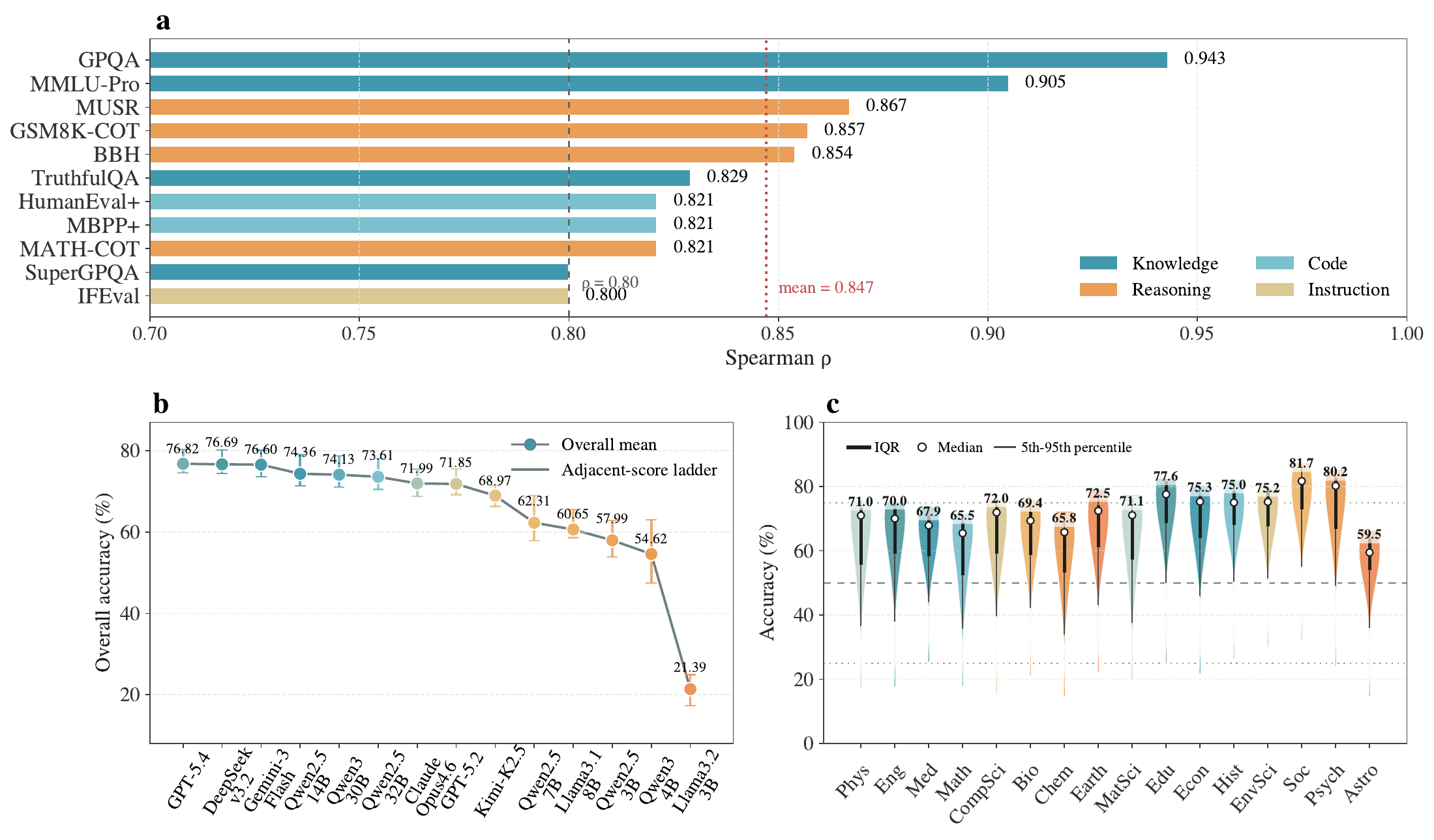}
  \vspace{-4mm}
  \caption{\textbf{Meta-evaluation of the ProDa-16 benchmark.}
\textbf{a}, Spearman rank correlation between ProDa-16 and 11
established benchmarks across models (dashed
line, $\rho = 0.80$; red dotted line, mean $\rho = 0.847$).
\textbf{b}, Overall accuracy by model, ranked in descending order;
error bars denote 95\% bootstrap confidence intervals across
disciplines.
\textbf{c}, Per-discipline accuracy distributions across all
models; thick bars, interquartile range; open circles, median;
dashed line, four-choice chance level (25\%).}
  \label{fig:benchmark_validation}
\end{figure}

A benchmark co-derived with training data from the same knowledge structure faces an inherent credibility challenge: does it measure genuine capability, or does it merely reward familiarity with the extraction source? Before using ProDa-16 as the test suite for the debugging cycle, we must establish that it behaves as a legitimate evaluation instrument whose judgments generalize beyond its own construction. To validate this construct validity, we compared model performance on ProDa-16 against 11 established international benchmarks spanning complex reasoning, mathematics, and coding (Figure~\ref{fig:benchmark_validation}a). The results demonstrate that ProDa-16 exhibits exceptional statistical consistency with mainstream evaluation paradigms, achieving a high overall mean Spearman's rank correlation coefficient of $\rho=0.847$. Notably, the benchmark displays remarkably strong positive correlations with GPQA ($\rho=0.943$) and MMLU-Pro ($\rho=0.905$), which represent frontier complex knowledge reasoning. \textbf{This high degree of external alignment establishes that ProDa-16 is not an isolated evaluation tool, but a reliable metric capable of precisely mapping to recognized industry capability standards.}

Beyond external alignment, an effective benchmark must possess sufficient internal discriminative power to clearly delineate the capability boundaries of models across varying parameter scales, without triggering floor or ceiling effects. Figure~\ref{fig:benchmark_validation}b depicts the score distribution of over a dozen open- and closed-source models on ProDa-16, revealing a smooth and highly structured ``adjacent-score ladder.'' Frontier closed-source models anchor the performance ceiling with an accuracy of approximately 76\%; concurrently, the homologous Qwen series models strictly adhere to scaling laws, exhibiting a monotonically increasing trajectory in scores from the 3B to 32B parameter scales. This well-stratified discriminative capability proves that \textbf{ProDa-16 possesses the requisite sensitivity to accurately capture the marginal capability leaps generated during iterative data fine-tuning}.

We verify that the benchmark provides consistent diagnostic coverage across all 16 disciplines rather than being dominated by a subset of trivially easy or intractably hard domains (Figure~\ref{fig:benchmark_validation}c). Per-discipline accuracy distributions, aggregated across all evaluated models, show that every discipline produces a median well above the 25\% four-choice chance baseline, confirming that the automatically constructed items are answerable by capable models rather than being degenerate or ill-formed. At the same time, no discipline exhibits a ceiling effect: even the highest-median disciplines remain below 82\%, leaving ample room for the debugging cycle to produce measurable gains. The interquartile ranges are wide in every discipline, indicating that ProDa-16 discriminates effectively within each domain across the model capability spectrum. 

These results establish that ProDa-16 provides a robust evaluation foundation for the framework. It strictly aligns with authoritative international standards at the macro level, possesses strong model discriminative power at the meso level, and provides sufficient multidisciplinary resolution at the micro level. This multi-level validation is the essential prerequisite ensuring that model failures can subsequently be precisely traced back to underlying knowledge nodes (L1/L2) to drive targeted repair.

\subsection{Implementation: first-pass compilation}

Having validated the test suite, we compiled the first version of the training corpus. The Builder synthesized 160K supervised fine-tuning instances from L1 concepts and L2 relations across all 16 disciplines, without incorporating L3 reasoning chains. We fine-tuned base models from two families at multiple scales: Llama-3.1-8B, Qwen-2.5 at 3B/7B/14B/32B, and Qwen-3 at 4B/8B/14B/32B, using identical hyperparameters for all runs. We denote these first-pass models ProDa-V1 and evaluate them on ProDa-16 alongside the corresponding official Instruct checkpoints, which incorporate RLHF, proprietary preference data, and substantially larger training budgets (Table~\ref{tab:full_experimental_results}, Panels A and B).

The central finding is that \textbf{a single round of fine-tuning on automatically synthesized data already produces domain competence competitive with industrial alignment pipelines across most model scales}. Among the Qwen-2.5 family, V1 models match or exceed their Instruct counterparts at two of four scales: Qwen-2.5-7B-V1 scores 65.86\% versus 62.31\% for the official Instruct (+3.55), and Qwen-2.5-32B-V1 scores 76.54\% versus 73.61\% (+2.93). The pattern is more pronounced in the Qwen-3 family, where Qwen-3-4B-V1 surpasses its Instruct counterpart by 11.17 points (65.79\% vs 54.62\%) and Qwen-3-14B-V1 exceeds the larger Qwen-3-30B-A3B-Instruct by 2.31 points (76.44\% vs 74.13\%). At the 32B scale, Qwen-3-32B-V1 reaches 77.35\%, the highest V1 score in the table and above every Instruct model except the closed-source frontier systems. These results are achieved using only the knowledge structure extracted in \S\ref{sec:specification}, with no human annotation, no preference optimization, and no data beyond the 160K automatically generated instances.

However, this superiority is not universally maintained across all parameter scales. For specific model sizes, ProDa-V1 fails to eclipse the officially aligned versions. This performance divergence exposes a fundamental mechanistic constraint: the inherent limitations of static data synthesis. Irrespective of the rigor applied to corpus filtering and structured extraction, a one-off static data injection (first-pass compilation) inevitably leaves conceptual blind spots. Unlike human experts in RLHF, static synthesis cannot dynamically rectify model-specific ``concept gaps,'' nor can it adequately cover the long-tail errors inherent in multi-step reasoning. 

This dichotomy of outperformance and underperformance offers profound mechanistic insights. It demonstrates that while high-quality, static structured data (V1) is sufficient to establish a formidable baseline, it simultaneously represents the upper bound of the traditional unidirectional data synthesis paradigm. To systematically shatter the performance ceiling set by official Instruct models across all parameter scales, we must pivot from unidirectional static injection to a dynamic paradigm. By exploiting the structured traceability intrinsic to ProDa-16, we can systematically diagnose these residual errors. This imperative directly motivates the core mechanism of our framework: diagnostic-driven targeted repair (the V2 stage).

\begin{table*}[htbp]
\centering
\caption{Full benchmark results across all 16 disciplines. This table provides a comprehensive breakdown of the performance of Instruct models (Panel A), Base models after the 1st fine-tuning iteration (Panel B), Base models after the 3rd iteration (Panel C), and the absolute performance gains achieved (Panel D). \textbf{Discipline Codes:} 001: Physics, 002: Engineering, 003: Medicine, 004: Mathematics, 005: Computer Sci., 006: Biology, 007: Chemistry, 008: Earth Sci., 009: Materials Sci., 010: Education, 011: Economics, 012: History, 013: Environmental Sci., 014: Sociology, 015: Psychology, 016: Astronomy.}
\label{tab:full_experimental_results}
% 适当增加行高，避免数据过于拥挤
\renewcommand{\arraystretch}{1.5} 
% 自动缩放表格以适应页面宽度
\resizebox{\textwidth}{!}{%
\begin{tabular}{lccccccccccccccccc}
\toprule
\textbf{Model} & \textbf{001} & \textbf{002} & \textbf{003} & \textbf{004} & \textbf{005} & \textbf{006} & \textbf{007} & \textbf{008} & \textbf{009} & \textbf{010} & \textbf{011} & \textbf{012} & \textbf{013} & \textbf{014} & \textbf{015} & \textbf{016} & \textbf{Avg} \\
\midrule
\multicolumn{18}{@{}l}{\textit{\textbf{Panel A: Instruct Models (Upper Bound Reference)}}} \\
\midrule
GPT-5.4          & 75.50 & 74.90 & 73.80 & 70.20 & 75.80 & 76.00 & 72.40 & 79.20 & 77.60 & 81.06 & 78.10 & 81.30 & 80.20 & 82.51 & 82.95 & 63.50 & \textbf{76.82} \\
Gemini-3-flash    & 75.80 & 73.60 & 73.60 & 68.80 & 76.40 & 75.20 & 71.20 & 75.50 & 75.30 & 82.22 & 80.50 & 80.00 & 79.50 & 87.46 & 85.45 & 65.00 & \textbf{76.60} \\
Claude Opus 4.6    & 72.70 & 70.00 & 67.70 & 66.20 & 72.90 & 69.20 & 65.70 & 71.70 & 70.90 & 77.37 & 76.90 & 75.00 & 73.60 & 81.80 & 80.23 & 60.00 & \textbf{71.99} \\
Kimi K2.5         & 68.10 & 69.70 & 66.10 & 64.00 & 66.40 & 68.60 & 65.10 & 67.90 & 67.10 & 74.71 & 71.30 & 71.40 & 72.00 & 75.09 & 78.18 & 56.00 & \textbf{68.97} \\
DeepSeekv3.2         & 75.20 & 75.40 & 73.90 & 71.80 & 74.70 & 75.00 & 70.60 & 77.00 & 75.70 & 82.68 & 77.90 & 82.20 & 79.50 & 85.51 & 83.41 & 66.50 & \textbf{76.69} \\
Llama-3.1-8B     & 54.80 & 58.40 & 60.70 & 50.60 & 59.10 & 59.90 & 54.60 & 60.70 & 58.70 & 66.40 & 63.00 & 66.60 & 67.30 & 72.79 & 65.23 & 59.00 & \textbf{60.65} \\
Qwen-2.5-3B    & 53.80 & 57.10 & 57.30 & 49.10 & 56.10 & 57.30 & 49.10 & 59.20 & 53.90 & 63.39 & 60.40 & 67.50 & 63.90 & 67.49 & 62.73 & 53.50 & \textbf{57.99} \\
Qwen-2.5-7B    & 58.40 & 60.90 & 57.60 & 58.00 & 59.50 & 58.40 & 52.70 & 62.60 & 56.70 & 73.56 & 67.10 & 69.70 & 68.70 & 73.50 & 71.36 & 49.50 & \textbf{62.31} \\
Qwen-2.5-14B    & 71.80 & 74.50 & 71.40 & 68.00 & 72.50 & 72.10 & 67.30 & 74.40 & 71.40 & 82.45 & 76.90 & 78.90 & 79.70 & 83.22 & 81.59 & 66.00 & \textbf{74.36} \\
Qwen-2.5-32B    & 71.60 & 71.50 & 69.30 & 68.80 & 72.90 & 70.90 & 66.10 & 73.20 & 72.00 & 80.95 & 77.00 & 77.90 & 78.80 & 81.63 & 83.64 & 64.00 & \textbf{73.61} \\
Qwen-3-4B    & 47.10 & 48.90 & 54.10 & 45.60 & 52.70 & 53.80 & 44.40 & 54.30 & 47.20 & 67.09 & 59.10 & 63.20 & 62.90 & 69.26 & 65.00 & 47.50 & \textbf{54.62} \\
Qwen-3-30B-A3B    & 73.60 & 72.40 & 70.20 & 68.90 & 73.90 & 72.20 & 66.50 & 73.50 & 71.40 & 81.87 & 78.70 & 77.20 & 79.20 & 82.51 & 81.36 & 60.50 & \textbf{74.13} \\
\midrule
\multicolumn{18}{@{}l}{\textit{\textbf{Panel B: Base Models - Round 2 (Initial Fine-Tuning)}}} \\
\midrule
Llama-3.1-8B         & 27.80 & 29.00 & 29.00 & 23.10 & 29.10 & 30.80 & 26.10 & 32.10 & 27.30 & 33.03 & 30.00 & 38.00 & 35.20 & 39.75 & 28.64 & 29.50 & \textbf{30.35} \\
Qwen-2.5-3B        & 52.10 & 51.50 & 49.30 & 47.70 & 46.90 & 49.00 & 43.70 & 49.00 & 45.30 & 55.20 & 51.90 & 55.50 & 53.30 & 59.19 & 51.14 & 53.00 & \textbf{50.42} \\
Qwen-2.5-7B        & 64.60 & 65.00 & 63.00 & 63.00 & 63.40 & 64.60 & 60.60 & 65.70 & 61.30 & 74.02 & 66.40 & 71.00 & 70.00 & 74.03 & 67.50 & 62.50 & \textbf{65.86} \\
Qwen-2.5-14B        & 68.70 & 70.80 & 65.20 & 64.70 & 66.30 & 68.30 & 67.40 & 69.20 & 67.70 & 76.79 & 73.60 & 73.70 & 74.80 & 76.15 & 77.50 & 63.50 & \textbf{70.12} \\
Qwen-2.5-32B        & 75.20 & 75.90 & 73.10 & 73.80 & 74.50 & 73.30 & 73.80 & 75.50 & 74.20 & 81.99 & 80.90 & 79.40 & 79.40 & 82.69 & 81.59 & 72.00 & \textbf{76.54} \\
Qwen-3-4B        & 65.40 & 65.30 & 60.00 & 62.80 & 63.20 & 63.00 & 62.80 & 63.40 & 61.40 & 73.33 & 69.50 & 68.70 & 71.60 & 71.02 & 74.77 & 60.50 & \textbf{65.79} \\
Qwen-3-8B        & 70.80 & 71.70 & 68.60 & 68.50 & 68.00 & 72.00 & 68.50 & 73.60 & 68.30 & 77.71 & 75.20 & 75.10 & 75.70 & 79.51 & 79.55 & 67.50 & \textbf{72.25} \\
Qwen-3-14B        & 76.00 & 76.00 & 73.10 & 70.90 & 74.60 & 74.70 & 73.50 & 77.40 & 73.10 & 82.68 & 79.50 & 79.50 & 78.90 & 82.16 & 81.59 & 72.50 & \textbf{76.44} \\
Qwen-3-32B        & 76.90 & 76.50 & 75.70 & 71.50 & 75.90 & 77.30 & 73.60 & 76.50 & 74.90 & 82.45 & 78.70 & 81.60 & 80.50 & 84.10 & 80.68 & 72.00 & \textbf{77.35} \\
\midrule
\multicolumn{18}{@{}l}{\textit{\textbf{Panel C: Base Models - Round 3 (Iterative Knowledge Injection \& Scaffolding)}}} \\
\midrule
Llama-3.1-8B         & 59.60 & 65.90 & 59.80 & 56.80 & 55.60 & 61.70 & 58.90 & 63.20 & 60.80 & 68.71 & 66.00 & 72.10 & 68.40 & 69.79 & 64.32 & 53.50 & \textbf{63.02} \\
Qwen-2.5-3B        & 65.20 & 67.10 & 65.50 & 61.70 & 61.40 & 65.60 & 60.60 & 67.30 & 64.50 & 74.71 & 70.30 & 75.30 & 76.10 & 77.03 & 76.14 & 63.00 & \textbf{67.87} \\
Qwen-2.5-7B        & 71.20 & 69.60 & 68.20 & 66.30 & 69.80 & 67.30 & 64.90 & 68.20 & 68.70 & 76.44 & 74.30 & 75.00 & 76.40 & 78.09 & 75.45 & 63.50 & \textbf{70.79} \\
Qwen-2.5-14B        & 76.00 & 76.00 & 74.90 & 72.10 & 73.00 & 74.50 & 72.00 & 75.20 & 75.30 & 80.48 & 78.70 & 80.10 & 79.50 & 83.22 & 81.82 & 68.00 & \textbf{76.30} \\
Qwen-2.5-32B        & 77.40 & 77.80 & 74.20 & 73.00 & 76.50 & 77.30 & 75.30 & 80.50 & 76.20 & 84.64 & 82.50 & 82.40 & 81.80 & 86.22 & 86.59 & 71.50 & \textbf{78.84} \\
Qwen-3-4B        & 66.00 & 66.20 & 63.60 & 62.00 & 63.10 & 65.60 & 60.60 & 67.80 & 64.80 & 75.87 & 70.60 & 72.50 & 73.80 & 76.68 & 74.77 & 53.50 & \textbf{67.46} \\
Qwen-3-8B        & 73.70 & 74.20 & 71.60 & 70.90 & 73.50 & 75.50 & 70.50 & 75.90 & 73.80 & 82.45 & 78.80 & 80.70 & 80.80 & 84.98 & 82.50 & 66.50 & \textbf{75.97} \\
Qwen-3-14B        & 75.40 & 75.80 & 74.70 & 70.00 & 73.50 & 77.60 & 72.10 & 77.80 & 76.00 & 83.26 & 80.60 & 81.30 & 81.70 & 83.92 & 85.00 & 65.50 & \textbf{77.21} \\
Qwen-3-32B        & 77.40 & 79.30 & 77.50 & 72.40 & 77.60 & 79.60 & 75.40 & 79.30 & 79.00 & 84.87 & 81.60 & 84.20 & 82.20 & 86.93 & 85.23 & 67.00 & \textbf{79.52} \\
\midrule
\multicolumn{18}{@{}l}{\textit{\textbf{Panel D: Performance Gain ($\Delta$ Round 3 - Round 2)}}} \\
\midrule
Llama-3.1-8B         & +31.80 & +36.90 & +30.80 & +33.70 & +26.50 & +30.90 & +32.80 & +31.10 & +33.50 & +35.68 & +36.00 & +34.10 & +33.20 & +30.04 & +35.68 & +24.00 & \textbf{+32.67} \\
Qwen-2.5-3B        & +13.10 & +15.60 & +16.20 & +14.00 & +14.50 & +16.60 & +16.90 & +18.30 & +19.20 & +19.51 & +18.40 & +19.80 & +22.80 & +17.84 & +25.00 & +10.00 & \textbf{+17.45} \\
Qwen-2.5-7B        & +6.60 & +4.60 & +5.20 & +3.30 & +6.40 & +2.70 & +4.30 & +2.50 & +7.40 & +2.42 & +7.90 & +4.00 & +6.40 & +4.06 & +7.95 & +1.00 & \textbf{+4.93} \\
Qwen-2.5-14B        & +7.30 & +5.20 & +9.70 & +7.40 & +6.70 & +6.20 & +4.60 & +6.00 & +7.60 & +3.69 & +5.10 & +6.40 & +4.70 & +7.07 & +4.32 & +4.50 & \textbf{+6.18} \\
Qwen-2.5-32B        & +2.20 & +1.90 & +1.10 & -0.80 & +2.00 & +4.00 & +1.50 & +5.00 & +2.00 & +2.65 & +1.60 & +3.00 & +2.40 & +3.53 & +5.00 & -0.50 & \textbf{+2.30} \\
Qwen-3-4B        & +0.60 & +0.90 & +3.60 & -0.80 & -0.10 & +2.60 & -2.20 & +4.40 & +3.40 & +2.54 & +1.10 & +3.80 & +2.20 & +5.66 & +0.00 & -7.00 & \textbf{+1.67} \\
Qwen-3-8B        & +2.90 & +2.50 & +3.00 & +2.40 & +5.50 & +3.50 & +2.00 & +2.30 & +5.50 & +4.74 & +3.60 & +5.60 & +5.10 & +5.47 & +2.95 & -1.00 & \textbf{+3.72} \\
Qwen-3-14B        & -0.60 & -0.20 & +1.60 & -0.90 & -1.10 & +2.90 & -1.40 & +0.40 & +2.90 & +0.58 & +1.10 & +1.80 & +2.80 & +1.76 & +3.41 & -7.00 & \textbf{+0.77} \\
Qwen-3-32B        & +0.50 & +2.80 & +1.80 & +0.90 & +1.70 & +2.30 & +1.80 & +2.80 & +4.10 & +2.42 & +2.90 & +2.60 & +1.70 & +2.83 & +4.55 & -5.00 & \textbf{+2.17} \\
\bottomrule
\end{tabular}
}
\end{table*}

\subsection{Debug: traceable repair and the value of structure}

We now test whether the shared knowledge enables targeted repair of these errors. The Debugger analyzes every incorrect V1 response, classifies each failure as a concept gap or a capability deficit, and traces the root cause to specific L1/L2/L3 knowledge nodes. The Synthesizer then generates targeted training instances anchored to these nodes. We fine-tune each V1 model on the combined original corpus plus the diagnostic patches to produce V2 models, corresponding to Panel C in Table~\ref{tab:full_experimental_results}. 

\paragraph{Systematic gains on ProDa-16.}

Panel D reports the accuracy change from V1 to V2 across all nine models and 16 disciplines. Every model improves on average. Gains range from +0.77 points for Qwen-3-14B to +32.67 points for Llama-3.1-8B, and the magnitude of improvement is inversely related to V1 performance: models with lower starting points receive larger gains, consistent with the expectation that weaker models harbor more diagnosable knowledge gaps. Llama-3.1-8B is the most dramatic case. This model scored only 30.35\% at V1 due to its inability to follow the evaluation format, but reaches 63.02\% at V2, surpassing its official Instruct counterpart at 60.65\% for the first time. The diagnostic patches therefore address not only factual gaps but also the structured response patterns that the base model lacked. Among the Qwen-2.5 family, all four scales improve: +17.45 at 3B, +4.93 at 7B, +6.18 at 14B, and +2.30 at 32B. At the top end, Qwen-2.5-32B-V2 reaches 78.84\% and Qwen-3-32B-V2 reaches 79.52\%. Both scores exceed every Instruct model in Table~\ref{tab:full_experimental_results}.

\noindent\textbf{Preservation of general capabilities.} A targeted fine-tuning intervention risks degrading the model's pre-existing general knowledge, a concern commonly referred to as catastrophic forgetting. We evaluate Base, V1, and V2 checkpoints on discipline-relevant subsets of two established benchmarks: 12 subsets of MMLU and 6 subsets of C-Eval (Table~\ref{tab:benchmark_results}). We select subsets that overlap with ProDa-16's disciplinary coverage to ensure the comparison measures knowledge regions where interference is most likely. Two patterns emerge. First, V1 models show a small but consistent accuracy decrease relative to their Base checkpoints on both benchmarks. On MMLU, the median V1 drop is 0.48 points; on C-Eval it is 0.41 points. This indicates that the initial domain-focused injection introduces a modest general-capability tax. Second, the V2 debugging step recovers nearly all of this loss. On MMLU, seven of nine V2 models match or exceed their Base scores, and the median Base-to-V2 change is +0.27 points. These results indicate that the diagnostic patching mechanism not only repairs domain-specific errors on ProDa-16 but also restores the general-capability balance disrupted by the first-pass compilation.

\begin{table}[h]
\centering
\small
\caption{\textbf{Model Performance on MMLU and C-Eval Subsets.} Comparison of Base, Model V1 (first-pass compilation), and Model V2 (targeted repair) across various models. All scores represent accuracy (\%).}
\label{tab:benchmark_results}
\setlength{\tabcolsep}{15pt}
\begin{tabular}{@{}lcccccc@{}}
\toprule
\multirow{2}{*}{\textbf{Model}} & \multicolumn{3}{c}{\textbf{MMLU (12 Subsets)}} & \multicolumn{3}{c}{\textbf{C-Eval (6 Subsets)}} \\ \cmidrule(lr){2-4} \cmidrule(l){5-7} 
                                                & \textbf{Base}  & \textbf{Model V1} & \textbf{Model V2} & \textbf{Base}  & \textbf{Model V1} & \textbf{Model V2} \\ \midrule
Llama3.1-8B                                     & 66.20          & 64.68          & 66.21          & 60.64          & 49.54          & 50.23          \\ \hline
Qwen2.5-3B                                      & 70.16          & 68.26          & 69.11          & 75.45          & 75.25          & 76.97          \\
Qwen2.5-7B                                      & 76.17          & 75.22          & 76.44          & 83.18          & 82.70          & 83.38          \\
Qwen2.5-14B                                     & 80.18          & 78.76          & 80.34          & 86.69          & 86.28          & 86.70          \\
Qwen2.5-32B                                     & 82.50          & 82.87          & 83.18          & 92.21          & 92.21          & 92.28          \\ \hline
Qwen3-4B                                        & 72.27          & 72.58          & 72.85          & 74.84          & 74.42          & 75.18          \\
Qwen3-8B                                        & 76.12          & 75.65          & 77.49          & 80.35          & 79.87          & 80.56          \\
Qwen3-14B                                       & 78.86          & 78.44          & 79.39          & 84.97          & 84.83          & 85.87          \\
Qwen3-32B                                       & 82.29          & 81.81          & 82.39          & 87.66          & 87.52          & 87.93          \\ \bottomrule
\end{tabular}
\end{table}

\paragraph{Comparison with baseline synthesis methods.}

\begin{figure}[h]
  \centering
  % \vspace{-2mm}
  \includegraphics[width=\textwidth]{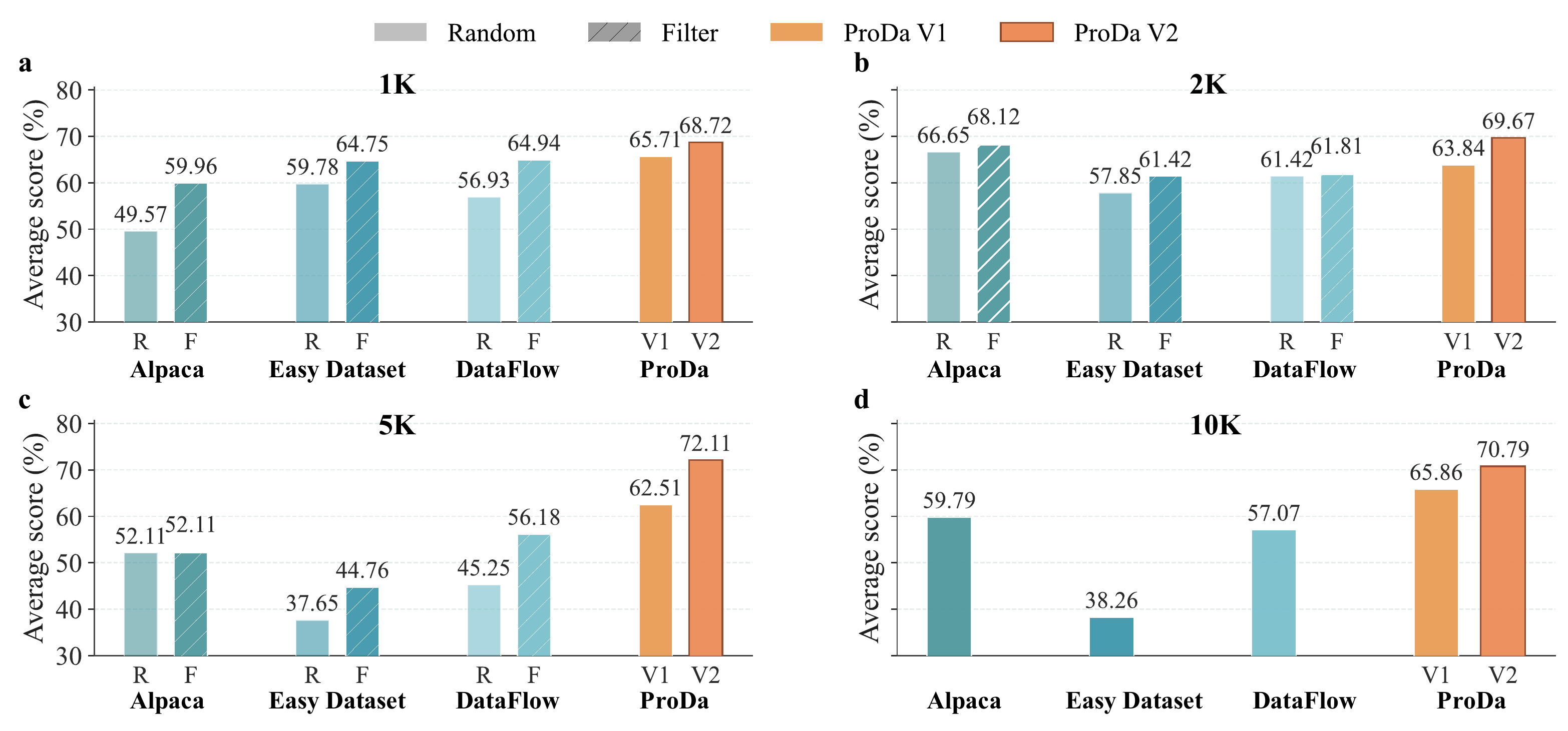}
  % \vspace{-2mm}
  \caption{\textbf{Performance comparison of data synthesis methods.} Average benchmark scores of Qwen2.5-7B fine-tuned on data generated by Alpaca, EasyDataset, DataFlow, and our ProDa framework across 1K--10K data scales. \textit{R} and \textit{F} denote random sampling and heuristic filtering baselines. ProDa V2, leveraging closed-loop diagnostic repair, exhibits exceptional sample efficiency at 1K and consistently outperforms all conventional methods across all scales.}
  % \vspace{-2mm}
  \label{fig:scaling_performance}
\end{figure}

To isolate the contribution of diagnostic targeting from the effect of additional data volume, we compare ProDa against three baseline synthesis methods at matched data scales. Alpaca, EasyDataset, and DataFlow each generate 1K, 2K, 5K, and 10K instances per discipline, and all methods are evaluated on the same Qwen-2.5-7B backbone (Figure~\ref{fig:scaling_performance}).

\begin{table}[h]
\centering
\small
\caption{Comparison of features across different data generation methods. \textit{Specification} indicates whether the data is processed into knowledge points; \textit{Traceability} indicates whether each training or evaluation instance can be traced back to specific knowledge points; \textit{Debugging loop} indicates whether a closed-loop debugging mechanism is supported.}
\label{tab:method_comparison}
\begin{tabular*}{\textwidth}{@{\extracolsep{\fill}} l m{0.36\textwidth} c c c}
\toprule
Method & \multicolumn{1}{c}{Features} & Specification & Traceability & Debugging Loop \\
\midrule
Alpaca~\cite{taori2023alpaca}       & Self-instruct generation from LLM             & \xmark & \xmark & \xmark \\
EasyDataset~\cite{miao2025easy}     & Direct generation from documents                & \xmark & \xmark & \xmark \\
DataFlow~\cite{liang2025dataflow}   & Data generation pipelines                      & \cmark & \xmark & \xmark \\
ProDa (V1) & Specification and implementation & \cmark & \cmark & \xmark \\
ProDa (V2)      & Complete closed loop                               & \cmark & \cmark & \cmark \\
\bottomrule
\vspace{-4mm}
\end{tabular*}
\end{table}

ProDa consistently establishes state-of-the-art performance across all evaluated scales, fundamentally circumventing the scaling stagnation and capability collapse inherent to conventional methods. Most notably, our framework exhibits exceptional \textit{sample efficiency}. Fine-tuning on merely 1K targeted repair samples (ProDa V2) yields an average score of 68.72\%. This effectively transcends the volumetric data barrier, strictly outperforming the absolute peak performance of all baseline methods regardless of their data scale (e.g., Alpaca's maximum of 68.12\% at the 2K scale).

Furthermore, as the data scale expands, the integration of the Debugger drives a definitive performance leap. At the 5K scale, ProDa V2 achieves a peak score of 72.11\%, surpassing the strongest baseline counterpart (DataFlow Filter, 56.18\%) by a massive absolute margin of nearly 16\%. This stark quantitative contrast explicitly validates a core conclusion: for enhancing the fundamental capabilities of large language models, high-quality, diagnostic-driven targeted data is fundamentally superior to blindly scaling up conventional data synthesis pipelines.

\subsection{Case studies}

To elucidate how the ProDa closed-loop debugging framework translates black-box model failures into transparent, actionable repair paths, we extract three representative intervention case studies from the benchmark. Spanning formal-reasoning physics, normative economics, and fact-intensive biomedicine, these cases encapsulate the two most intractable categories of systemic failure in model fine-tuning: concept gaps and capability deficits.

\begin{figure}[!htbp]
  \centering
  \includegraphics[width=\textwidth]{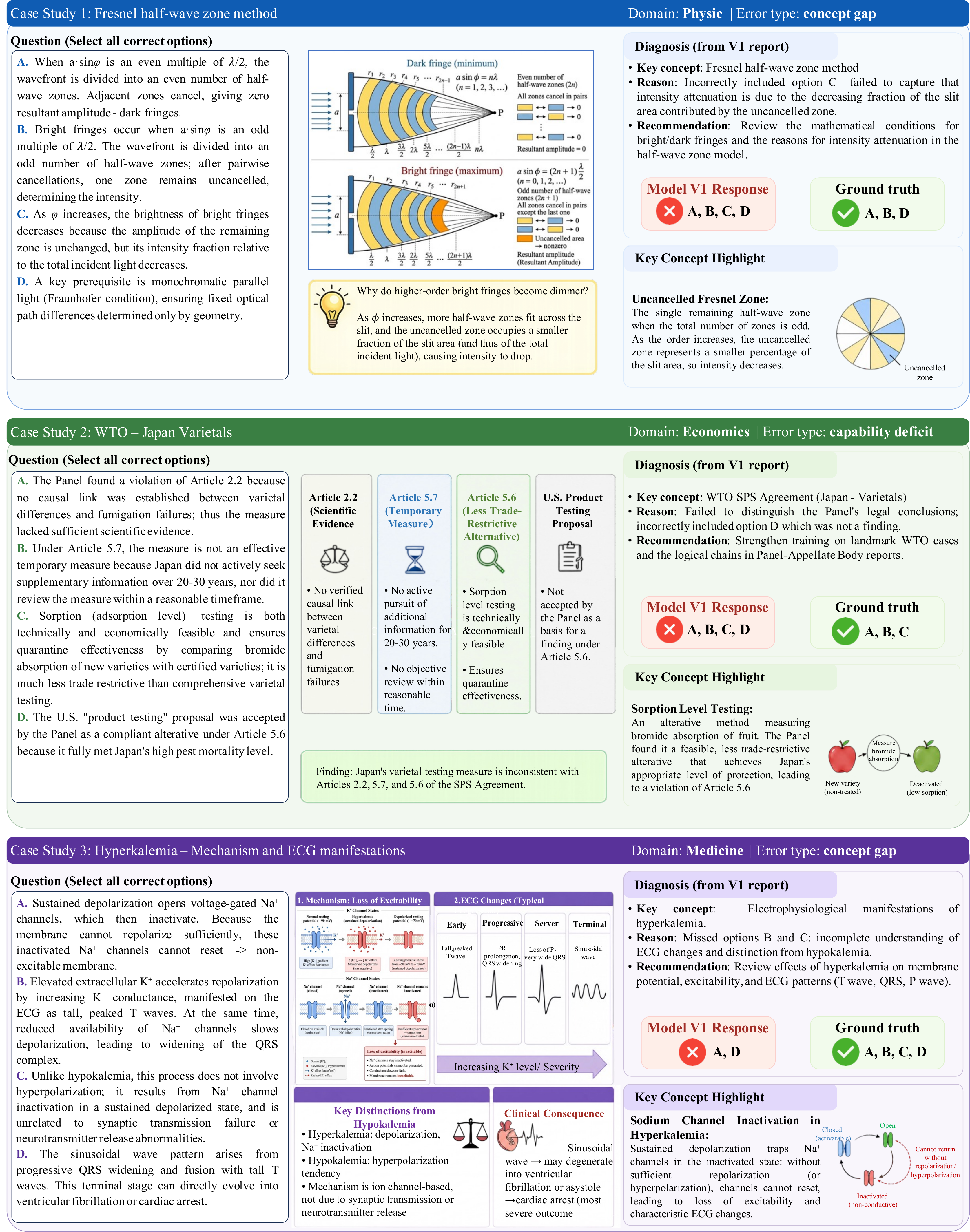}
  \caption{\textbf{Three diagnostic-repair case studies.}
  Each row shows the question (left), the relevant knowledge
  structure (centre), and the diagnostic report with V1/V2
  responses (right). Cases span Physics (concept gap),
  Economics (capability deficit), and Medicine (concept gap).
  In all cases the V2 model corrects the V1 error after
  training on patches anchored to the diagnosed knowledge nodes.}
  \label{fig:case_study}
\end{figure}

\paragraph{Case 1: Rectifying physical intuition in Optics.} Formal reasoning in physics demands not merely formulaic memorization, but an accurate geometric intuition of dynamic variables (a visual walkthrough of this specific diagnostic-repair loop is detailed in Figure~\ref{fig:case_study}). When evaluating the bright and dark fringe formation using the Fresnel half-wave zone method, the V1 model hallucinated by endorsing a physical distractor that conflated the geometric root causes of intensity attenuation. The Debugger recognized this as a \texttt{concept\_gap} and anchored the failure to the L1 concept \textit{Uncancelled Fresnel Zone}, specifically isolating the flawed L2 logic regarding how the uncancelled zone's fractional area dictates intensity drops. By synthesizing corrective data that quantitatively compared the energy distribution of different wave zones, the system reconstructed the model's geometric intuition of the ``zone area ratio.'' Consequently, the V2 model successfully eliminated the hallucinated distractor, achieving a profound alignment with the underlying geometric and physical laws.

\paragraph{Case 2: Reconstructing legal logic chains in Economics and Law.} In normative sciences intersecting international law and economics, such as the WTO SPS Agreement, models frequently succumb to logical disorientation amidst verbose judicial rulings. Evaluated on the WTO Panel report regarding Japan's ``Varietal Testing'' measure, the V1 model failed to differentiate between unverified proposals and final judicial logic, erroneously accepting an unadopted alternative proposed by the United States as the Panel's definitive ruling. Diagnosed as a \texttt{capability\_deficit}, this logical fracture was pinpointed to the L1 concept \textit{Sorption Level Testing} and its L2 judicial chain, which explicitly established it as the less restrictive alternative triggering an Article 5.6 violation. To address this legal drift, the Synthesizer generated jurisprudence-focused learning samples emphasizing the ``three-pronged test'' to enforce a rigorous argumentation scaffold. Empowered by this reconstructed logical chain, the V2 model precisely stripped away the non-official conclusions, yielding an accurate legal interpretation.

\paragraph{Case 3: Concept-level targeted reinforcement in Biomedicine.} In fact-intensive domains, the omission of granular knowledge often precipitates critical reasoning errors. For instance, when tasked with identifying the mechanisms underlying the loss of cellular excitability induced by hyperkalemia, the V1 model exhibited an incomplete cognitive representation. While it recognized superficial electrocardiographic symptoms, it fatally omitted the underlying sodium channel inactivation mechanism. Utilizing our hierarchical graph, the Debugger classified this failure as a \texttt{concept\_gap} and precisely traced it to the L1 concept \textit{Sodium Inactivation} and its governing L2 proposition: ``Lack of membrane hyperpolarization prevents inactivated sodium channels from resetting.'' To rectify this, the Synthesizer generated targeted patch entries forcing the model to internalize the biophysical constraints of the inactivation gate. Following this focused injection, the V2 model successfully bridged its knowledge blind spot, producing a comprehensively correct analysis of the electrophysiological mechanisms and aligning perfectly with the ground truth.

All three cases follow the same closed-loop path: incorrect option identification, diagnostic classification with L1/L2/L3 grounding, targeted patch generation, and verified correction at V2. The tracing granularity enabled by the shared knowledge structure is what distinguishes this repair mechanism from generic data augmentation.

\subsection{ProDa Studio: an IDE for any raw corpora}

\begin{figure}[h]
  \centering
  % \vspace{-2mm}
  \includegraphics[width=\textwidth]{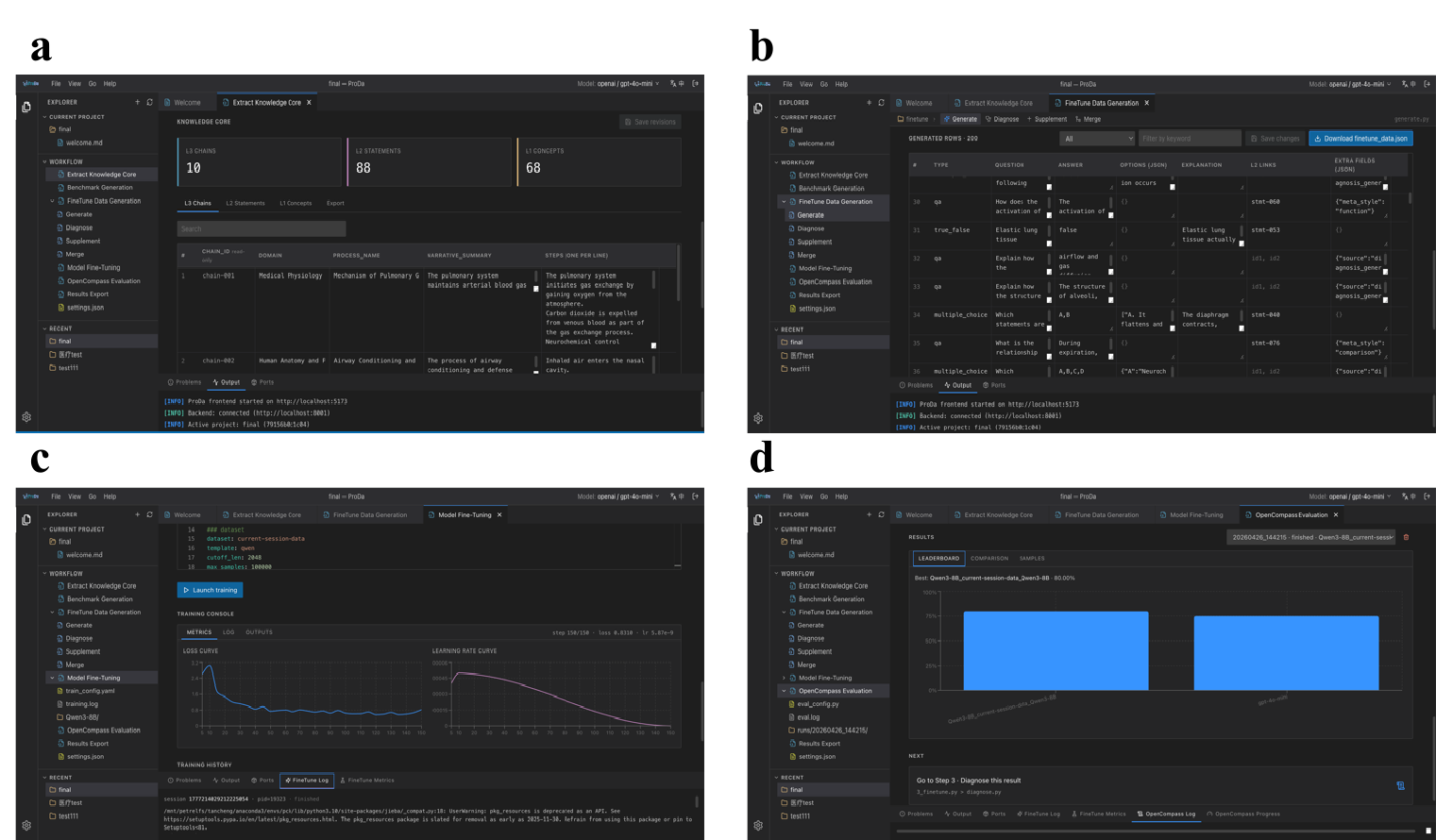}
  \caption{\textbf{The ProDa Studio integrated development environment.} \textbf{a,} Knowledge extraction interface showing L3 chains,
  L2 statements, and L1 concepts.
  \textbf{b,} Data generation interface displaying individual
  training instances with their type, source chain, linked
  knowledge nodes, and generation metadata.
  \textbf{c,} Fine-tuning console with real-time loss and
  learning rate monitoring.
  \textbf{d,} Evaluation dashboard with per-discipline scores
  and a link to initiate the next diagnostic cycle.}
  \label{fig:proda_studio}
\end{figure}

To instantiate the \textit{Programming with Data} paradigm as a systematic engineering practice, we developed ProDa Studio, an integrated development environment (IDE) that encapsulates the full ProDa pipeline into a single interactive platform (Figure~\ref{fig:proda_studio}). The environment is designed to make the extraction, synthesis, training, and debugging stages executable in sequence without switching between disconnected scripts or manual file management.

The left sidebar organizes the pipeline as a linear workflow: Extract Knowledge Core, Benchmark Generation, FineTune Data Generation (with sub-steps for Generate, Diagnose, Supplement, and Merge), Model Fine-Tuning, and Evaluation. Each step reads the output of its predecessor and writes structured artifacts to a shared project directory, ensuring full provenance from raw corpus to final evaluation score. Figure~\ref{fig:proda_studio}a shows the knowledge extraction interface, where users inspect the extracted L3 chains, L2 statements, and L1 concepts, and Figure~\ref{fig:proda_studio}b shows the data generation interface, where each training instance is displayed with its type, source chain, linked L2 statements, and L1 concepts.

The training and evaluation stages are similarly integrated. Figure~\ref{fig:proda_studio}c shows the fine-tuning console with real-time loss and learning rate curves, and Figure~\ref{fig:proda_studio}d shows the evaluation dashboard with per-discipline scores from OpenCompass and a direct link to initiate the next diagnostic cycle. This last feature closes the loop: after reviewing evaluation results, the user can trigger the Debugger on the current model's errors and proceed to the next V1-to-V2 iteration without leaving the environment.

\section{Methods}

\subsection{Knowledge structure extraction}
\label{sec:methods_extraction}

The source collection comprises textbook-level documents spanning 16 disciplines across natural sciences, engineering, biomedical sciences and social sciences, selected for academic authority, disciplinary diversity and contamination control. Documents are classified by academic level and reasoning type, then segmented and scored to yield chunks that are both academically substantive and inferentially rich; curation details and retention statistics are provided in Supplementary Information~\ref{app:prodalib_corpus}.

From the selected chunks we extract the three-level knowledge structure $\mathcal{K} = (\mathcal{K}_1, \mathcal{K}_2, \mathcal{K}_3)$ that serves as the shared specification for all downstream artifacts. Consistent with the CORE \textit{Organized} standard, $\mathcal{K}$ stratifies domain knowledge into three levels of increasing compositional complexity.

\textbf{L1 Key Concepts}
$\mathcal{K}_1 = \{e_1, e_2, \ldots\}$ is a set of atomic domain concepts, where each $e_i = (\textit{term}_i, \textit{type}_i, \textit{def}_i, \textit{src}_i)$ carries a canonical term, a category type, a concise definition grounded in its source text, and a provenance link to the source chunk.

\textbf{L2 Knowledge Relations}
$\mathcal{K}_2 = \{r_1, r_2, \ldots\}$ is a set of typed relational triples, where each $r_j = (e_s, \phi_j, e_o)$ connects a subject concept $e_s \in \mathcal{K}_1$ and an object concept $e_o \in \mathcal{K}_1$ through a relation type $\phi_j \in \Phi$, with $\Phi = \{\textsc{specialization}, \textsc{causal}, \textsc{prerequisite}, \textsc{contrast}, \ldots\}$.

\textbf{L3 Reasoning Chains}
$\mathcal{K}_3 = \{g_1, g_2, \ldots\}$ is a set of multi-step inferential pathways, where each $g_k = (e_1 \xrightarrow{\phi_{1,2}} e_2 \xrightarrow{\phi_{2,3}} \cdots \xrightarrow{\phi_{T-1,T}} e_T)$ is an ordered sequence of L1 concepts $e_t \in \mathcal{K}_1$ connected by L2 relations $\phi_{t,t+1}$, with each transition annotated by the nature of the inferential step.

Extraction proceeds top-down through three stages, each performed by a strong language model with level-specific prompts (Supplementary Information~\ref{app:prodalib_key_concept}). First, L3 reasoning chains are extracted directly from each selected chunk. Each chain encodes a multi-step inferential pathway and is decomposed into discrete steps annotated by the concepts invoked and the nature of each inferential transition. Second, each L3 chain $g_k$ of length $T$ is decomposed into $T{-}1$ L2 triples via a sliding-window over adjacent steps, producing triples of the form $(v_t, \phi_{t,t+1}, v_{t+1})$. Contextualization is enforced: if a step contains a pronoun or vague referent, it is resolved to the specific antecedent from preceding steps before the triple is formed. Third, L1 concepts are harvested from the subjects and objects of all L2 triples, with cross-chunk occurrences merged through string normalization and semantic deduplication.

This top-down extraction order is a deliberate design choice that guarantees a structural property essential for debuggability:
\begin{equation}
    \forall\, e \in \mathcal{K}_1,\; \exists\, g \in \mathcal{K}_3 \;\text{s.t.}\; e \in \mathrm{nodes}(g); \qquad
    \forall\, r \in \mathcal{K}_2,\; \exists\, g \in \mathcal{K}_3 \;\text{s.t.}\; r \in \mathrm{edges}(g),
    \label{eq:reachability}
\end{equation}
where $\mathrm{nodes}(g)$ and $\mathrm{edges}(g)$ denote the concepts and relations traversed by chain $g$. This reachability property eliminates orphan entries that would be untestable by the benchmark and therefore undebuggable by the pipeline. The conventional bottom-up order does not provide this guarantee, because entities and relations extracted in isolation may never participate in any testable reasoning pattern. Extraction prompts and cross-discipline examples are provided in Supplementary Information~\ref{app:cross-discipline_examples}.

\subsection{Benchmark and training data synthesis}
\label{sec:methods_synthesis}

The shared knowledge structure $\mathcal{K}$ supports the synthesis of two complementary artifacts: a benchmark $\mathcal{B}$ that defines target capabilities and a training corpus $\mathcal{S}$ that encodes domain knowledge. Consistent with the test-first principle established in \S\ref{sec:pipeline}, we describe benchmark construction first.

\paragraph{Benchmark construction from L3 chains.}

For each L3 chain $g_k \in \mathcal{K}_3$, a benchmark synthesis function produces an item $b_k = (x_k, \mathcal{A}_k, y_k, \mu_k)$, where $x_k$ is the question stem, $\mathcal{A}_k$ is the option set, $y_k \subseteq \mathcal{A}_k$ is the correct answer subset, and $\mu_k = (\textit{chain\_id}, \{r_j\}, \{e_i\})$ is structural metadata linking the item to its source chain, constituent L2 triples, and L1 concepts. Items are formatted as multiple-choice questions, predominantly multi-select, requiring the model to evaluate the correctness of each reasoning step independently rather than relying on elimination.

The CORE \textit{Rigorous} standard imposes two requirements on benchmark construction. First, each item must contain adversarial distractors generated from the knowledge structure itself. We employ three perturbation operators:
\begin{align}
    \textsc{SubstAdj}&: \quad \text{replace } e_i \in \mathrm{nodes}(g_k) \text{ with } e_i' \in \mathcal{N}(e_i), \label{eq:perturb_subst} \\
    \textsc{InvRel}&: \quad \text{replace } \phi_{t,t+1} \text{ with } \bar{\phi}_{t,t+1} \text{ (semantic inverse)}, \label{eq:perturb_inv} \\
    \textsc{Trunc}&: \quad \text{truncate } g_k \text{ at step } t < T, \text{ yielding an incomplete conclusion}, \label{eq:perturb_trunc}
\end{align}
where $\mathcal{N}(e_i) \subset \mathcal{K}_1$ denotes the set of concepts semantically adjacent to $e_i$ (sharing at least one L2 relation type) and $\bar{\phi}$ denotes the semantic inverse of relation $\phi$ (for example, ``promotes'' $\to$ ``inhibits''). These operators ensure that correct responses demand discrimination between closely related concepts rather than elimination of implausible options.

Second, benchmark items and training samples must maintain instance-level orthogonality. Because benchmark items are constructed from L3 chains while training data is synthesized from L1 and L2 entries, the two artifact sets are structurally separated:
\begin{equation}
    \mathcal{B} = f_{\mathrm{bench}}(\mathcal{K}_3), \qquad \mathcal{S} = f_{\mathrm{syn}}(\mathcal{K}_1, \mathcal{K}_2).
    \label{eq:orthogonality}
\end{equation}
No benchmark item is answerable by verbatim recall of a training sample, because answering any $b_k$ requires composing multiple L1/L2 entries along the specific inferential pathway encoded in the source L3 chain---a composition not present in any individual training sample. The knowledge-level connection through $\mathcal{K}$ enables diagnostic traceability; the instance-level separation ensures that benchmark performance reflects genuine capability. Benchmark construction prompts and distractor generation details are provided in Supplementary Information~S3.

\paragraph{Initial training data synthesis from L1 and L2 entries.}

For each discipline, a sliding window selects batches of L2 triples along with their associated L1 definitions. Each batch is passed to a synthesis model that generates training samples in three formats: \textit{open-ended} questions requiring explanation of mechanisms and definitions, \textit{multiple-choice} items testing relational knowledge with plausible distractors, and \textit{true-false judgments} probing boundary conditions and common misconceptions---at a prescribed ratio that emphasizes open-ended reasoning. The CORE \textit{Contextualized} standard is enforced by supplying the synthesis model with the full L2 context and parent L1 definitions, ensuring that generated samples preserve the dependencies present in the source text. Each sample $s_i$ carries metadata $(\textit{l2\_ids}, \textit{l1\_ids})$ preserving full traceability to $\mathcal{K}$. The initial training corpus constitutes the first version of the source code to be compiled. Synthesis prompts and per-discipline statistics are provided in Supplementary Information~\ref{app:data_synthesis}.

\subsection{Failure diagnosis and data repair}
\label{sec:methods_debugging}

After the model $M_{\mathrm{ft}} = f_{\mathrm{train}}(M_{\mathrm{base}}, \mathcal{S})$ is compiled and evaluated against $\mathcal{B}$, the Debugger processes the error set $\mathcal{E} = \{b_k \in \mathcal{B} : M_{\mathrm{ft}}(x_k) \neq y_k\}$ to produce diagnostic reports and targeted data patches.

\paragraph{Diagnostic classification.}

For each error $b_k \in \mathcal{E}$, the Debugger receives the question $x_k$, the model's prediction $\hat{y}_k$, the correct answer $y_k$, and the structural metadata $\mu_k$. An LLM judge classifies the error into one of two categories:
\begin{itemize}[leftmargin=2em]
    \item A \textit{concept gap} indicates that $M_{\mathrm{ft}}$ lacks or confuses a specific piece of domain knowledge: $\exists\, e \in \mathrm{nodes}(g_k)$ or $\exists\, r \in \mathrm{edges}(g_k)$ that is insufficiently represented in $\mathcal{S}$.
    \item A \textit{reasoning deficit} indicates that $M_{\mathrm{ft}}$ possesses the requisite knowledge components but fails to compose them correctly: the individual L1/L2 entries along $g_k$ are adequately represented in $\mathcal{S}$, but the compositional pattern of $g_k$ is not.
\end{itemize}
Each diagnosis identifies the core concept or reasoning step involved, provides a natural-language explanation of the failure mechanism, and specifies a repair recommendation. The diagnostic prompt and a complete example report are provided in Supplementary Information~\ref{app:diagnostic}.

\paragraph{Patch generation.}

Each diagnosis drives the generation of a targeted repair batch, with the strategy conditioned on the error type:
\begin{equation}
    \mathcal{S}^{\mathrm{patch}}_k =
    \begin{cases}
        f_{\mathrm{refine}}(\kappa_k,\; \mathcal{N}(\kappa_k),\; \mathcal{K}) & \text{if concept gap at } \kappa_k \in \mathcal{K}_1, \\[6pt]
        f_{\mathrm{cot}}(g_k,\; \mathcal{K}) & \text{if reasoning deficit along } g_k \in \mathcal{K}_3,
    \end{cases}
    \label{eq:patch_gen}
\end{equation}
where $f_{\mathrm{refine}}$ is a knowledge reinforcement function that produces samples explicitly contrasting the confused concept $\kappa_k$ with its semantically adjacent alternatives $\mathcal{N}(\kappa_k)$, providing precise definitions, distinguishing attributes, and contrastive examples. The function $f_{\mathrm{cot}}$ is a chain-of-thought scaffolding function that decomposes the failed reasoning pathway into explicit intermediate steps, with each step justified by reference to the relevant L1 and L2 knowledge. Both functions generate samples in the same three-format mixture as the initial synthesis. 

\paragraph{Data mixing and replay.}

The repair corpus for each debugging cycle combines newly generated patches with a replay subset of the original training data. For each discipline $d$, the number of repair samples is proportional to that discipline's share of total errors, concentrating debugging effort where failures are most prevalent. To prevent catastrophic forgetting of previously acquired capabilities, the replay subset is drawn from the initial training data under a diversity constraint:
\begin{equation}
    \mathrm{L2\_ids}\bigl(\mathcal{S}^{\mathrm{replay},(d)}\bigr) \;\cap\; \mathrm{L2\_ids}\bigl(\mathcal{S}^{\mathrm{patch},(d)}\bigr) = \varnothing,
    \label{eq:replay_disjoint}
\end{equation}
ensuring that replayed samples cover knowledge regions complementary to those targeted by the patches rather than redundantly reinforcing the same entries. The augmented corpus
\begin{equation}
    \mathcal{S}' = \bigcup_{d=1}^{D} \bigl(\mathcal{S}^{\mathrm{patch},(d)} \;\cup\; \mathcal{S}^{\mathrm{replay},(d)}\bigr)
    \label{eq:augmented_corpus}
\end{equation}
is scaled to match the volume of the initial training data, and the model is retrained from the same base checkpoint: $M_{\mathrm{ft}}' = f_{\mathrm{train}}(M_{\mathrm{base}}, \mathcal{S}')$, completing the debugging loop described in \S\ref{sec:pipeline}. The \textit{Evolving} standard of the CORE principle is thus operationalized: training data is not a static artifact but evolves through empirically driven iteration in which each benchmark failure generates a traceable data repair. The mixing procedure is detailed in Supplementary Information~\ref{app:data_mixing}.

\section{Related Work}

\subsection{Data Synthesis for LLMs}
% Context (现状): 介绍合成数据在大模型训练中的统治地位。
% 提及 "Textbooks Are All You Need" (Phi series) 证明了高质量合成数据可以训练出小而强的模型。
% 提及 Self-Instruct 和 WizardLM (Evol-Instruct)，它们通过种子指令进化，展示了 LLM 自生成数据的潜力。
% Despite (肯定): 尽管这些方法成功地降低了对人工标注的依赖，并显著提升了模型的通用指令遵循能力（General Instruction Following）。
% However (Gap & Critique):
% Lack of Domain Rigor: 现有的合成方法大多侧重于“多样性”或“难度增加”，缺乏针对特定垂直领域（Vertical Domains）的严谨性与正确性保证。它们往往是“幻觉的放大器”。
% Open-loop Generation: 绝大多数合成过程是开环的（Open-loop）。数据是一次性生成的，生成后就扔进模型训练，一旦模型表现不佳，没有机制去回溯修正数据。
% We 引入了 CORE 原则，将合成数据视为需要符合 Spec（原始语料）的“源代码”，并通过 Debug 闭环保证质量。

Synthetic data generation has emerged as a dominant paradigm in training LLMs~\cite{honovich2023unnatural,koksal2024longform,mukherjee2023orca}, fundamentally shifting the focus from data quantity to data quality. Early works such as Self-Instruct~\cite{wang2023self} and Alpaca~\cite{taori2023alpaca} pioneered the approach of bootstrapping off-the-shelf LLMs to generate diverse instruction-following data. WizardLM~\cite{xu2023wizardlm} introduced Evol-Instruct to iteratively complicate instructions, while Orca~\cite{mukherjee2023orca} showed that learning from the rich explanation traces of stronger teachers significantly boosts reasoning. LIMA~\cite{zhou2023lima} demonstrated that superficial alignment requires only a few thousand carefully curated examples, while the Phi series \cite{gunasekar2023textbooks,li2023textbooks} demonstrated that textbook-quality synthetic data could enable compact models to outperform much larger counterparts. 

Despite these successes, existing approaches face critical limitations when applied to specialized domains. As highlighted by AgoraBench~\cite{kim2025evaluating}, not all generators are created equal, and capability in general NLP does not guarantee high-quality data synthesis~\cite{gudibande2023false}. Most synthesis pipelines remain open-loop without ground-truth anchoring. Our work treats synthetic data as \textit{source code} derived from corpora and enforces a closed-loop debugging cycle to ensure domain fidelity.

\vspace{-2mm}
\subsection{LLM Benchmarking and Evaluation}
% Context (现状): 介绍主流评估范式。
% Static Benchmarks: 如 MMLU, GSM8K, C-Eval。它们定义了通用能力的标尺。
% LLM-as-a-Judge: 如 AlpacaEval, MT-Bench。利用强模型评估弱模型。
% Despite (肯定): 这些 Benchmark 为社区提供了标准化的度量衡，推动了模型的快速迭代。
% However (Gap & Critique):
% Data Contamination: 静态 Benchmark 面临严重的数据泄露问题（Test set leakage），导致分数虚高。
% Disconnect from Training: 目前的 Benchmark 仅仅是“计分卡”（Scorecard），而不是“诊断报告”（Diagnostic Report）。它们告诉你模型得了 60 分，但不能告诉你具体哪条逻辑链断了，更无法直接指导训练数据的修复。
% Your Solution: ProDa 将 Benchmark 重定义为 "Unit Tests"。它们是基于 Raw Corpus 动态生成的，且与训练数据正交（Orthogonal）。最重要的是，它们直接触发 Data Debugging 流程，实现了“评测即指导”。
The evaluation landscape for LLMs has evolved to keep pace with their capabilities. Static benchmarks serve as the primary barometer for model performance~\cite{huang2023c}. Pioneering suites such as MMLU~\cite{hendrycks2021measuring} and Big-Bench~\cite{srivastava2023beyond} provide comprehensive assessments across diverse domains~\cite{liang2022holistic}, while specialized datasets like GSM8K~\cite{cobbe2021training} focus on multi-step mathematical reasoning~\cite{lin2022truthfulqa}. To evaluate open-ended generation, the community has increasingly adopted the \textit{LLM-as-a-Judge} paradigm~\cite{aiyappa2023can}, exemplified by MT-Bench~\cite{zheng2023judging} and AlpacaEval~\cite{li2023alpacaeval}, which leverage stronger models to score weaker models. Despite providing standardized metrics, data contamination has become a pervasive issue~\cite{deng2024investigating}. As static benchmarks are widely disseminated, they inadvertently leak into the pre-training corpora, reflecting memorization rather than genuine reasoning \cite{golchin2024time,zhou2023don}. More fundamentally, existing benchmarks function primarily as scorecards rather than diagnostic reports~\cite{liu2021explainaboard}. They yield scalar performance metrics but provide little insight into why models fail~\cite{kiela2021dynabench}. In this work, we reconceptualize benchmarking through a software engineering lens, framing evaluations as unit tests~\cite{ribeiro2022adaptive}. Instead of relying on fixed datasets, our benchmarks are dynamically synthesized from raw domain corpora. Crucially, these evaluations are designed not merely to assign scores, but to activate a data debugging loop~\cite{kim2023prometheus}.

\subsection{Self-Improving LLM}
% Context (现状): 介绍模型自我进化的尝试。
% RLHF / RLAIF: 通过反馈信号优化模型。
% Self-Play / Zero-Shot Reasoning: 如 AlphaZero/R-Zero、SPIN (Self-Play Fine-Tuning)、Self-Rewarding LMs。
% STaR (Self-Taught Reasoner): 让模型在推理正确的样本上微调。
% Despite (肯定): 这些工作展示了模型可以通过自我博弈或自我筛选，在没有外部人类监督的情况下实现性能提升（Zero-human intervention）。
% However (Gap & Critique):
% Lack of External Grounding: 大多数 Self-Evolving 方法是在优化“对齐”（Alignment）或“解题技巧”，而非系统的“知识习得”。对于垂直领域（如医学、法律），模型不能仅靠自己“冥想”（Self-play）来获得知识，必须Grounding 到外部的权威语料（Raw Corpus）上。
% Weight vs. Data: 现有的进化主要体现在更新权重（RL）或筛选现有数据。
% Your Solution: ProDa 的进化是 "Code-level Evolution"。我们不是简单地筛选，而是主动编写（Synthesize）新的“补丁数据”。这是一种更符合 Data-Centric AI 理念的进化方式——通过完善“源代码”（数据）来完善软件（模型）。
The pursuit of autonomous model improvement has shifted focus from human-annotated supervision to self-generated signals. Foundational work in RLHF~\cite{ouyang2022training} demonstrated the efficacy of feedback loops, which was subsequently scaled by RLAIF~\cite{lee2024rlaif, bai2022constitutional} to replace human evaluators. More recently, this line of research has evolved into fully iterative self-improvement frameworks. Representative approaches such as STaR~\cite{zelikman2022star} and ReST~\cite{gulcehre2023reinforced} follow a generate–filter–finetune pipeline, whereby a model produces candidate solutions, selectively retains high-quality trajectories, and retrains on its own verified reasoning processes. In parallel, self-play-inspired methods—including SPIN~\cite{chen2024self}, Absolute Zero~\cite{zhao2025absolute}, and R-Zero~\cite{huang2025r}—as well as Self-Rewarding Language Models~\cite{yuan2024self}, further relax external supervision by allowing models to compete against themselves or act as autonomous judges.

However, domain-specific expertise cannot be reliably acquired through self-play in the absence of authoritative grounding. Emerging work such as LANCE~\cite{wang2025language} and EVOLVE~\cite{zeng2025evolving} has begun to adopt a data-centric perspective, reframing the model as an autonomous data engineer. Our work reframes self-evolution as code-level data engineering, synthesizing executable data patches from authoritative corpora to address diagnosed failures beyond what sample-level self-filtering can achieve.

\section{Discussion}

\textit{Programming with Data} is built on the observation that fine-tuning and pre-training pose fundamentally different data-engineering problems. Pre-training ingests broad corpora to acquire general linguistic competence, where tracing specific behaviours to specific documents is neither feasible nor necessary. Fine-tuning targets well-defined competencies, making it both possible and desirable to know exactly what knowledge the data encodes and where the model falls short. By introducing a shared knowledge structure that simultaneously specifies training data, evaluation, and diagnosis, the framework establishes a direct correspondence with software-engineering practice: knowledge structure as requirements specification, synthesised data as implementation, benchmark as test suite, and diagnostic module as fault localiser.

Existing approaches each address fragments of this loop while leaving it structurally open. Synthetic data methods~\cite{wang2023self,mukherjee2023orca,gunasekar2023textbooks,li2023textbooks} define quality through prior heuristics without feedback from the trained model; self-improvement frameworks~\cite{zelikman2022star,yuan2024self} refine outputs at inference time but leave knowledge-gap-producing training data untouched; data-centric approaches~\cite{wang2025language,zeng2025evolving} rewrite samples without diagnosing which knowledge components are deficient; and diagnostic benchmarks~\cite{liu2021explainaboard,kiela2021dynabench} identify failure patterns but cannot trace them back to data. The common deficiency is the absence of a shared specification that closes the loop at the knowledge level. 

The present work establishes the macro-level architecture of \textit{Programming with Data}; it does not exhaust the design space it opens. Every module constitutes a research problem in its own right, each admitting improvement as the community brings domain-specific techniques to bear. We expect particularly intersections with retrieval-augmented generation for grounding synthesized data in primary sources, with mechanistic interpretability for fine-grained diagnosis.

The relationship between fine-tuning data and model behaviour has been widely treated as too entangled to engineer systematically, justifying a culture in which scale substitutes for understanding. Our results demonstrate that this entanglement is not irreducible: once mediated by a shared knowledge specification, it becomes transparent, diagnosable, and repairable. What we offer is not a finished system but a general-purpose blueprint—a discipline-agnostic method for converting raw textual knowledge into verifiable model competence, one that any field can instantiate, extend, and refine.

\section*{Data Availability}

The ProDa-16 benchmark, the extracted knowledge structures,
and all training data generated in this study are publicly
available at \href{https://huggingface.co/datasets/OpenRaiser/ProDalib}{https://huggingface.co/datasets/OpenRaiser/ProDalib}.

\section*{Code Availability}

The source code of ProDa Studio is
publicly available at \href{https://github.com/OpenRaiser/ProDa}{https://github.com/OpenRaiser/ProDa}.

% \section*{Acknowledgments}
% \label{sec:ack}
% We thank xxx for their contributions and support for the project.

\clearpage
\newpage
\bibliographystyle{plainnat}
\setcitestyle{numbers}
\bibliography{ref}

\clearpage
\newpage
\beginappendix

\section{Corpus Curation}
\label{app:prodalib_corpus}

\subsection{Document classification prompt}
\label{app:prodalib_document}

\begin{center}
\captionsetup{type=figure}
  \begin{promptbox}[Document Classification \& Curation Judge]
  \textbf{Role.} You are a \emph{Scientific Data Curator} for a large-scale science reasoning dataset.
  Given a document's title and summary, your task is to \textbf{classify} the document along four axes and \textbf{decide} whether it should be retained for downstream knowledge extraction.

  \medskip
  \textbf{Scientific Domain}

  Select up to \textbf{two} domains that best describe the core scientific content.
  Choose \emph{only} from:
  \texttt{physics}, \texttt{chemistry}, \texttt{biology}, \texttt{medicine},
  \texttt{materials\_science}, \texttt{computer\_science}, \texttt{mathematics},
  \texttt{engineering}, \texttt{earth\_science}, \texttt{astronomy},
  \texttt{interdisciplinary}, \texttt{other}.

  \medskip
  \textbf{Academic Level Rules}

  Classify the \emph{highest} level required to understand the document:
  \begin{itemize}\setlength{\itemsep}{2pt}
    \item \textbf{introductory}: Popular science or high-school level; general descriptions with no assumed technical background.
    \item \textbf{undergraduate}: Standard university-level concepts; basic models, introductory methods, limited formalism.
    \item \textbf{graduate}: Advanced theoretical or methodological content; assumes strong background knowledge.
    \item \textbf{research}: Original research contributions, novel methods, or frontier topics.
  \end{itemize}

  \medskip
  \textbf{Reasoning Type Rules}

  Identify the \emph{primary} reasoning type used in the document:
  \begin{itemize}\setlength{\itemsep}{2pt}
    \item \textbf{descriptive}: Mainly narrative or factual description; minimal reasoning.
    \item \textbf{procedural}: Step-by-step methods or workflows without deep theoretical analysis.
    \item \textbf{conceptual}: Explanation of mechanisms, principles, or causal reasoning.
    \item \textbf{mathematical}: Formal derivations, equations, proofs, or quantitative modeling.
    \item \textbf{experimental}: Experimental design, analysis of results, or hypothesis testing.
  \end{itemize}

  \medskip
  \textbf{Inclusion Decision Rules}

  Set \texttt{keep} to \texttt{true} \textbf{only if all} of the following are satisfied:
  \begin{itemize}\setlength{\itemsep}{2pt}
    \item The content is scientific or technical.
    \item The academic level is \texttt{undergraduate}, \texttt{graduate}, or \texttt{research}.
    \item The reasoning type is \emph{not} \texttt{descriptive}.
    \item The document contains non-trivial reasoning (conceptual, mathematical, experimental, or procedural).
  \end{itemize}
  Otherwise, set \texttt{keep} to \texttt{false}.

  \medskip
  \textbf{Confidence}

  Provide a confidence score ($0.0$--$1.0$) indicating certainty about the classification, based on the clarity of the title and summary.

  \medskip
  \textbf{Inputs}
  \begin{itemize}\setlength{\itemsep}{2pt}
    \item Title: \texttt{\{TITLE\}}
    \item Summary: \texttt{\{SUMMARY\}}
  \end{itemize}

  \medskip
  \textbf{Output requirements (must be strictly followed)}

  Return \textbf{only} valid JSON in the following format; do not output any other text:

  \vspace{2pt}
  \texttt{\{}\\
  \hspace*{1.2em}\texttt{"domains": ["domain\_1", "domain\_2"],}\\
  \hspace*{1.2em}\texttt{"level": "introductory | undergraduate | graduate | research",}\\
  \hspace*{1.2em}\texttt{"reasoning\_type": "descriptive | procedural | conceptual | mathematical | experimental",}\\
  \hspace*{1.2em}\texttt{"keep": true | false,}\\
  \hspace*{1.2em}\texttt{"confidence": 0.0--1.0}\\
  \texttt{\}}
  \end{promptbox}
  \vspace{-2mm}
  \captionof{figure}{Prompt template used for \textbf{corpus-level document triage} in ProDa's preprocessing stage.}
  \vspace{2mm}
  \label{prompt:document_curation_prompt}
\end{center}

\subsection{Chunk quality scoring rubrics}
\label{app:prodalib_chunk_quality_matrix}
This appendix provides the complete scoring rubrics for the six-dimensional quality matrix used in chunk-level quality assessment (\S X.4 of the main text). Each dimension is scored on an integer scale from 1 to 5 by a language model evaluator. Below we define each dimension, state its purpose within the ProDa pipeline, and provide detailed anchor descriptions for all five score levels.
% ============================================================
% Dimension 1: Reasoning Depth
% ============================================================
\subsubsection{Dimension 1: Reasoning Depth}
Reasoning Depth measures the number of logically dependent inferential steps present in the text. It captures how many intermediate conclusions must be established before the final claim is reached. Chunks scoring highly on this dimension are prioritized for L3 Reasoning Chain extraction and for constructing multi-hop benchmark items.
\begin{table*}[htbp]
\centering
\caption{Scoring rubric for Reasoning Depth.}
\label{tab:rubric_reasoning_depth}
\begin{tabular}{c p{0.88\linewidth}}
\toprule
\textbf{Score} & \textbf{Anchor Description} \\
\midrule
1 & The text consists entirely of flat declarative statements or enumerations. No logical dependencies exist between sentences. Typical pattern: ``A is B. C has property D.'' \\
\midrule
2 & The text contains isolated single-step inferences, such as a direct cause-effect pair or a simple if-then statement, but these do not chain into longer arguments. \\
\midrule
3 & The text contains at least one reasoning sequence spanning two to three dependent steps. Intermediate conclusions serve as premises for subsequent claims, but the chain is short and linear. \\
\midrule
4 & The text contains multi-hop reasoning spanning four or more steps, or includes branching logic where a conclusion depends on the conjunction of multiple independent premises. Conditional qualifications appear. \\
\midrule
5 & The text contains extended inferential chains with five or more dependent steps, potentially involving nested conditionals, iterative refinement, proof by contradiction, or convergence of multiple independent argument threads toward a unified conclusion. \\
\bottomrule
\end{tabular}
\end{table*}
% ============================================================
% Dimension 2: Prerequisite Density
% ============================================================
\subsubsection{Dimension 2: Prerequisite Density}
Prerequisite Density estimates the volume and specificity of domain knowledge a reader must already possess to understand the chunk. It distinguishes self-contained introductory expositions from advanced discussions that presuppose fluency with specialized terminology and conceptual frameworks. This dimension controls the difficulty distribution of synthesized training data and helps distinguish introductory from advanced Key Concepts during L1 extraction.
\begin{table*}[htbp]
\centering
\caption{Scoring rubric for Prerequisite Density.}
\label{tab:rubric_prerequisite_density}
\begin{tabular}{c p{0.88\linewidth}}
\toprule
\textbf{Score} & \textbf{Anchor Description} \\
\midrule
1 & The text is fully self-contained. All terms are defined upon introduction. A general audience with no domain training can follow the content without difficulty. \\
\midrule
2 & The text assumes basic literacy in the broad field but defines most specialized terms. An undergraduate student in the first year of study could follow with minimal external reference. \\
\midrule
3 & The text assumes working knowledge of core concepts and standard notation in the discipline. Key terms are used without definition. A student midway through an undergraduate program would find the content accessible. \\
\midrule
4 & The text presupposes mastery of multiple advanced topics and employs specialized jargon, notation, or formalism without explanation. A graduate student or practitioner in the specific subfield is the implied reader. \\
\midrule
5 & The text operates at the frontier of the discipline, assuming deep expertise across multiple subfields. It references advanced theorems, experimental paradigms, or methodological frameworks that only active researchers in the area would recognize. \\
\bottomrule
\end{tabular}
\end{table*}
% ============================================================
% Dimension 3: Scenario Applicability
% ============================================================
\subsubsection{Dimension 3: Scenario Applicability}
Scenario Applicability assesses whether theoretical knowledge in the chunk is grounded in concrete problem-solving contexts. It measures the degree to which abstract principles are instantiated through cases, experiments, diagnostic workflows, or engineering applications. High-scoring chunks provide raw material for synthesizing application-oriented training samples that test whether a model can deploy knowledge in context, not merely recall definitions.
\begin{table*}[htbp]
\centering
\caption{Scoring rubric for Scenario Applicability.}
\label{tab:rubric_scenario_applicability}
\begin{tabular}{c p{0.88\linewidth}}
\toprule
\textbf{Score} & \textbf{Anchor Description} \\
\midrule
1 & The text is purely abstract. It presents definitions, axioms, or formal derivations with no reference to any concrete instance, experiment, or real-world situation. \\
\midrule
2 & The text is predominantly theoretical but includes passing mention of a potential application or a generic illustrative example without substantive detail. \\
\midrule
3 & The text integrates at least one concrete scenario with moderate detail, such as a worked numerical example, a simplified case study, or a standard laboratory procedure, but the scenario serves primarily as illustration rather than as the organizing structure. \\
\midrule
4 & The text is substantially organized around one or more concrete scenarios. Theoretical principles are introduced in service of solving a specific problem, diagnosing a specific condition, or analyzing a specific case. \\
\midrule
5 & The text is deeply embedded in real-world problem-solving. It presents detailed clinical diagnoses, engineering failure analyses, legal case reasoning, experimental troubleshooting, or multi-step design workflows where domain knowledge is exercised under realistic constraints and trade-offs. \\
\bottomrule
\end{tabular}
\end{table*}
% ============================================================
% Dimension 4: Counter-Intuitive Index
% ============================================================
\subsubsection{Dimension 4: Counter-Intuitive Index}
The Counter-Intuitive Index captures the presence of content that contradicts naive expectations or surface-level reasoning. It identifies exceptions to general rules, common misconceptions and their corrections, boundary conditions where standard models fail, and paradoxes requiring careful resolution. This dimension is the primary source for constructing hard negative distractors in benchmark items. Content that exposes common errors provides the most discriminative test of genuine understanding versus superficial pattern matching.
\begin{table*}[htbp]
\centering
\caption{Scoring rubric for Counter-Intuitive Index.}
\label{tab:rubric_counter_intuitive}
\begin{tabular}{c p{0.88\linewidth}}
\toprule
\textbf{Score} & \textbf{Anchor Description} \\
\midrule
1 & The content is entirely consistent with common intuition. All statements follow predictably from general principles, and no exceptions, caveats, or surprising results are mentioned. \\
\midrule
2 & The text contains minor qualifications or edge cases, but these are noted in passing and do not challenge the dominant narrative. A reader relying on surface-level heuristics would not be misled. \\
\midrule
3 & The text includes at least one substantive exception or non-obvious result that a careful reader would need to note. However, the counter-intuitive element is not the central focus of the passage. \\
\midrule
4 & The text contains multiple counter-intuitive elements, or one element that is sufficiently important to restructure understanding of the topic. Explicit correction of a common misconception is present. \\
\midrule
5 & The text is centrally organized around counter-intuitive phenomena: paradoxes, systematic misconceptions, failure modes of standard approaches, or results that require abandoning a widely held assumption. The passage would generate significant confusion if read carelessly. \\
\bottomrule
\end{tabular}
\end{table*}
% ============================================================
% Dimension 5: Knowledge Synthesis
% ============================================================
\subsubsection{Dimension 5: Knowledge Synthesis}
Knowledge Synthesis evaluates how effectively the chunk constructs a connected knowledge framework. The core criterion is the ability to link isolated concepts into logical sequences and further elevate these sequences into systematic theoretical or applied architectures. This encompasses bridging theory and practice, unifying micro-level mechanisms with macro-level phenomena, and integrating perspectives from multiple subfields. High-scoring chunks feed L2 Key Concept Relation extraction, since they make inter-concept connections explicit. They also provide the structural scaffolding for multi-concept training samples.
\begin{table*}[htbp]
\centering
\caption{Scoring rubric for Knowledge Synthesis.}
\label{tab:rubric_knowledge_synthesis}
\begin{tabular}{c p{0.88\linewidth}}
\toprule
\textbf{Score} & \textbf{Anchor Description} \\
\midrule
1 & The text presents isolated knowledge fragments. Individual facts or definitions appear without any explicit connection to each other or to a broader framework. \\
\midrule
2 & The text groups related facts under a common theme, but connections between items remain implicit or superficial. The passage reads as a list rather than an argument. \\
\midrule
3 & The text makes explicit connections between multiple concepts, establishing local structure such as cause-effect pairs, classification hierarchies, or procedural sequences. However, the scope remains within a single narrow topic. \\
\midrule
4 & The text constructs a coherent framework that spans multiple related topics. It bridges levels of abstraction, connects theoretical models to empirical observations, or synthesizes findings from different experimental paradigms into a unified account. \\
\midrule
5 & The text achieves deep synthesis across multiple dimensions: it unifies macro and micro perspectives, bridges theory and practice, integrates evidence from different subfields or methodologies, and articulates a systematic framework that organizes the full scope of the discussed material. \\
\bottomrule
\end{tabular}
\end{table*}
% ============================================================
% Dimension 6: Breakpoint Smoothness
% ============================================================
\subsubsection{Dimension 6: Breakpoint Smoothness}
Breakpoint Smoothness assesses the semantic integrity of chunk boundaries. It evaluates whether the chunking process has introduced hard truncations that sever ongoing arguments or created dangling references to content that falls outside the chunk. Evaluation focuses on the first 500 tokens and the last 500 tokens of each chunk. This dimension serves as a mandatory quality gate. All chunks selected for downstream processing must achieve a minimum score of 4, ensuring that no boundary artifact propagates into Key Concept extraction or training data synthesis.
\begin{table*}[htbp]
\centering
\caption{Scoring rubric for Breakpoint Smoothness.}
\label{tab:rubric_breakpoint_smoothness}
\begin{tabular}{c p{0.88\linewidth}}
\toprule
\textbf{Score} & \textbf{Anchor Description} \\
\midrule
1 & Both boundaries are severely damaged. The opening begins mid-sentence or with unresolvable anaphoric references. The closing truncates an argument, proof, or procedure before its conclusion. Critical information is lost at both ends. \\
\midrule
2 & One boundary is severely damaged while the other is partially intact, or both boundaries exhibit moderate truncation. Key context is missing, requiring a reader to consult adjacent chunks to follow the content. \\
\midrule
3 & Both boundaries are partially intact. The opening may begin with a brief context-dependent reference that is quickly resolved, and the closing may end at a natural pause point that leaves a minor thread open. The chunk is largely self-contained but not fully clean. \\
\midrule
4 & Both boundaries are clean. The opening begins at a clear paragraph or section transition. The closing ends at a natural stopping point. Minor residual artifacts may exist but do not affect comprehension of the core content. \\
\midrule
5 & Both boundaries are perfectly clean. The opening starts with a section heading, chapter title, or the first sentence of a clearly delineated topic. The closing completes all open arguments and ends with a logically closed statement. The chunk reads as a self-contained unit. \\
\bottomrule
\end{tabular}
\end{table*}

\clearpage
\subsection{Corpus statistics}

\begin{figure}[h]
\centering
\includegraphics[width=\textwidth]{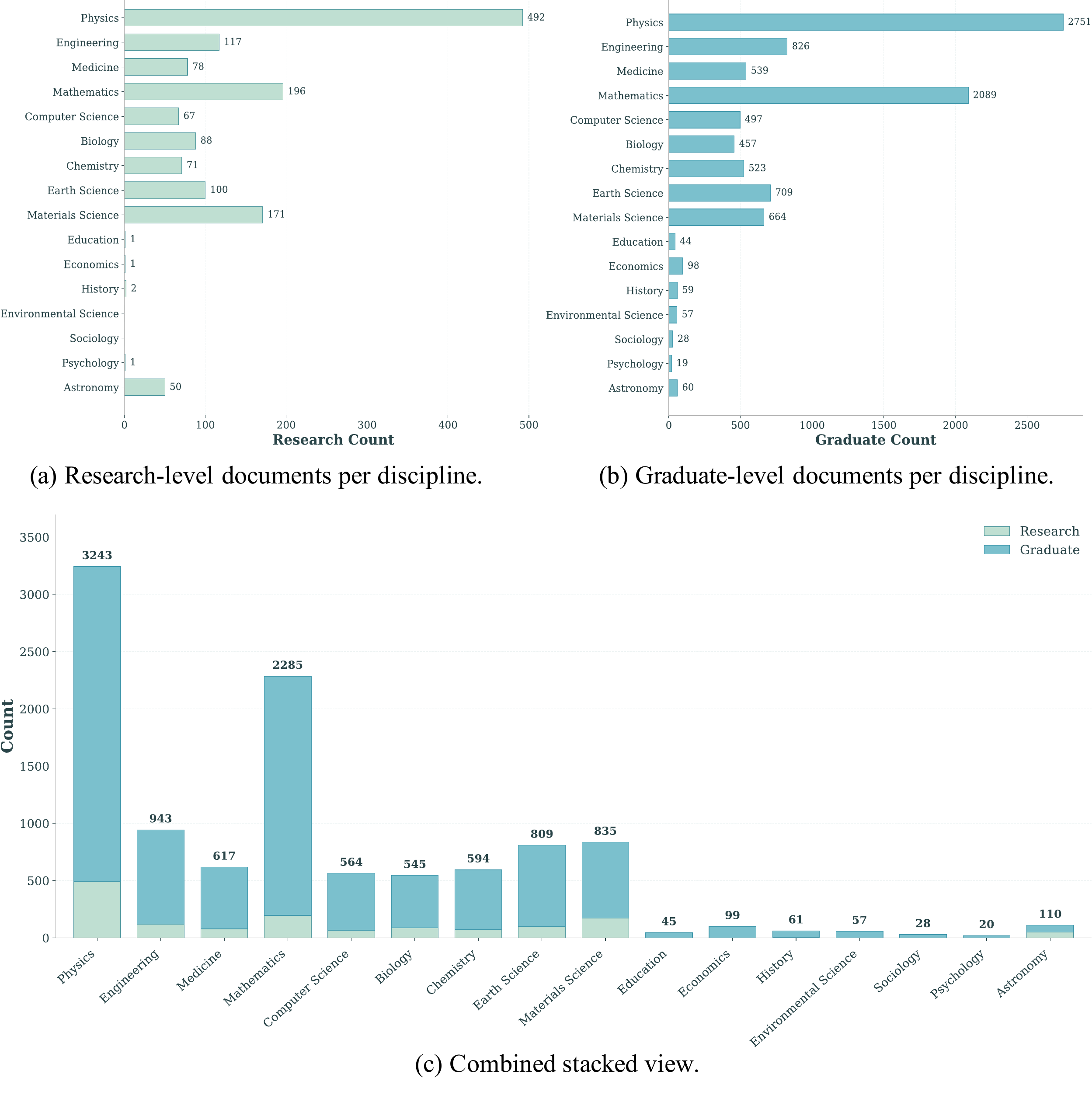}
\caption{Distribution of documents by discipline and academic level after document-level curation.
}
\label{fig:prodalib_domain_stat}
\end{figure}

\begin{figure}[h]
\centering
\includegraphics[width=\textwidth]{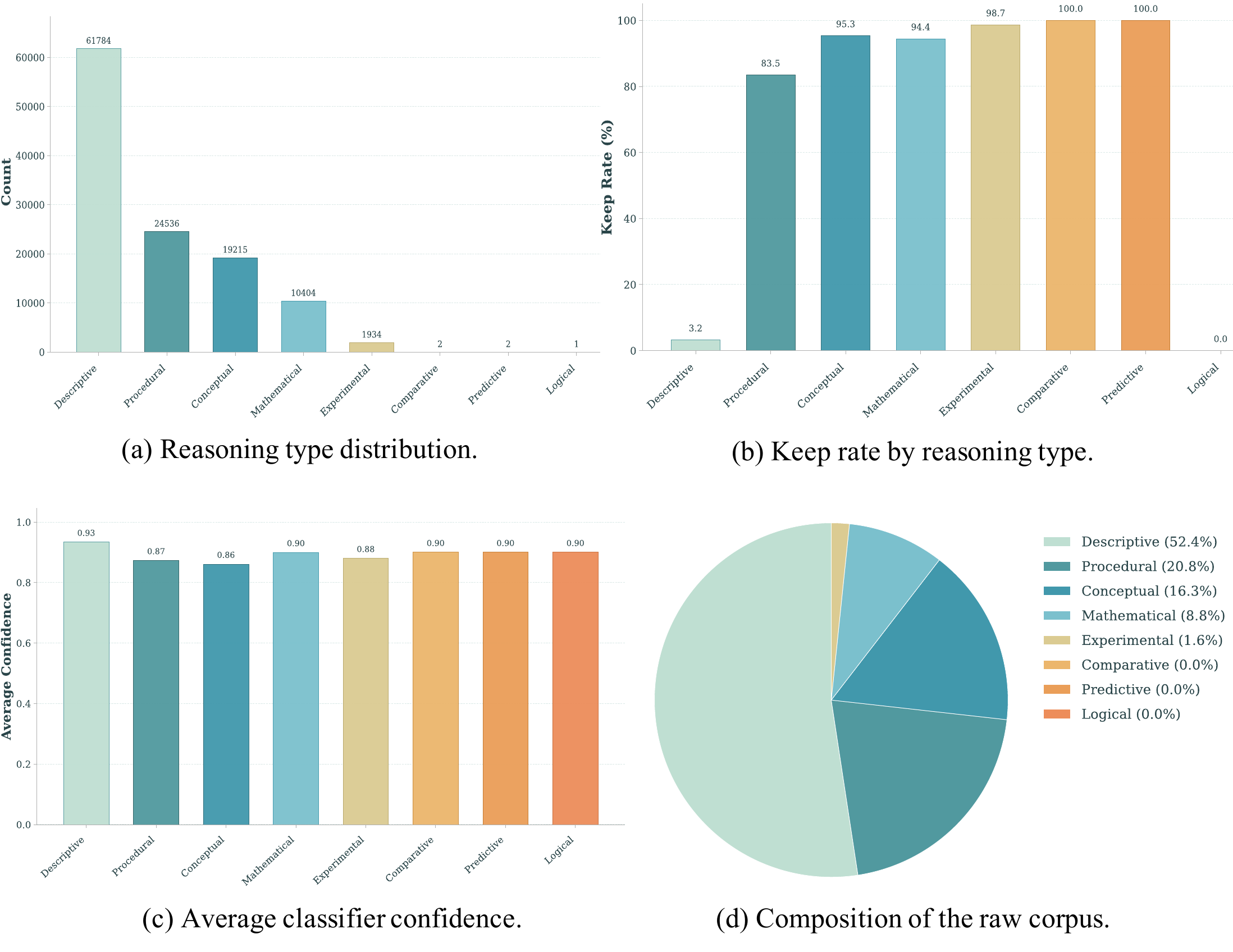}
\caption{
Reasoning type analysis of the raw corpus during document-level curation.
}
\label{fig:prodalib_reasoning}
\end{figure}

\begin{figure}[h]
\centering
\includegraphics[width=\textwidth]{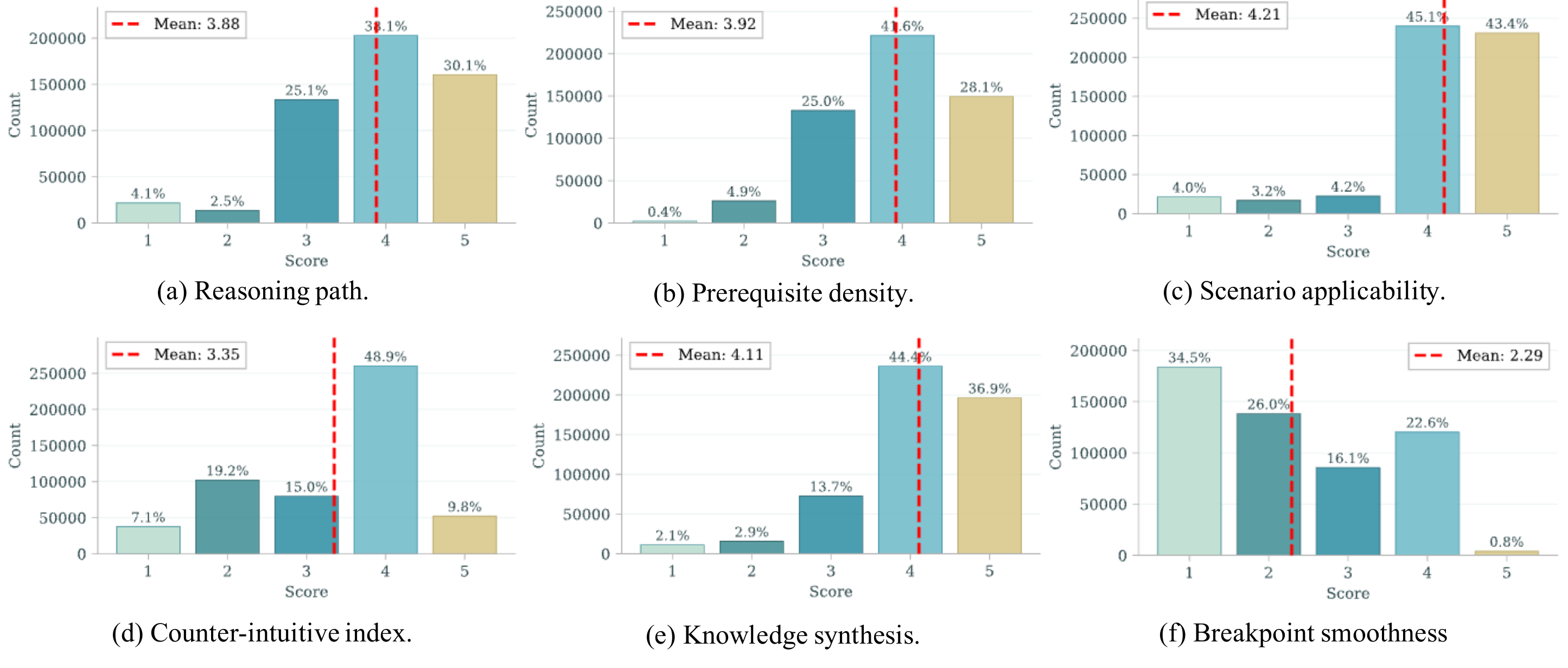}
\caption{
Score distributions of the six-dimensional quality matrix across all corpus chunks.
}
\label{fig:prodalib_score_distribution}
\end{figure}

\clearpage
\section{Knowledge Structure Extraction}
\label{app:prodalib_key_concept}

\subsection{L3 reasoning chain extraction prompt}
\begin{center}
\captionsetup{type=figure}
  \begin{promptbox}[L3 Reasoning Chain Extraction]
  \textbf{Role.} You are an \emph{Expert Knowledge Engineer and Logic Analyst}.
  You possess the ability to analyze complex texts across various domains
  (including Medicine, Law, Chemistry, Engineering, and Humanities)
  to identify underlying logical structures and processes.
  \medskip
  \textbf{Goal}
  Read the input text and identify the single most important
  \textbf{Core Reasoning Chain} (L3 Knowledge).
  A Reasoning Chain is a continuous, coherent sequence of events,
  arguments, or steps where one element logically leads to,
  causes, or precedes the next.
  These chains represent the narrative backbone or mechanism
  of the knowledge.
  \medskip
  \textbf{Extraction Procedure}
  \begin{enumerate}\setlength{\itemsep}{2pt}
    \item \textbf{Domain Analysis.}
      Determine the domain of the text
      (e.g., Legal Argument, Chemical Reaction, Medical Diagnosis,
      Historical Sequence).
      Adapt extraction logic to fit the domain's reasoning style.
    \item \textbf{Chain Extraction.}
      Identify distinct, multi-step processes or logical arguments:
      \begin{itemize}\setlength{\itemsep}{1pt}
        \item \emph{Causal chains}: A causes B, which causes C.
        \item \emph{Procedural chains}: Step~1, then Step~2, then Step~3.
      \end{itemize}
    \item \textbf{Validation.}
      Ensure the chain is continuous.
      Every step must have a direct connection to the next.
      Do not skip intermediate steps mentioned in the text.
    \item \textbf{Narrative Synthesis.}
      For each chain, write a paragraph-length summary
      explaining the mechanism or logic behind it.
    \item \textbf{Step-by-Step Breakdown.}
      List the exact sequence of nodes in text format.
  \end{enumerate}
  \medskip
  \textbf{Key Constraints}
  \begin{itemize}\setlength{\itemsep}{2pt}
    \item \textbf{Strict Logic}: Every step must connect directly to the next.
    \item \textbf{No Disconnected Facts}: Do not list static facts
      (e.g., ``Water is H\textsubscript{2}O'').
      Only extract flows
      (e.g., ``Hydrogen bonds with Oxygen to form Water'').
    \item \textbf{Context Independence}: Step descriptions should be
      understandable without reading the full source text.
    \item \textbf{Closed-Book Policy}: Extract logic \emph{only}
      from the provided text.
      Do not use external knowledge to fill gaps.
  \end{itemize}
  \medskip
  \textbf{Logical Scope and Granularity}
  Use \textbf{logical completeness} rather than step count as the boundary:
  \begin{itemize}\setlength{\itemsep}{2pt}
    \item \textbf{Natural Closure}: A chain should start at a clear
      initiation point (cause or start condition) and end at a logical
      conclusion (effect or end state).
      Do not arbitrarily cut off an unfinished process.
    \item \textbf{Topic Cohesion}: If the text shifts to a completely
      different mechanism, topic, or argument, end the current chain
      and start a new one. Do not merge unrelated processes.
    \item \textbf{Granularity Level}:
      Avoid micro-steps that are synonym repetitions.
      Avoid macro-steps that skip large chunks of logic.
      Each step should represent a distinct state change
      or logical progression.
  \end{itemize}
  \medskip
  \textbf{Extraction Dimensions for Benchmark Support}
  For each chain, also identify the following to support
  downstream benchmark construction:
  \begin{itemize}\setlength{\itemsep}{2pt}
    \item \textbf{Preconditions}: What must be true for this chain
      to be valid?
      (e.g., ``Occurs only in high-temperature environments.'')
    \item \textbf{Negative Constraints and Confusions}:
      What does the text explicitly exclude?
      (e.g., ``Unlike X, this process does not involve Y.'')
  \end{itemize}
  \medskip
  \textbf{Single-Chain Selection Rule}
  You must output \textbf{exactly one} reasoning chain per input text.
  Select the mechanism that represents the primary narrative backbone
  of the entire passage: the core topic with the longest causal arc
  and highest knowledge value.
  Ignore secondary or supporting mechanisms.
  If multiple candidates seem equally important, choose the one with
  the most complete start-to-end progression and the richest
  step-by-step detail grounded in the text.
  \medskip
  \textbf{Input}
  \texttt{\{INPUT\_TEXT\}}
  \medskip
  \textbf{Output requirements (must be strictly followed)}
  Return \textbf{only} a valid JSON array containing exactly one object;
  do not output any other text:
  \vspace{2pt}
  \texttt{[}\\
  \hspace*{0.6em}\texttt{\{}
  \hspace*{1.8em}\texttt{"chain\_id": "chain-001",}\\
  \hspace*{1.8em}\texttt{"domain\_context": "String",}\\
  \hspace*{1.8em}\texttt{"process\_name": "String",}\\
  \hspace*{1.8em}\texttt{"narrative\_summary": "String",}\\
  \hspace*{1.8em}\texttt{"preconditions": ["String", ...],}\\
  \hspace*{1.8em}\texttt{"negative\_constraints": ["String", ...],}\\
  \hspace*{1.8em}\texttt{"steps": ["Step 1", "Step 2", ..., "Step N"]}\\
  \hspace*{0.6em}\texttt{\}}
  \texttt{]}
  \end{promptbox}
  \vspace{-2mm}
  \captionof{figure}{Prompt template used for \textbf{L3 Reasoning Chain extraction} from high-quality corpus chunks. Each chunk yields exactly one chain representing its primary inferential pathway.}
  \label{prompt:l3_extraction_prompt}
  \vspace{-2mm}
\end{center}

\subsection{L2 atomic statement decomposition prompt}
% ============================================================
%  L2 KC Relations: Atomic Statement Decomposition from L3 Chains
% ============================================================

\begin{center}
\captionsetup{type=figure}
  \begin{promptbox}[L2 Atomic Statement Decomposition]
  \textbf{Role.} You are a \emph{Structural Logic Decomposer}.
  Your task is to break down complex, continuous reasoning chains
  into discrete, atomic factual statements (L2 Knowledge).

  \medskip
  \textbf{Goal}

  Translate the continuous flow of L3 Reasoning Chains into
  distinct \textbf{Atomic Links}.
  Each link captures the precise relationship between
  one adjacent pair of steps in a chain,
  serving as a high-quality relational unit
  for downstream training data synthesis.

  \medskip
  \textbf{Decomposition Procedure}

  \begin{enumerate}\setlength{\itemsep}{2pt}
    \item \textbf{Process Chains.}
      Iterate through every \texttt{chain\_id} in the input.
      All chains must be processed with equal attention.

    \item \textbf{Sliding Window Decomposition.}
      For every adjacent pair of steps
      (\texttt{Step[i]} and \texttt{Step[i+1]}),
      generate one Factual Statement.\\
      Example chain: A $\rightarrow$ B $\rightarrow$ C\\
      Target output: Statement(A $\rightarrow$ B)
      \textsc{and} Statement(B $\rightarrow$ C).

    \item \textbf{Formulate the Statement.}
      Each statement must contain:
      \begin{itemize}\setlength{\itemsep}{1pt}
        \item \emph{Subject}: The concept or entity in \texttt{Step[i]}.
        \item \emph{Object}: The concept or entity in \texttt{Step[i+1]}.
        \item \emph{Predicate}: A specific verb phrase describing
          the relationship
          (e.g., ``causes'', ``is followed by'', ``implies'',
          ``reacts with'', ``legally binds'').
        \item \emph{Source Quote}: The exact short phrase from
          the original text that evidences this specific link.
      \end{itemize}

    \item \textbf{Contextualization.}
      If a step contains a pronoun or vague reference
      (e.g., ``It expands''),
      replace it with the specific noun based on preceding steps
      (e.g., ``The lung expands'').
      Both Subject and Object must be understandable in isolation.
  \end{enumerate}

  \medskip
  \textbf{Key Constraints}

  \begin{itemize}\setlength{\itemsep}{2pt}
    \item \textbf{Atomicity}: Each statement must describe
      exactly one logical step.
    \item \textbf{Strict Adjacency}: Only link immediate neighbors.
      Do not create links between \texttt{Step[i]}
      and \texttt{Step[i+2]}.
    \item \textbf{Hallucination Prevention}: If the relationship
      between two adjacent steps is not explicitly supported
      by the original text, skip it.
  \end{itemize}

  \medskip
  \textbf{Balance Requirement}

  Statements must be distributed across \emph{all} input chains.
  No single chain may contribute more than 20\% of total statements
  unless the input contains fewer than five chains.
  Before finalizing output, verify that no subset of three
  or fewer chains contributes more than 50\% of all statements.

  \medskip
  \textbf{Inputs}

  \begin{itemize}\setlength{\itemsep}{2pt}
    \item L3 Chains: \texttt{\{L3\_CHAINS\_INPUT\}}
    \item Original Text (for context verification and quote extraction):
      \texttt{\{ORIGINAL\_TEXT\}}
  \end{itemize}

  \medskip
  \textbf{Output requirements (must be strictly followed)}

  Return \textbf{only} a valid JSON array;
  do not include markdown code blocks, explanations,
  or any text before or after the JSON.
  Start directly with \texttt{[} and end with \texttt{]}.

  \vspace{2pt}
  \texttt{[}\\
  \hspace*{0.6em}\texttt{\{}\\
  \hspace*{1.8em}\texttt{"statement\_id": "stmt-001",}\\
  \hspace*{1.8em}\texttt{"parent\_chain\_id": "chain-001",}\\
  \hspace*{1.8em}\texttt{"subject": "String (Standardized Concept A)",}\\
  \hspace*{1.8em}\texttt{"predicate": "String (Relationship Verb)",}\\
  \hspace*{1.8em}\texttt{"object": "String (Standardized Concept B)",}\\
  \hspace*{1.8em}\texttt{"source\_quote": "String (Evidence from text)"}\\
  \hspace*{0.6em}\texttt{\}}\\
  \texttt{]}
  \end{promptbox}
  \vspace{-2mm}
  \captionof{figure}{Prompt template used for \textbf{L2 atomic statement decomposition}. Each adjacent step pair in an L3 chain is converted into a single typed relational triple with textual evidence.}
  \label{prompt:l2_decomposition_prompt}
  \vspace{-2mm}
\end{center}

\subsection{L1 key concept extraction prompt}
% ============================================================
%  L1 Key Concepts: Concept Extraction from L2 Statements
% ============================================================

\begin{center}
\captionsetup{type=figure}
  \begin{promptbox}[L1 Key Concept Extraction]
  \textbf{Role.} You are a \emph{Cross-Domain Ontologist and Terminologist}.
  Your task is to extract, standardize, and define the core concepts
  found in a set of atomic factual statements.

  \medskip
  \textbf{Goal}

  Create a \textbf{Dictionary of Concepts} (L1 Knowledge)
  derived from the L2 statements.
  This ensures that all entities used in the knowledge base
  are clearly defined, categorized, and traceable
  to their source statements.

  \medskip
  \textbf{Extraction Procedure}

  \begin{enumerate}\setlength{\itemsep}{2pt}
    \item \textbf{Term Collection.}
      Scan all \texttt{subject} and \texttt{object} fields
      in the L2 statements.
      Collect every unique term.

    \item \textbf{Deduplication and Canonicalization.}
      Merge lexical variations into a single canonical entry
      (e.g., ``The defendant'' and ``Defendant party''
      $\rightarrow$ ``Defendant'').
      Choose the most standard, professional name
      for the given domain.

    \item \textbf{Context-Aware Definition.}
      Provide a concise definition (1--2 sentences) for each concept.
      The definition must fit the context of the source statements.
      For example, if the domain is Law, define ``Bond''
      as a financial instrument, not a chemical connection.
      Use the \texttt{predicate} and \texttt{source\_quote}
      fields in L2 as context clues.

    \item \textbf{Typing.}
      Assign a category type relevant to the domain
      (e.g., Legal Entity, Chemical Element,
      Physiological Structure, Abstract Concept).

    \item \textbf{Traceability Annotation.}
      For each concept, record:
      \begin{itemize}\setlength{\itemsep}{1pt}
        \item \texttt{parent\_statement\_ids}:
          All \texttt{statement\_id} values from input statements
          where this term appears as subject or object.
        \item \texttt{CID}:
          All unique CID values from the corresponding
          source statements.
      \end{itemize}
  \end{enumerate}

  \medskip
  \textbf{Key Constraints}

  \begin{itemize}\setlength{\itemsep}{2pt}
    \item \textbf{Relevance}: Only define concepts that
      actually appear in the L2 statements.
      Do not introduce external terms.
    \item \textbf{Conciseness}: Each definition must be
      1--2 sentences. Do not provide encyclopedic entries.
    \item \textbf{Completeness}: \texttt{parent\_statement\_ids}
      must contain \emph{all} statement IDs where the term appears.
      \texttt{CID} must contain \emph{all} unique CID values
      from those statements.
      If a concept appears in multiple statements,
      include every corresponding ID.
  \end{itemize}

  \medskip
  \textbf{Input}

  L2 Statements: \texttt{\{L2\_STATEMENTS\_INPUT\}}

  \medskip
  \textbf{Output requirements (must be strictly followed)}

  Return \textbf{only} a valid JSON array;
  do not include markdown code blocks, explanations,
  or any text before or after the JSON.
  Start directly with \texttt{[} and end with \texttt{]}.

  \vspace{2pt}
  \texttt{[}\\
  \hspace*{0.6em}\texttt{\{}\\
  \hspace*{1.8em}\texttt{"concept\_id": "concept-001",}\\
  \hspace*{1.8em}\texttt{"term": "String (Canonical Name)",}\\
  \hspace*{1.8em}\texttt{"type": "String (Category)",}\\
  \hspace*{1.8em}\texttt{"definition": "String (Context-aware definition)",}\\
  \hspace*{1.8em}\texttt{"parent\_statement\_ids": ["stmt-001", "stmt-003"],}\\
  \hspace*{1.8em}\texttt{"CID": ["cid-001", "cid-002"]}\\
  \hspace*{0.6em}\texttt{\}}\\
  \texttt{]}
  \end{promptbox}
  \vspace{-2mm}
  \captionof{figure}{Prompt template used for \textbf{L1 Key Concept extraction}. Concepts are harvested from L2 statement subjects and objects, then deduplicated, defined in context, and linked back to their source statements for traceability.}
  \vspace{2mm}
  \label{prompt:l1_concept_prompt}
\end{center}

\vspace{-6mm}
\subsection{Per-discipline statistics — L3/L2/L1}
\label{app:prodalib_key_concept_stat}

% ============================================================
%  L3 Reasoning Chains — Per-Discipline Statistics
% ============================================================

\begin{table}[h]
\centering
\vspace{-2mm}
\caption{L3 Reasoning Chain statistics by discipline. \textit{Avg Steps} reports the mean number of inferential steps per chain; \textit{Range} gives the observed minimum and maximum.}
\label{tab:l3_stats}
\begin{tabular*}{\textwidth}{@{\extracolsep{\fill}} r l r r r r}
\toprule
CID & Discipline & Chains & Unique Chunks & Avg Steps & Range \\
\midrule
001 & Physics              & 10,622 & 10,621 & 9.31 & 4--16 \\
002 & Engineering          &  6,686 &  6,683 & 9.30 & 4--17 \\
003 & Medicine             &  7,242 &  7,242 & 9.40 & 4--19 \\
004 & Mathematics          &  3,874 &  3,873 & 8.96 & 3--15 \\
005 & Computer Science     &  1,904 &  1,904 & 9.41 & 5--16 \\
006 & Biology              &  1,764 &  1,764 & 9.30 & 4--16 \\
007 & Chemistry            &  1,550 &  1,550 & 9.03 & 4--17 \\
008 & Earth Science        &  2,535 &  2,535 & 9.04 & 5--17 \\
009 & Materials Science    &  1,709 &  1,709 & 9.17 & 5--14 \\
010 & Education            &    866 &    866 & 8.73 & 5--14 \\
011 & Economics            &  1,726 &  1,726 & 9.07 & 5--18 \\
012 & History              &  1,148 &  1,148 & 9.05 & 5--14 \\
013 & Environmental Science &  1,126 &  1,126 & 9.10 & 5--15 \\
014 & Sociology            &    566 &    566 & 8.86 & 6--15 \\
015 & Psychology           &    440 &    440 & 8.70 & 5--14 \\
016 & Astronomy            &    200 &    200 & 9.45 & 6--13 \\
\midrule
--- & Other                &  1,898 &  1,898 & 8.64 & 5--13 \\
\midrule
    & \textbf{Total}       & \textbf{45,856} & \textbf{45,851} & \textbf{9.14} & \textbf{3--19} \\
\bottomrule
\end{tabular*}
\vspace{-14mm}
\end{table}

% ============================================================
%  L2 KC Relations — Per-Discipline Statistics
% ============================================================

\begin{table}[h]
\centering
\caption{L2 KC Relation statistics by discipline. \textit{Avg Stmts} is the mean number of atomic statements decomposed from each L3 chain. \textit{Predicate Types} counts the number of distinct relation verbs observed.}
\label{tab:l2_stats}
\begin{tabular*}{\textwidth}{@{\extracolsep{\fill}} r l r r r l r}
\toprule
CID & Discipline & Statements & Chains & Avg Stmts & Top Predicate & Predicate Types \\
\midrule
001 & Physics              & 16,535 & 2,000 & 8.27 & results in    & 10,281 \\
002 & Engineering          & 16,590 & 2,000 & 8.29 & is followed by & 9,704 \\
003 & Medicine             & 16,932 & 2,000 & 8.47 & is followed by & 8,183 \\
004 & Mathematics          & 15,492 & 1,992 & 7.78 & is followed by & 9,980 \\
005 & Computer Science     & 16,005 & 1,903 & 8.41 & is followed by & 10,712 \\
006 & Biology              & 14,648 & 1,763 & 8.31 & results in    & 9,078 \\
007 & Chemistry            & 12,380 & 1,548 & 8.00 & results in    & 7,809 \\
008 & Earth Science        & 16,247 & 1,998 & 8.13 & results in    & 9,303 \\
009 & Materials Science    & 14,000 & 1,709 & 8.19 & results in    & 7,870 \\
010 & Education            &  6,698 &   866 & 7.73 & leads to      & 4,240 \\
011 & Economics            & 13,970 & 1,725 & 8.10 & results in    & 7,981 \\
012 & History              &  9,272 & 1,148 & 8.08 & results in    & 6,568 \\
013 & Environmental Science &  9,108 & 1,125 & 8.10 & leads to      & 5,637 \\
014 & Sociology            &  4,461 &   566 & 7.88 & results in    & 3,081 \\
015 & Psychology           &  3,393 &   440 & 7.71 & leads to      & 2,246 \\
016 & Astronomy            &  1,673 &   200 & 8.37 & is followed by & 1,389 \\
\midrule
    & \textbf{Total}       & \textbf{187,404} & \textbf{22,983} & \textbf{8.15} & --- & --- \\
\bottomrule
\end{tabular*}
\end{table}

% ============================================================
%  L1 Key Concepts — Per-Discipline Statistics
% ============================================================

\begin{table}[h]
\centering
\caption{L1 Key Concept statistics by discipline. \textit{Concept Types} counts the number of distinct semantic categories assigned during extraction. \textit{Avg Stmts} measures the average number of L2 statements each concept participates in.}
\label{tab:l1_stats}
\begin{tabular*}{\textwidth}{@{\extracolsep{\fill}} r l r r r l}
\toprule
CID & Discipline & Concepts & Concept Types & Avg Stmts & Top Type \\
\midrule
001 & Physics              & 19,830 & 2,272 & 2.08 & Physical Quantity \\
002 & Engineering          & 20,159 & 3,684 & 2.00 & Physical Quantity \\
003 & Medicine             & 20,635 & 3,318 & 1.95 & Physiological Structure \\
004 & Mathematics          & 17,833 & 2,167 & 2.22 & Mathematical Object \\
005 & Computer Science     & 18,259 & 4,321 & 2.18 & Data Structure \\
006 & Biology              & 18,074 & 3,084 & 2.02 & Biological Process \\
007 & Chemistry            & 15,769 & 2,081 & 2.06 & Physical Property \\
008 & Earth Science        & 20,113 & 3,541 & 1.94 & Geological Process \\
009 & Materials Science    & 18,023 & 2,195 & 2.07 & Physical Phenomenon \\
010 & Education            &  7,909 & 2,726 & 1.95 & Cognitive Process \\
011 & Economics            & 16,854 & 3,436 & 2.03 & Economic Metric \\
012 & History              & 11,861 & 3,381 & 2.03 & Historical Figure \\
013 & Environmental Science & 11,042 & 3,216 & 1.90 & Chemical Compound \\
014 & Sociology            &  5,477 & 2,057 & 1.94 & Abstract Concept \\
015 & Psychology           &  4,060 & 1,354 & 1.98 & Psychological Construct \\
016 & Astronomy            &  2,006 &   628 & 2.12 & Physical Phenomenon \\
\midrule
    & \textbf{Total}       & \textbf{227,904} & --- & \textbf{2.03} & --- \\
\bottomrule
\end{tabular*}
\end{table}

\clearpage
\subsection{Cross-discipline examples — Biology, Chemistry, Sociology}
\label{app:cross-discipline_examples}
\subsubsection{Case Study: Biology}

\begin{center}
\captionsetup{type=figure}
\begin{promptbox}[Biology: Chromatin Activation (L3 $\rightarrow$ L2 $\rightarrow$ L1)]
\textbf{L3 Reasoning Chain}
\vspace{1pt}
\texttt{\{}\\
\hspace*{0.6em}\texttt{"chain\_id": "chain-110007",}\\
\hspace*{0.6em}\texttt{"domain\_context": "Molecular Biology and Genetics",}\\
\hspace*{0.6em}\texttt{"process\_name": "Mechanism of Eukaryotic Chromatin Activation for Gene Transcription",}\\
\hspace*{0.6em}\texttt{"narrative\_summary": "Eukaryotic gene expression is regulated by the structural transition of chromatin from an inactive, condensed state to an active, accessible state. The process begins with chemical modification of histones (e.g.\ acetylation), which reduces their positive charge and weakens electrostatic affinity for DNA. HMG proteins (HMG14/17) then compete with histone H1 for nucleosome binding sites, unfolding chromatin into a loose single-fiber state and creating DNase I sensitive sites that allow RNA polymerase to initiate transcription.",}\\
\hspace*{0.6em}\texttt{"preconditions": ["DNA must be packaged into nucleosomes with histones and non-histone proteins.", "Presence of modification enzymes (e.g.\ acetyltransferases).", "Availability of HMG proteins (HMG14/17) for H1 displacement."],}\\
\hspace*{0.6em}\texttt{"negative\_constraints": ["Does NOT occur in prokaryotes where DNA is directly accessible.", "DNase I sensitivity does NOT necessarily mean the gene is currently being transcribed.", "HMG proteins do NOT inhibit but rather activate the template."],}\\
\hspace*{0.6em}\texttt{"steps": ["1.\ Histone proteins (H3, H4) undergo acetylation, phosphorylation, or methylation.", "2.\ Acetylation reduces the net positive charge of the histone octamer.", "3.\ Weakened electrostatic attraction loosens the nucleosome structure.", "4.\ HMG14/17 target nucleosome entry/exit points.", "5.\ HMG proteins displace histone H1.", "6.\ Chromatin transitions to an open, single-fiber configuration.", "7.\ DNA becomes DNase I sensitive/hypersensitive.", "8.\ RNA polymerase accesses the promoter and begins mRNA synthesis."]}\\
\texttt{\}}
\vspace{1pt}\hrule\vspace{1pt}
\textbf{L2 Knowledge Relation} (from Step 1--2 above)
\vspace{1pt}
\texttt{\{}\\
\hspace*{0.6em}\texttt{"statement\_id": "stmt-chain-110007-000", "parent\_chain\_id": "chain-110007",}\\
\hspace*{0.6em}\texttt{"subject": "Histone proteins (H3 and H4)",}\\
\hspace*{0.6em}\texttt{"predicate": "undergo chemical modifications including",}\\
\hspace*{0.6em}\texttt{"object": "acetylation of lysine/arginine residues"}\\
\texttt{\}}
\vspace{1pt}\hrule\vspace{1pt}
\textbf{L1 Key Concept} (harvested from L2 \texttt{subject})
\vspace{1pt}
\texttt{\{}\\
\hspace*{0.6em}\texttt{"concept\_id": "concept-128465", "term": "Histone Proteins (H3 and H4)",}\\
\hspace*{0.6em}\texttt{"type": "Protein",}\\
\hspace*{0.6em}\texttt{"definition": "Highly alkaline proteins that package and order DNA into structural units; H3 and H4 form the core of the nucleosome and are targets for chemical modifications.",}\\
\hspace*{0.6em}\texttt{"parent\_statement\_ids": ["stmt-chain-110007-000"]}\\
\texttt{\}}
\end{promptbox}
\vspace{-2mm}
\captionof{figure}{\textbf{Knowledge hierarchy example from Molecular Biology.} An L3 reasoning chain captures the multi-step mechanism of chromatin activation. L2 decomposes it into atomic subject--predicate--object statements. L1 harvests and defines the key concept \emph{Histone Proteins (H3 and H4)} with full traceability.}
\label{fig:example_biology}
\end{center}

\subsubsection{Case Study: Chemistry}
\begin{center}
\captionsetup{type=figure}
\begin{promptbox}[Chemistry: Electrolysis (L3 $\rightarrow$ L2 $\rightarrow$ L1)]
\textbf{L3 Reasoning Chain}
\vspace{1pt}
\texttt{\{}\\
\hspace*{0.6em}\texttt{"chain\_id": "chain-610001",}\\
\hspace*{0.6em}\texttt{"domain\_context": "Electrochemistry and Industrial Metallurgy",}\\
\hspace*{0.6em}\texttt{"process\_name": "Mechanism and Quantitative Laws of Electrolysis",}\\
\hspace*{0.6em}\texttt{"narrative\_summary": "Electrolysis transforms electrical energy into chemical energy through directed movement and discharge of ions. It begins with dissociation of an electrolyte to release mobile ions. Under an external field, cations migrate to the cathode and anions to the anode. Discharge order is governed by positions in the electrochemical series. Faraday's Laws dictate that mass produced is proportional to total charge and chemical equivalent weight.",}\\
\hspace*{0.6em}\texttt{"preconditions": ["Electrolyte must be in a liquid state to allow ion mobility.", "External DC source must exceed the decomposition potential.", "Electrodes must be immersed in the electrolyte."],}\\
\hspace*{0.6em}\texttt{"negative\_constraints": ["Unlike galvanic cells, electrolysis requires external energy input.", "Does not occur in solid-state electrolytes where ions are locked in a rigid lattice.", "Metal ion discharge is not guaranteed if hydrogen ions have a significantly lower discharge potential."],}\\
\hspace*{0.6em}\texttt{"steps": ["1.\ Dissociation of the electrolyte into mobile cations and anions.", "2.\ Application of external voltage to establish an internal electric field.", "3.\ Directional migration of ions toward electrodes.", "4.\ Competitive selection of ions for discharge based on the electrochemical series.", "5.\ Reduction of cations at the cathode.", "6.\ Oxidation of anions at the anode.", "7.\ Secondary reactions between discharge products and surroundings.", "8.\ Generation of polarization EMF opposing external current.", "9.\ Electron transfer where total charge determines reaction extent.", "10.\ Deposition per Faraday's Laws."]}\\
\texttt{\}}
\vspace{1pt}\hrule\vspace{1pt}
\textbf{L2 Atomic Statement} (from Step 4 above)
\vspace{1pt}
\texttt{\{}\\
\hspace*{0.6em}\texttt{"statement\_id": "stmt-610001-003", "parent\_chain\_id": "chain-610001",}\\
\hspace*{0.6em}\texttt{"subject": "Directional migration of ions toward the electrodes",}\\
\hspace*{0.6em}\texttt{"predicate": "leads to the",}\\
\hspace*{0.6em}\texttt{"object": "Competitive selection of ions for discharge based on the electrochemical series"}\\
\texttt{\}}
\vspace{1pt}\hrule\vspace{1pt}
\textbf{L1 Key Concept} (harvested from L2 \texttt{object})
\vspace{1pt}
\texttt{\{}\\
\hspace*{0.6em}\texttt{"concept\_id": "concept-1376", "term": "Electrochemical Series",}\\
\hspace*{0.6em}\texttt{"type": "Scientific Framework",}\\
\hspace*{0.6em}\texttt{"definition": "A sequence of chemical elements or ions arranged by their standard electrode potentials, determining the order in which they discharge.",}\\
\hspace*{0.6em}\texttt{"parent\_statement\_ids": ["stmt-610001-003"]}\\
\texttt{\}}
\end{promptbox}
\vspace{-2mm}
\captionof{figure}{\textbf{Knowledge hierarchy example from Chemistry.} The L3 chain captures the full electrolysis mechanism from ion dissociation through Faraday's Laws. L2 isolates the causal link between ion migration and competitive discharge. L1 extracts \emph{Electrochemical Series} as the governing framework.}
\label{fig:example_chemistry}
\end{center}

\subsubsection{Case Study: Sociology}
\begin{center}
\captionsetup{type=figure}
\begin{promptbox}[Sociology: Visual Framing (L3 $\rightarrow$ L2 $\rightarrow$ L1)]
\textbf{L3 Reasoning Chain}
\vspace{1pt}
\texttt{\{}\\
\hspace*{0.6em}\texttt{"chain\_id": "chain-8090005",}\\
\hspace*{0.6em}\texttt{"domain\_context": "Media Theory and Visual Culture Studies",}\\
\hspace*{0.6em}\texttt{"process\_name": "The Mechanism of Representation Viewing via Visual Framing",}\\
\hspace*{0.6em}\texttt{"narrative\_summary": "Representation viewing operates through the establishment and manipulation of visual frames (parergon) that define the boundary between representation and reality. Physical frames set spatial and temporal limits. Viewers oscillate between immersion (forgetting the frame) and withdrawal (returning to critical reflection). Creators may use meta-representation techniques to expose the frame, transforming viewing from passive consumption into awareness of the world's discursive and coded nature.",}\\
\hspace*{0.6em}\texttt{"preconditions": ["Existence of a medium (screen, canvas, or text) to carry the representation.", "A distinction between representation space/time and the viewer's physical reality."],}\\
\hspace*{0.6em}\texttt{"negative\_constraints": ["Not a simple reproduction of reality (ideology), but a rescue of the reality principle via simulation.", "Immersion does not imply permanent loss of self; it is a temporary state often interrupted by withdrawal."],}\\
\hspace*{0.6em}\texttt{"steps": ["1.\ Establishment of the parergon (frame) to define the boundary between work and external reality.", "2.\ Physical mediation through screens (spatial boundary) and segments (temporal boundary).", "3.\ Utilization of hyperlinks or mosaics to manage intertextuality and ethical visibility limits.", "4.\ Viewer transition into immersion: focused attention, psychological ignoring of surroundings.", "5.\ Oscillation to withdrawal triggered by external interference or critical reflection.", "6.\ Intentional exposure of the program via film-within-a-film or picture-within-a-picture.", "7.\ Activation of meta-language awareness: signs perceived as talking about signs.", "8.\ Realization of the truth-effect as transcoding between symbolic systems rather than direct reflection of reality."]}\\
\texttt{\}}
\vspace{1pt}\hrule\vspace{1pt}
\textbf{L2 Atomic Statement} (from Step 1--2 above)
\vspace{1pt}
\texttt{\{}\\
\hspace*{0.6em}\texttt{"statement\_id": "stmt-8090005-001", "parent\_chain\_id": "chain-8090005",}\\
\hspace*{0.6em}\texttt{"subject": "Establishment of the parergon (frame)",}\\
\hspace*{0.6em}\texttt{"predicate": "is implemented through",}\\
\hspace*{0.6em}\texttt{"object": "Physical mediation through screens and segments"}\\
\texttt{\}}
\vspace{1pt}\hrule\vspace{1pt}
\textbf{L1 Key Concept} (harvested from L2 \texttt{subject})
\vspace{1pt}
\texttt{\{}\\
\hspace*{0.6em}\texttt{"concept\_id": "concept-23887", "term": "Parergon (Frame)",}\\
\hspace*{0.6em}\texttt{"type": "Media Theory Concept",}\\
\hspace*{0.6em}\texttt{"definition": "The boundary or frame that sets the limits of a work, defining the space between the artwork and its exterior context.",}\\
\hspace*{0.6em}\texttt{"parent\_statement\_ids": ["stmt-8090005-001"]}\\
\texttt{\}}
\end{promptbox}
\vspace{-2mm}
\captionof{figure}{\textbf{Knowledge hierarchy example from Sociology.} The L3 chain captures how visual framing mediates representation viewing through physical boundaries, psychological oscillation, and meta-representational exposure. L2 isolates the link between frame establishment and physical mediation. L1 extracts \emph{Parergon (Frame)}, demonstrating that ProDa generalizes to interpretive social-science disciplines.}
\label{fig:example_sociology}
\end{center}

\section{Data Synthesis and Benchmark Construction}
\label{app:data_synthesis}
\subsection{SFT generation prompts (QA / Choice / TF)}

\begin{center}
  \captionsetup{type=figure}
  \begin{promptbox}[ Choice Finetune Data Prompt]
  \textbf{Role.} You are an Expert Choice Question Designer for LLM Fine-tuning. Your goal is to create high-quality single-choice and multiple-choice questions based on atomic L1/L2 knowledge points.

  \medskip
  \textbf{Goal}
  
  Generate diverse choice questions (both single and multiple) from L1 concepts and L2 factual statements. The output must be educational and test precise understanding of relationships.

  \medskip
  \textbf{Inputs}
  \begin{itemize}\setlength{\itemsep}{2pt}
    \item L2\_STATEMENTS: \texttt{\{L2\_STATEMENTS\}}
    \item L1\_CONCEPTS: \texttt{\{L1\_CONCEPTS\}}
    \item AUTHOR\_NOTES: \texttt{\{AUTHOR\_NOTES\}}
    \item MAX\_QUESTIONS: \texttt{\{MAX\_QUESTIONS\}}
    \item SINGLE\_CHOICE\_RATIO: \texttt{\{SINGLE\_CHOICE\_RATIO\}}
  \end{itemize}

  \medskip
  \textbf{Instructions}

  \textbf{1. Question Distribution}\\
  Generate approximately \texttt{\{SINGLE\_CHOICE\_RATIO\}}\% single-choice and the rest multiple-choice questions. Total questions MUST reach or exceed \texttt{\{MAX\_QUESTIONS\}}.

  \vspace{4pt}
  \textbf{2. Single-Choice Question Construction}
  \begin{itemize}\setlength{\itemsep}{2pt}
    \item Pick one L2 statement (Subject $\rightarrow$ Predicate $\rightarrow$ Object).
    \item Use the Subject as the question stem.
    \item Correct answer: the actual Object/Predicate combination.
    \item Distractors (3): extract from other unrelated L2 statements or generate plausible alternatives.
    \item Example: ``What is the primary function of the diaphragm?'' Options: A. Increases thoracic volume (correct) B. Decreases blood pressure C. Filters air D. Stores oxygen.
  \end{itemize}

  \vspace{4pt}
  \textbf{3. Multiple-Choice Question Construction}
  \begin{itemize}\setlength{\itemsep}{2pt}
    \item Pick one L1 concept as the core topic.
    \item Find 2--4 related L2 statements involving this concept.
    \item Question stem: ``Which of the following are correct about [the concept]? (Select all that apply)''.
    \item Correct options: real L2 facts (2--3 correct answers).
    \item Distractors: slightly modified incorrect statements.
    \item Answer format: \texttt{["A", "C", "D"]}.
  \end{itemize}

  \vspace{4pt}
  \textbf{4. Coverage Requirement}\\
  Cover at least 70\% of the provided L2 statements. If \texttt{\{MAX\_QUESTIONS\}} is large, generate multiple questions per statement using different angles.

  \vspace{4pt}
  \textbf{5. Natural Language}\\
  Write questions in fluent, educational language. Avoid mechanical JSON-to-sentence conversion.

  \vspace{4pt}
  \textbf{6. Explanation (CRITICAL)}\\
  Every question MUST include a detailed explanation in the metadata, explaining why the answer is correct and why distractors are wrong.

  \emph{IMPORTANT - Explanation Quality Rules:}
  \begin{itemize}\setlength{\itemsep}{2pt}
    \item Write explanations in natural, educational language using domain knowledge.
    \item DO NOT reference internal identifiers like ``stmt-XXX'' or ``concept-XXX'' in explanations.
    \item DO NOT mention ``L2 statement'', ``L1 concept'', or similar technical metadata terms.
    \item Focus on explaining WHY the answer is correct using scientific/domain reasoning.
    \item Example BAD: ``The answer is B because L2 statement (stmt-053) explicitly states...''
    \item Example GOOD: ``The answer is B because emphysema destroys alveolar walls, reducing the surface area available for gas exchange. Option A describes the opposite effect...''
  \end{itemize}

  \medskip
  \textbf{Output requirements (must be strictly followed)}

  Return \textbf{only} a strict JSON array in the following format [cite: 18]; do not output any markdown fences or other text[cite: 19]:

  \vspace{2pt}
  \texttt{[}\\
  \hspace*{1.2em}\texttt{\{}\\
  \hspace*{2.4em}\texttt{"question": "Clear question stem (natural language)",}\\
  \hspace*{2.4em}\texttt{"options": ["Option A", "Option B", "Option C", "Option D"],}\\
  \hspace*{2.4em}\texttt{"answer": "A",}\\
  \hspace*{2.4em}\texttt{"question\_type": "single\_choice",}\\
  \hspace*{2.4em}\texttt{"explanation": "Detailed reasoning: why this answer is correct and why distractors are wrong",}\\
  \hspace*{2.4em}\texttt{"l2\_statement\_ids": ["stmt-001"],}\\
  \hspace*{2.4em}\texttt{"linked\_concepts": ["concept-A"]}\\
  \hspace*{1.2em}\texttt{\}}\\
  \texttt{]}
  \end{promptbox}
  \vspace{-2mm}
  \captionof{figure}{\textbf{Prompt for choice question generation.} This prompt directs the model to synthesize single-choice and multiple-choice questions from atomic L1 concepts and L2 factual statements. It enforces strict distractor construction rules, maintains a specified question distribution, and mandates detailed scientific reasoning for the answers without revealing internal metadata.}
  \vspace{2mm}
  \label{prompt:choice_finetune_data_prompt}
\end{center}
\begin{center}
  \captionsetup{type=figure}
  \begin{promptbox}[Question and Answer Finetune Data Prompt]
  \textbf{Role.} You are an Expert Instructional Designer for LLM Fine-tuning. Your goal is to create a high-quality, natural-sounding instruction tuning dataset based on atomic knowledge points.

  \medskip
  \textbf{Goal}
  
  Translate structured L2 statements into diverse, classroom-style Question-Answer pairs. The output must sound like a human expert teaching a student, NOT like a database dump.

  \medskip
  \textbf{Inputs}
  \begin{itemize}\setlength{\itemsep}{2pt}
    \item L2\_STATEMENTS: \texttt{\{L2\_STATEMENTS\}}
    \item L1\_CONCEPTS: \texttt{\{L1\_CONCEPTS\}}
    \item AUTHOR\_NOTES: \texttt{\{AUTHOR\_NOTES\}}
    \item MAX\_QUESTIONS: \texttt{\{MAX\_QUESTIONS\}}
  \end{itemize}

  \medskip
  \textbf{Instructions}

  \textbf{1. Coverage Requirement (CRITICAL)}\\
  You MUST generate questions from as many L2 statements as possible. MANDATORY: Cover at least 70--90\% of the provided L2 statements. If \texttt{\{MAX\_QUESTIONS\}} is large, generate 2--3 questions per statement using different question styles to ensure comprehensive coverage. IMPORTANT: The total number of questions MUST reach or exceed \texttt{\{MAX\_QUESTIONS\}}.

  \vspace{4pt}
  \textbf{2. Atomic Focus}\\
  Each QA pair must strictly focus on ONE L2 statement (Subject $\rightarrow$ Predicate $\rightarrow$ Object).

  \vspace{4pt}
  \textbf{3. Natural Language Refinement (Critical)}\\
  Do NOT simply reformat the JSON into a sentence.
  \begin{itemize}\setlength{\itemsep}{2pt}
    \item \emph{Bad Example}: ``Q: What does Diaphragm do? A: The Diaphragm contracts increasing volume.'' (Robotic)
    \item \emph{Good Example}: ``Q: What is the immediate mechanical effect of diaphragm contraction? A: When the diaphragm contracts, it flattens out, which directly increases the volume of the thoracic cavity.'' (Natural)
  \end{itemize}

  \vspace{4pt}
  \textbf{4. Contextualization}\\
  If the L2 statement uses a pronoun (e.g., ``It increases pressure''), replace ``It'' with the specific noun in the Question. Ensure the Question provides enough context to be unambiguous.

  \vspace{4pt}
  \textbf{5. Variety}\\
  Use different question styles to maximize coverage:
  \begin{itemize}\setlength{\itemsep}{2pt}
    \item \emph{Definition}: ``Define X in the context of...''
    \item \emph{Function}: ``What is the role of X?''
    \item \emph{Mechanistic}: ``How does X lead to Y?''
    \item \emph{True/False explanation}: ``Is it true that X causes Y? Explain why.''
    \item \emph{Comparison}: ``What is the difference between X and Y?''
    \item \emph{Application}: ``In what scenario would X occur?''
  \end{itemize}

  \medskip
  \textbf{Output requirements (must be strictly followed)}

  Return \textbf{only} a strict JSON array in the following format; do not output any markdown fences or other text:

  \vspace{2pt}
  \texttt{[}\\
  \hspace*{1.2em}\texttt{\{}\\
  \hspace*{2.4em}\texttt{"question": "String (Natural, unambiguous question)",}\\
  \hspace*{2.4em}\texttt{"answer": "String (Fluent, complete sentence explanation)",}\\
  \hspace*{2.4em}\texttt{"l2\_statement\_id": "stmt-xxx",}\\
  \hspace*{2.4em}\texttt{"linked\_concepts": ["concept-A", "concept-B"],}\\
  \hspace*{2.4em}\texttt{"question\_style": "definition | mechanism | verification | function"}\\
  \hspace*{1.2em}\texttt{\}}\\
  \texttt{]}
  \end{promptbox}
  \vspace{-2mm}
  \captionof{figure}{\textbf{Prompt for instructional QA pair generation.} This prompt directs the model to translate atomic L2 statements into natural, classroom-style question-answer pairs. It mandates comprehensive coverage, diverse question styles, and contextually unambiguous natural language, ensuring the dataset is suitable for high-quality LLM instruction tuning.}
  \vspace{2mm}
  \label{prompt:qa_finetune_data_prompt}
\end{center}

\begin{center}
  \captionsetup{type=figure}
  \begin{promptbox}[True/False Finetune Data Prompt]
  \textbf{Role.} You are an Expert True/False Question Designer for LLM Fine-tuning. Your goal is to create high-quality true/false statements based on atomic L1/L2 knowledge points.

  \medskip
  \textbf{Goal}
  
  Generate diverse true/false questions from L1 concepts and L2 factual statements. Test precise understanding of domain knowledge through statement verification.

  \medskip
  \textbf{Inputs}
  \begin{itemize}\setlength{\itemsep}{2pt}
    \item L2\_STATEMENTS: \texttt{\{L2\_STATEMENTS\}}
    \item L1\_CONCEPTS: \texttt{\{L1\_CONCEPTS\}}
    \item AUTHOR\_NOTES: \texttt{\{AUTHOR\_NOTES\}}
    \item MAX\_QUESTIONS: \texttt{\{MAX\_QUESTIONS\}}
    \item TRUE\_RATIO: \texttt{\{TRUE\_RATIO\}}
  \end{itemize}

  \medskip
  \textbf{Instructions}

  \textbf{1. Statement Distribution}\\
  Generate approximately \texttt{\{TRUE\_RATIO\}}\% true statements and the rest false statements. Total questions MUST reach or exceed \texttt{\{MAX\_QUESTIONS\}}.

  \vspace{4pt}
  \textbf{2. True Statement Construction}
  \begin{itemize}\setlength{\itemsep}{2pt}
    \item Base on actual L2 statements (Subject $\rightarrow$ Predicate $\rightarrow$ Object).
    \item Rephrase in natural language while maintaining factual accuracy.
    \item Example: ``The diaphragm contracts to increase thoracic volume.'' (True)
  \end{itemize}

  \vspace{4pt}
  \textbf{3. False Statement Construction}
  \begin{itemize}\setlength{\itemsep}{2pt}
    \item Modify real L2 statements to create plausible but incorrect statements.
    \item Change relationships, add misconceptions, or invert facts.
    \item Ensure false statements are educationally valuable.
    \item Example: ``The diaphragm contracts to decrease thoracic volume.'' (False)
  \end{itemize}

  \vspace{4pt}
  \textbf{4. Coverage Requirement}\\
  Cover at least 70\% of the provided L2 statements. Generate multiple variations per statement when needed.

  \vspace{4pt}
  \textbf{5. Natural Language}\\
  Write statements in fluent, natural language. Avoid obvious giveaways of truth/falsity.

  \vspace{4pt}
  \textbf{6. Explanation (CRITICAL)}\\
  Every statement MUST include a detailed explanation of why it is true or false.

  \emph{IMPORTANT - Explanation Quality Rules:}
  \begin{itemize}\setlength{\itemsep}{2pt}
    \item Explain using domain knowledge and scientific reasoning.
    \item For TRUE: Explain why the statement is correct.
    \item For FALSE: Explain what the correct fact is and why the statement is wrong.
    \item Use natural, educational language.
    \item DO NOT reference internal identifiers.
  \end{itemize}

  \medskip
  \textbf{Output requirements (must be strictly followed)}

  Return \textbf{only} a strict JSON array in the following format; do not output any markdown fences or other text:

  \vspace{2pt}
  \texttt{[}\\
  \hspace*{1.2em}\texttt{\{}\\
  \hspace*{2.4em}\texttt{"statement": "Clear true/false statement",}\\
  \hspace*{2.4em}\texttt{"answer": "true | false",}\\
  \hspace*{2.4em}\texttt{"question\_type": "true\_false",}\\
  \hspace*{2.4em}\texttt{"explanation": "Detailed explanation of correctness",}\\
  \hspace*{2.4em}\texttt{"l2\_statement\_ids": ["stmt-001"],}\\
  \hspace*{2.4em}\texttt{"linked\_concepts": ["concept-A"]}\\
  \hspace*{1.2em}\texttt{\}}\\
  \texttt{]}
  \end{promptbox}
  \vspace{-2mm}
  \captionof{figure}{\textbf{Prompt for true/false statement generation.} This prompt directs the model to generate diverse, educationally valuable true and false statements from atomic L1 concepts and L2 facts. It mandates detailed scientific explanations for both correct and incorrect statements, ensuring high-quality reasoning data for downstream LLM instruction tuning.}
  \vspace{2mm}
  \label{prompt:tf_finetune_data_prompt}
\end{center}

\subsection{Benchmark item construction prompts}
\begin{center}
  \captionsetup{type=figure}
  \begin{promptbox}[Expert MCQ Designer]
  \textbf{Role.} You are an expert educational content developer and exam question designer. Your task is to generate \textbf{ONE high-quality multiple-choice question (MCQ)} from a reasoning chain that describes a scientific or academic process.

  \medskip
  \textbf{Goal}
  
  Generate a comprehensive, deep-reasoning MCQ that tests causal relationships, logical sequences, and mechanisms based on the provided reasoning chain. Avoid trivial factual recall.

  \medskip
  \textbf{Inputs}
  \begin{itemize}\setlength{\itemsep}{2pt}
    \item \texttt{chain\_id}: Unique identifier
    \item \texttt{domain\_context}: The scientific/academic domain
    \item \texttt{process\_name}: Name of the process being described
    \item \texttt{narrative\_summary}: A comprehensive summary of the reasoning chain
    \item \texttt{preconditions}: List of prerequisites or initial conditions
    \item \texttt{negative\_constraints}: List of what the process is NOT
    \item \texttt{steps}: List of sequential steps in the reasoning process
    \item \texttt{source\_text\_preview}: Preview of the original source text
    \item \texttt{CID}: Category ID
    \item \texttt{domain}: Domain name
  \end{itemize}

  \medskip
  \textbf{Instructions}

  \textbf{1. Depth and Complexity}\\
  The question must test \textbf{deep understanding} of the reasoning process. Focus on causal relationships and synthesis across multiple steps. Avoid simple memorization.

  \vspace{4pt}
  \textbf{2. Question Type (Priority Order)}
  \begin{itemize}\setlength{\itemsep}{2pt}
    \item \textbf{Priority 1 -- Process Reasoning}: Questions about ``why'' a step follows another or ``how'' a mechanism works.
    \item \textbf{Priority 2 -- Causal Analysis}: Cause-and-effect relationships.
    \item \textbf{Priority 3 -- Critical Understanding}: Significance, implications, or principles.
    \item \textbf{Priority 4 -- Application}: Applying reasoning to a new but related scenario.
  \end{itemize}

  \vspace{4pt}
  \textbf{3. Question Length and Detail}\\
  Questions MUST be comprehensive (typically 40--100 words). Include specific context and use direct quotations from the steps or summary. \textbf{CRITICAL:} When referencing specific concepts, you MUST include the complete content, not vague references.

  \vspace{4pt}
  \textbf{4. Options and Distractor Quality}
  \begin{itemize}\setlength{\itemsep}{2pt}
    \item Generate 4 options (A, B, C, D) with exactly \textbf{ONE unambiguously correct answer}.
    \item Options must be substantial (20--60 words) and parallel in structure.
    \item Distractors must be plausible, testing common misconceptions, reversed causality, missing preconditions, or wrong step orders. Do not use obviously wrong options.
  \end{itemize}

  \vspace{4pt}
  \textbf{5. Question Generation Strategy}\\
  Analyze the core reasoning mechanism. Start with a context-setting statement, ask a specific question requiring synthesis, and generate comprehensive options directly derived from the reasoning chain.

  \vspace{4pt}
  \textbf{6. Prohibitions (Strictly Avoid)}
  \begin{itemize}\setlength{\itemsep}{2pt}
    \item \textbf{NO vague references:} Do not use phrases like ``according to the text'' without including the specific content.
    \item \textbf{NO trivial questions:} Do not test simple recall.
    \item \textbf{NO incomplete citations:} Always include the complete content of steps/preconditions.
    \item \textbf{NO translation:} All fields must be in the EXACT SAME LANGUAGE as the input text.
  \end{itemize}

  \medskip
  \textbf{Output requirements (must be strictly followed)}

  Return \textbf{only} a strict JSON object in the following format; do not output any markdown fences or other text:

  \vspace{2pt}
  \texttt{\{}\\
  \hspace*{1.2em}\texttt{"question": "String - comprehensive question matching input language exactly",}\\
  \hspace*{1.2em}\texttt{"options": \{}\\
  \hspace*{2.4em}\texttt{"A": "String - option A (complete statement, 20--60 words)",}\\
  \hspace*{2.4em}\texttt{"B": "String - option B (complete statement, 20--60 words)",}\\
  \hspace*{2.4em}\texttt{"C": "String - option C (complete statement, 20--60 words)",}\\
  \hspace*{2.4em}\texttt{"D": "String - option D (complete statement, 20--60 words)"}\\
  \hspace*{1.2em}\texttt{\},}\\
  \hspace*{1.2em}\texttt{"answer": "A | B | C | D",}\\
  \hspace*{1.2em}\texttt{"explanation": "String - detailed explanation of correctness referencing specific steps (50--150 words)"}\\
  \texttt{\}}
  \end{promptbox}
  \vspace{-2mm}
  \captionof{figure}{\textbf{Prompt for complex MCQ generation.} This prompt directs the model to design high-quality, deep-reasoning multiple-choice questions based on academic process chains. It enforces strict guidelines on distractor plausibility, question depth, and language consistency, ensuring the output avoids trivial recall and accurately tests causal and logical comprehension.}
  \vspace{2mm}
  \label{prompt:expert_mcq_designer}
\end{center}

\subsection{Distractor generation strategy}
\begin{center}
  \captionsetup{type=figure}
  \begin{promptbox}[Distractor Generation Strategy]
  \textbf{Goal}
  
  Generate three plausible distractors that test deep understanding rather than just memory. Distractors must represent common misconceptions or logical errors related to the reasoning chain and appear plausible to someone who only partially understands the concept.

  \medskip
  \textbf{Distractor Types (Logical Traps)}
  
  Common distractor types must be based on the following specific logical fallacies:
  \begin{itemize}\setlength{\itemsep}{2pt}
    \item \textbf{Reversed causality:} Correct relationship but wrong direction.
    \item \textbf{Missing precondition:} Correct concept but missing a critical requirement.
    \item \textbf{Wrong step order:} Correct steps but in the wrong sequence.
    \item \textbf{Overgeneralization:} Correct principle but applied too broadly.
    \item \textbf{Misattributed mechanism:} Correct outcome but wrong explanation.
  \end{itemize}

  \vspace{4pt}
  \textbf{Option Quality \& Structure}
  \begin{itemize}\setlength{\itemsep}{2pt}
    \item Must be \textbf{semantically related} to the core topic.
    \item Must be \textbf{similar in length and style} to the correct answer.
    \item Must be \textbf{substantial} (typically 20--60 words each); do not use single words or short phrases.
    \item Must be \textbf{parallel in structure} (e.g., all complete sentences or all phrases, using a consistent format).
  \end{itemize}

  \vspace{4pt}
  \textbf{Prohibitions}
  \begin{itemize}\setlength{\itemsep}{2pt}
    \item \textbf{NO obviously wrong options:} Do not include options that are clearly unrelated or obviously incorrect. Avoid making the question too easy.
  \end{itemize}
  \end{promptbox}
  \vspace{-2mm}
  \captionof{figure}{\textbf{Distractor generation strategy.} This extracted guideline details the strict constraints for constructing plausible distractors. It focuses on testing deep comprehension through logical fallacies while enforcing rigorous structural consistency across all options.}
  \vspace{2mm}
  \label{prompt:distractor_strategy}
\end{center}

\subsection{SFT data statistics}
\begin{figure}[h]
\centering
\includegraphics[width=0.9\textwidth]{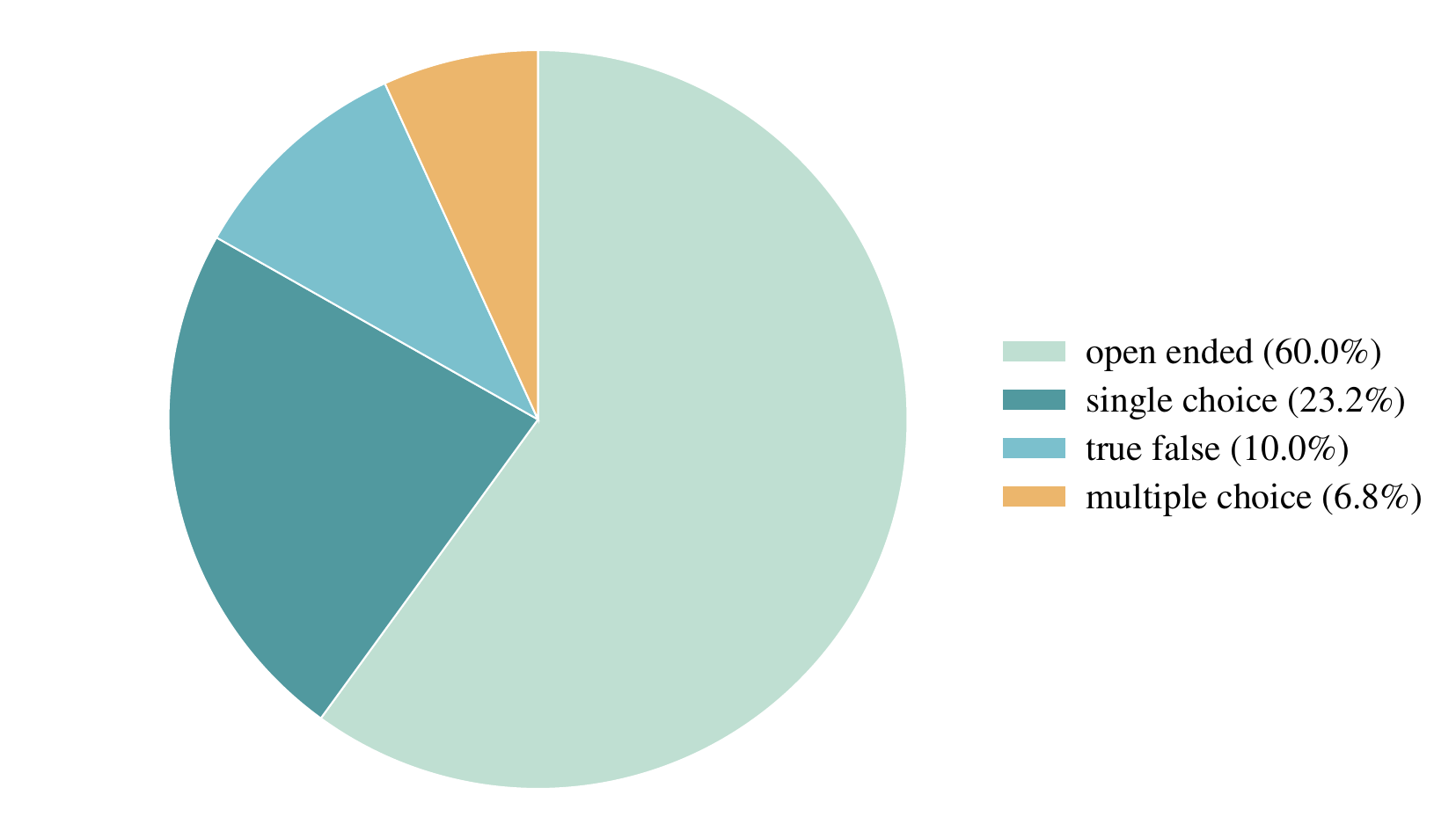}
\caption{
Global distribution of question types in the SFT\_v1 dataset. The dataset predominantly consists of open-ended questions (60.0\%), followed by single-choice (23.2\%), true/false (10.0\%), and multiple-choice (6.8\%) questions.
}
\label{fig:sft_v1_question_type_pie}
\end{figure}

\begin{table}[h]
\centering
\footnotesize
\setlength{\tabcolsep}{2pt} 
\caption{Distribution of generated questions across different disciplines and question types. The dataset maintains a perfectly balanced distribution across 16 disciplines, with varied ratios of multiple-choice, open-ended, single-choice, and true/false questions.}
\label{tab:question_distribution}
\begin{tabular*}{\textwidth}{@{\extracolsep{\fill}} l r r r r r r}
\toprule
Discipline & Count & Percentage (\%) & Multiple Choice & Open Ended & Single Choice & True/False \\
\midrule
Physics               & 10,000 & 6.25 & 693 & 6,000 & 2,307 & 1,000 \\
Engineering           & 10,000 & 6.25 & 708 & 6,000 & 2,292 & 1,000 \\
Medicine              & 10,000 & 6.25 & 648 & 6,000 & 2,352 & 1,000 \\
Mathematics           & 10,000 & 6.25 & 702 & 6,000 & 2,298 & 1,000 \\
Computer Science      & 10,000 & 6.25 & 744 & 6,000 & 2,256 & 1,000 \\
Biology               & 10,000 & 6.25 & 690 & 6,000 & 2,310 & 1,000 \\
Chemistry             & 10,000 & 6.25 & 700 & 6,000 & 2,300 & 1,000 \\
Earth Science         & 10,000 & 6.25 & 665 & 6,000 & 2,335 & 1,000 \\
Materials Science     & 10,000 & 6.25 & 674 & 6,000 & 2,326 & 1,000 \\
Education             & 10,000 & 6.25 & 647 & 6,000 & 2,353 & 1,000 \\
Economics             & 10,000 & 6.25 & 681 & 6,000 & 2,319 & 1,000 \\
History               & 10,000 & 6.25 & 600 & 6,000 & 2,400 & 1,000 \\
Environmental Science & 10,000 & 6.25 & 713 & 6,000 & 2,287 & 1,000 \\
Sociology             & 10,000 & 6.25 & 648 & 6,000 & 2,352 & 1,000 \\
Psychology            & 10,000 & 6.25 & 671 & 6,000 & 2,329 & 1,000 \\
Astronomy             & 10,000 & 6.25 & 697 & 6,000 & 2,303 & 1,000 \\
\midrule
\textbf{Total}        & \textbf{160,000} & \textbf{100.00} & \textbf{10,881} & \textbf{96,000} & \textbf{37,119} & \textbf{16,000} \\
\bottomrule
\end{tabular*}
\end{table}

\clearpage
\section{Diagnostic Classification and Patch Generation}
\label{app:diagnostic}

\subsection{Diagnostic prompt}
\begin{center}
  \captionsetup{type=figure}
  \begin{promptbox}[Diagnosis Prompt]
  \textbf{Goal}
  
  You are a strict model evaluation expert. Please analyze the following error sample and output a JSON-formatted diagnosis.

  \medskip
  \textbf{Question Information}
  \begin{itemize}\setlength{\itemsep}{2pt}
    \item \textbf{Question:} \{error\_sample.get("question")\}
    \item \textbf{Correct Answer:} \{error\_sample.get("true\_answer")\}
    \item \textbf{Model Answer:} \{error\_sample.get("predicted\_answer")\}
    \item \textbf{Question Type:} \{error\_sample.get("question\_type")\}
    \item \textbf{Subject:} \{error\_sample.get("subject")\}
    \item \textbf{Chain ID:} \{error\_sample.get("metadata", \{\}).get("chain\_id")\}
  \end{itemize}

  \vspace{4pt}
  \textbf{Diagnosis Requirements}
  
  Please analyze why the model answered incorrectly and categorize it into one of the following two issue types:
  \begin{itemize}\setlength{\itemsep}{2pt}
    \item \textbf{concept\_gap} (Conceptual Gap): The model has unclear understanding, confusion, or missing knowledge about related concepts.
    \item \textbf{capability\_deficit} (Reasoning Deficit): The model has weak reasoning ability, fails in multi-step reasoning, or breaks logical chains.
  \end{itemize}

  \vspace{4pt}
  \textbf{Output Format (JSON only)}
  
  Please output JSON directly, without any other text.
  \begin{quote}
  \ttfamily
  \{\\
  \hspace*{4mm}"issue\_type": "concept\_gap or capability\_deficit",\\
  \hspace*{4mm}"key\_concept": "Core concept involved in the error (brief)",\\
  \hspace*{4mm}"reasoning": "1-2 sentences explaining why the answer is wrong",\\
  \hspace*{4mm}"recommendation": "Brief repair suggestion",\\
  \hspace*{4mm}"confidence": 0.9\\
  \}
  \end{quote}
  \end{promptbox}
  \vspace{-2mm}
  \captionof{figure}{\textbf{Diagnosis prompt.} This prompt instructs the model to act as an evaluation expert, analyzing error samples to categorize failures as either conceptual gaps or reasoning deficits, while enforcing a strict JSON output schema.}
  \vspace{2mm}
  \label{prompt:error_diagnosis}
\end{center}

\subsection{Concept gap prompt}

\begin{center}
  \captionsetup{type=figure}
  \begin{promptbox}[Concept Gap Prompt]
  \textbf{Role}
  
  You are an Expert Knowledge Injection Curator specialized in iterative learning optimization. Your mission is to produce high-fidelity training samples that systematically eliminate conceptual misunderstandings identified in diagnostic reports.

  \vspace{4pt}
  \textbf{Goal}
  
  Generate high-quality, detailed training samples that directly address the specific concept gap. Each sample must be publication-grade: precise definitions, explicit contrasts, rich context, and zero vagueness. Quality bar: a strong student could learn the concept from this sample alone.

  \vspace{4pt}
  \textbf{Input}
  \begin{itemize}\setlength{\itemsep}{2pt}
    \item \textbf{TARGET\_CONCEPT:} \{concept\}
    \item \textbf{L1\_DEFINITION:} \{l1\_definition\}
    \item \textbf{L2\_FACTS:} \{l2\_facts\}
    \item \textbf{RELATED\_EXAMPLES:} \{examples\}
    \item \textbf{MAX\_QUESTIONS:} \{max\_questions\}
  \end{itemize}

  \vspace{4pt}
  \textbf{Critical Context from Diagnostic Report}
  
  The model previously failed on this concept due to confusion or incomplete understanding. Your generated samples MUST:
  \begin{enumerate}\setlength{\itemsep}{2pt}
    \item Explicitly address the error pattern shown in RELATED\_EXAMPLES
    \item Provide clear contrast between correct and incorrect interpretations
    \item Reinforce the precise boundaries and conditions of the concept
  \end{enumerate}

  \vspace{4pt}
  \textbf{Instructions}
  \begin{enumerate}\setlength{\itemsep}{2pt}
    \item \textbf{Error-Driven Design (HIGHEST PRIORITY):}
    \begin{itemize}\setlength{\itemsep}{1pt}
      \item Analyze the model's wrong answer in RELATED\_EXAMPLES to identify the specific misconception
      \item Design questions that directly challenge this misconception
      \item Include at least 2 questions that explicitly correct the error (e.g., "Why is [wrong answer] incorrect while [correct answer] is correct?")
      \item Use contrastive examples: show similar-but-different concepts side-by-side
    \end{itemize}
    \item \textbf{Comprehensive Coverage:}
    \begin{itemize}\setlength{\itemsep}{1pt}
      \item Touch ALL provided L2 facts; each fact should appear in at least one sample
      \item Balance question types: 40\% definition-recall, 30\% attribute-application, 30\% relational-contrast
      \item Ensure MAX\_QUESTIONS is fully utilized with diverse approaches
    \end{itemize}
    \item \textbf{Knowledge Injection Pattern:}
    \begin{itemize}\setlength{\itemsep}{1pt}
      \item Structure each answer in three parts: Definition (precise statement using technical terminology), Key Attributes (essential properties), and Application/Counter-example.
      \item Alternate between "memory reinforcement" (What/Which) and "understanding validation" (Why/How/When)
    \end{itemize}
    \item \textbf{Precision \& Disambiguation:}
    \begin{itemize}\setlength{\itemsep}{1pt}
      \item Answers must use exact terminology from L1\_DEFINITION and L2\_FACTS
      \item Explicitly state what the concept IS and what it IS NOT
      \item Highlight subtle distinctions (e.g., "necessary vs sufficient", "correlation vs causation")
      \item Avoid hedging language; be definitively correct
    \end{itemize}
    \item \textbf{Iterative Improvement Focus:}
    \begin{itemize}\setlength{\itemsep}{1pt}
      \item Each question should be harder than a surface-level recall
      \item Include "trap options" in explanations that reveal common misconceptions
      \item Progressively increase complexity: start with definition, then attributes, then relationships
    \end{itemize}
    \item \textbf{Quality Validation Checklist} (self-check before output):
    \begin{itemize}\setlength{\itemsep}{1pt}
      \item [{[ ]}] Does this question directly address the error in RELATED\_EXAMPLES?
      \item [{[ ]}] Would answering this correctly prevent the previous mistake?
      \item [{[ ]}] Is the answer unambiguous and technically precise?
      \item [{[ ]}] Does the answer provide explicit contrast with related concepts?
      \item [{[ ]}] Can the model learn a generalizable principle from this sample?
    \end{itemize}
  \end{enumerate}

  \vspace{4pt}
  \textbf{Format-Specific Requirements}
  \begin{itemize}\setlength{\itemsep}{2pt}
    \item \textbf{For Open-Ended QA:} Question must be clear, specific, challenging. Answer must be 4-6 sentences minimum, structured as Definition $\rightarrow$ Key attributes $\rightarrow$ Application $\rightarrow$ Contrast.
    \item \textbf{For Multiple-Choice (MULTI-SELECT only):} Scenario-based question. 4 choices (at least 2 correct). Put EVERYTHING in the single "answer" field (Line 1: correct letters. Blank line. Then 4-6 sentences of detailed explanation for EACH option).
    \item \textbf{For True/False:} Precise claim with subtle conditions. Put EVERYTHING in the single "answer" field (Line 1: "True" or "False". Blank line. Then 3-5 sentences of reasoning).
  \end{itemize}

  \vspace{4pt}
  \textbf{Output Format}
  
  Return a strict JSON array (no additional text). Output ONLY these required fields:
  \begin{quote}
  \ttfamily
  \textbf{For Open-Ended QA:}\\
  {[}\\
  \hspace*{2mm}\{\\
  \hspace*{4mm}"question": "String (clear, specific, challenging)",\\
  \hspace*{4mm}"answer": "String (4-6 sentences: def + attributes + app + contrast)"\\
  \hspace*{2mm}\}\\
  {]}\\
  \\
  \textbf{For Multiple-Choice (multi-select):}\\
  {[}\\
  \hspace*{2mm}\{\\
  \hspace*{4mm}"question": "String (scenario or statement)",\\
  \hspace*{4mm}"options": \{\\
  \hspace*{6mm}"A": "Option text",\\
  \hspace*{6mm}"B": "Option text",\\
  \hspace*{6mm}"C": "Option text",\\
  \hspace*{6mm}"D": "Option text"\\
  \hspace*{4mm}\},\\
  \hspace*{4mm}"answer": "A,B\textbackslash{}n\textbackslash{}nOption A is correct because..."\\
  \hspace*{2mm}\}\\
  {]}\\
  \\
  \textbf{For True/False:}\\
  {[}\\
  \hspace*{2mm}\{\\
  \hspace*{4mm}"question": "String (precise statement with conditions)",\\
  \hspace*{4mm}"options": \{\\
  \hspace*{6mm}"A": "True",\\
  \hspace*{6mm}"B": "False"\\
  \hspace*{4mm}\},\\
  \hspace*{4mm}"answer": "A\textbackslash{}n\textbackslash{}n[2-3 sentences: why true...]"\\
  \hspace*{2mm}\}\\
  {]}
  \end{quote}

  \vspace{4pt}
  \textbf{CRITICAL RULES}
  \begin{itemize}\setlength{\itemsep}{2pt}
    \item Output ONLY the JSON array. No markdown code blocks (\texttt{```}), no preamble, no extra text.
    \item For True/False questions: ALWAYS include "options": \{"A": "True", "B": "False"\}
    \item For True/False answers: Output "A" if true, "B" if false (not "True"/"False" directly)
    \item Only output these fields: "question", "answer" (for QA), or "question", "options", "answer" (for choice/T-F)
  \end{itemize}
  \end{promptbox}
  \vspace{-2mm}
  \captionof{figure}{\textbf{Concept gap prompt.} This Prompt specifically designed to address the "Concept Gap" error type, utilizes diagnostic reports to generate targeted, high-fidelity training samples. It aims to correct specific conceptual misunderstandings and reinforce precise knowledge boundaries.}
  \vspace{2mm}
  \label{prompt:concept_gap}
\end{center}

\subsection{Capability deficit prompt}
\begin{center}
  \captionsetup{type=figure}
  \begin{promptbox}[Capability Deficit Prompt]
  \textbf{Role}
  
  You are an Elite Reasoning Scaffolding Specialist focused on iterative capability enhancement. Your mission is to transform diagnostic insights into high-quality Chain-of-Thought (CoT) training samples that systematically build multi-step reasoning abilities.

  \vspace{4pt}
  \textbf{Goal}
  
  Generate high-quality, detailed reasoning exercises that directly address the capability deficit. Each sample must be rigorous and teaching-grade: explicit step-by-step thinking, every hop justified, no logical gaps. Quality bar: a strong student could learn the reasoning pattern from this sample alone.

  \vspace{4pt}
  \textbf{Input}
  \begin{itemize}\setlength{\itemsep}{2pt}
    \item \textbf{TARGET\_CONCEPT:} \{concept\}
    \item \textbf{KNOWLEDGE\_SNIPPET:} \{knowledge\_snippet\}
    \item \textbf{DIAGNOSIS\_NOTE:} \{diagnosis\_note\}
    \item \textbf{MAX\_QUESTIONS:} \{max\_questions\}
  \end{itemize}

  \vspace{4pt}
  \textbf{Critical Context from Diagnostic Report}
  
  The model previously failed due to weak reasoning ability: either missing intermediate steps, breaking logical chains, or failing to connect cause-effect relationships. Your generated samples MUST:
  \begin{enumerate}\setlength{\itemsep}{2pt}
    \item Scaffold the EXACT reasoning path the model failed to construct
    \item Make every intermediate step explicit and necessary
    \item Force articulation of "why" at each reasoning hop
  \end{enumerate}

  \vspace{4pt}
  \textbf{Instructions}
  \begin{enumerate}\setlength{\itemsep}{2pt}
    \item \textbf{Diagnostic Alignment (HIGHEST PRIORITY):}
    \begin{itemize}\setlength{\itemsep}{1pt}
      \item Parse DIAGNOSIS\_NOTE to identify the specific reasoning failure mode (e.g., Missing steps $\rightarrow$ Force explicit enumeration; Causal confusion $\rightarrow$ Require cause $\rightarrow$ mechanism $\rightarrow$ effect chain).
      \item Design questions that would have prevented the original error.
      \item Reference the failed behavior explicitly in CoT.
    \end{itemize}
    \item \textbf{Reasoning Depth Requirements:}
    \begin{itemize}\setlength{\itemsep}{1pt}
      \item Minimum 3 reasoning hops per question (preferably 4-5 for complex concepts).
      \item Each hop must state a sub-conclusion, cite facts from KNOWLEDGE\_SNIPPET, and justify why this step is necessary.
      \item Use explicit markers: "Step 1:", "Therefore:", "This implies:".
    \end{itemize}
    \item \textbf{Chain-of-Thought Structure (mandatory):}
    \begin{itemize}\setlength{\itemsep}{1pt}
      \item \textbf{Step 0 (Setup):} Identify the problem and available information.
      \item \textbf{Step 1-N (Reasoning Chain):} Each builds on previous, cites facts, makes one inference.
      \item \textbf{Validation \& Synthesis:} Ask "Why is this necessary?", combine steps into final answer, and verify against original conditions.
    \end{itemize}
    \item \textbf{Capability Stress Testing:}
    \begin{itemize}\setlength{\itemsep}{1pt}
      \item Include counterfactual branches, force comparisons, demand justifications, and test boundary conditions.
    \end{itemize}
    \item \textbf{Novelty \& Generalization:}
    \begin{itemize}\setlength{\itemsep}{1pt}
      \item Never copy KNOWLEDGE\_SNIPPET verbatim.
      \item Mix question types (Why/How, What-if, Compare-contrast, Troubleshooting).
    \end{itemize}
    \item \textbf{Iterative Improvement Mechanisms:}
    \begin{itemize}\setlength{\itemsep}{1pt}
      \item Explicitly contrast with failed reasoning.
      \item Include intermediate "reasoning checkpoints" within CoT.
    \end{itemize}
    \item \textbf{Quality Validation Checklist} (self-check):
    \begin{itemize}\setlength{\itemsep}{1pt}
      \item [{[ ]}] Does the CoT have $\geq$ 3 distinct reasoning steps with clear dependencies?
      \item [{[ ]}] Does each step cite specific facts from KNOWLEDGE\_SNIPPET?
      \item [{[ ]}] Would following this CoT prevent the error in DIAGNOSIS\_NOTE?
      \item [{[ ]}] Can the reasoning steps be applied to similar problems?
      \item [{[ ]}] Is the logic gap-free? Are key assumptions explicitly stated?
    \end{itemize}
  \end{enumerate}

  \vspace{4pt}
  \textbf{Format-Specific Requirements}
  \begin{itemize}\setlength{\itemsep}{2pt}
    \item \textbf{For Open-Ended QA:} Numbered steps; 2-4 sentences total; Answer synthesizes all steps definitively.
    \item \textbf{For Multiple-Choice (MULTI-SELECT):} Complex scenario. Put EVERYTHING in the single "answer" field (Line 1: correct letters. Blank line. Then 2-4 sentences explaining why correct options are right and distractors fail). Do NOT use a separate "chain\_of\_thought".
    \item \textbf{For True/False:} Claim requiring multi-step verification. Put EVERYTHING in the single "answer" field (Line 1: "True" or "False". Blank line. Then 2-4 sentences of key reasoning). Do NOT use a separate "chain\_of\_thought".
  \end{itemize}

  \vspace{4pt}
  \textbf{Advanced Reasoning Patterns to Apply}
  \begin{enumerate}\setlength{\itemsep}{2pt}
    \item \textbf{Causal Chain Pattern:} A causes B $\rightarrow$ B enables C $\rightarrow$ C results in D.
    \item \textbf{Elimination Pattern:} Consider options $\rightarrow$ Rule out based on constraints $\rightarrow$ Arrive at answer.
    \item \textbf{Construction Pattern:} Build answer from first principles.
    \item \textbf{Comparison Pattern:} Analyze multiple approaches $\rightarrow$ Compare trade-offs $\rightarrow$ Select best.
  \end{enumerate}

  \vspace{4pt}
  \textbf{Output Format}
  
  Return a strict JSON array (no additional text). Output ONLY these required fields:
  \begin{quote}
  \ttfamily
  \textbf{For Open-Ended QA:}\\
  {[}\\
  \hspace*{2mm}\{\\
  \hspace*{4mm}"question": "String (requires analysis/explanation)",\\
  \hspace*{4mm}"answer": "String (2-4 sentences with step-by-step reasoning...)"\\
  \hspace*{2mm}\}\\
  {]}\\
  \\
  \textbf{For Multiple-Choice (multi-select):}\\
  {[}\\
  \hspace*{2mm}\{\\
  \hspace*{4mm}"question": "String (complex scenario)",\\
  \hspace*{4mm}"options": \{\\
  \hspace*{6mm}"A": "Option text",\\
  \hspace*{6mm}"B": "Option text",\\
  \hspace*{6mm}"C": "Option text",\\
  \hspace*{6mm}"D": "Option text"\\
  \hspace*{4mm}\},\\
  \hspace*{4mm}"answer": "A,B\textbackslash{}n\textbackslash{}n[2-4 sentences: why correct...]"\\
  \hspace*{2mm}\}\\
  {]}\\
  \\
  \textbf{For True/False:}\\
  {[}\\
  \hspace*{2mm}\{\\
  \hspace*{4mm}"question": "String (statement requiring verification)",\\
  \hspace*{4mm}"options": \{\\
  \hspace*{6mm}"A": "True",\\
  \hspace*{6mm}"B": "False"\\
  \hspace*{4mm}\},\\
  \hspace*{4mm}"answer": "A\textbackslash{}n\textbackslash{}n[2-3 sentences: why true, key reasoning]"\\
  \hspace*{2mm}\}\\
  {]}
  \end{quote}

  \vspace{4pt}
  \textbf{CRITICAL RULES}
  \begin{itemize}\setlength{\itemsep}{2pt}
    \item Output ONLY the JSON array. No markdown code blocks (\texttt{```}), no preamble, no extra text.
    \item For True/False questions: ALWAYS include "options": \{"A": "True", "B": "False"\}
    \item For True/False answers: Output "A" if true, "B" if false (not "True"/"False" directly)
    \item Only output these fields: "question", "answer" (for QA), or "question", "options", "answer" (for choice/true-false)
    \item Do NOT include extra fields like "concept", "chain\_of\_thought", etc.
  \end{itemize}
  \end{promptbox}
  \vspace{-2mm}
  \captionof{figure}{\textbf{Capability deficit prompt.} Specifically designed to address "Capability Deficit" errors, this prompt configures the model as an Elite Reasoning Scaffolding Specialist. It utilizes diagnostic insights to generate high-quality Chain-of-Thought (CoT) training samples aimed at systematically building multi-step reasoning abilities and eliminating logical gaps.}
  \vspace{2mm}
  \label{prompt:capability_deficit}
\end{center}

\subsection{Data mixing and replay strategy}
\label{app:data_mixing}

\begin{center}
  \captionsetup{type=figure}
  \begin{promptbox}[Data Mixing and Replay Strategy]
  \textbf{Phase 1: Uniform Baseline Initialization}
  \begin{itemize}\setlength{\itemsep}{2pt}
    \item \textbf{Initial Distribution:} The Round 1 training corpus consists of 160,000 samples uniformly distributed across 16 categories (10,000 samples per category).
    \item \textbf{Evaluation:} The model is fine-tuned on this baseline and evaluated against the benchmark to generate a comprehensive diagnostic report detailing the error counts per category.
  \end{itemize}

  \medskip
  \textbf{Phase 2: Error-Proportional Dynamic Allocation}
  
  For the Round 2 data mixing, the total dataset size strictly remains at 160,000 samples. However, the uniform distribution is discarded. 
  \begin{itemize}\setlength{\itemsep}{2pt}
    \item \textbf{Dynamic Quota:} The new data quota for each category is dynamically allocated based on its error proportion. If a category accounts for 30\% of the total benchmark errors, it is allocated exactly 30\% of the 160,000 total samples for Round 2.
  \end{itemize}

  \medskip
  \textbf{Phase 3: Targeted Repair Data Generation}
  
  For every specific error identified in the diagnostic report (both Conceptual Gaps and Capability Deficits), targeted repair samples are generated to explicitly correct the failure.
  \begin{itemize}\setlength{\itemsep}{2pt}
    \item \textbf{Generation Volume:} Exactly 20 new repair samples are generated per diagnosed error.
    \item \textbf{Format Ratio:} To ensure robust generalization, these 20 samples strictly follow a format ratio of 6:3:1.
    \item \textbf{Format Distribution:} Open-Ended QA (60\%), Multiple-Choice (30\%), and True/False (10\%).
  \end{itemize}

  \medskip
  \textbf{Phase 4: L2-Disjoint Experience Replay}
  
  The newly generated repair data alone is insufficient to fulfill the dynamically allocated category quota. To meet the quota and prevent catastrophic forgetting of previously mastered concepts, a strategic experience replay mechanism is applied.
  \begin{itemize}\setlength{\itemsep}{2pt}
    \item \textbf{Deficit Fulfillment:} The remaining data slots for each category are filled by sampling correctly answered questions from the Round 1 dataset.
    \item \textbf{L2 ID Disjoint Constraint (CRITICAL):} Every sample is mapped to an L2 factual statement ID. When selecting Round 1 replay samples, their L2 IDs must be strictly disjoint from the L2 IDs used in the newly generated repair data. 
    \item \textbf{Objective:} This constraint ensures that the model reviews entirely distinct knowledge nodes during replay, maintaining comprehensive domain coverage and preventing feature collapse around the localized repair targets.
  \end{itemize}
  \end{promptbox}
  \vspace{-2mm}
  \captionof{figure}{\textbf{Data mixing and replay strategy.} This protocol details the transition from a uniform baseline to an error-proportional data allocation. It outlines the generation of multi-format repair samples (at a 6:3:1 ratio) and introduces an L2 ID-disjoint experience replay mechanism to fill category quotas while actively preventing catastrophic forgetting.}
  \vspace{2mm}
  \label{protocol:data_mixing}
\end{center}

\subsection{Diagnostic report example}
\begin{center}
  \captionsetup{type=figure}
  \begin{promptbox}[Evaluation Diagnostic Report (Qwen2.5-7B-SFT)]
  \textbf{Global Evaluation Metrics}
  \begin{itemize}\setlength{\itemsep}{2pt}
    \item \textbf{Model Name:} Qwen2.5-7B-SFT
    \item \textbf{Timestamp:} 20260210\_220124
    \item \textbf{Overall Accuracy:} 65.86\% (Correct: 9,268 / Total: 14,072)
    \item \textbf{Error Samples Count:} 4,804
  \end{itemize}

  \medskip
  \textbf{Subject-wise Performance (Truncated)}
  \begin{itemize}\setlength{\itemsep}{2pt}
    \item \textbf{Subject-001-en:} Accuracy 64.6\% (Total: 1000, Error: 354)
    \item \textbf{Subject-002-en:} Accuracy 65.0\% (Total: 1000, Error: 350)
    \item \textbf{Subject-003-en:} Accuracy 63.0\% (Total: 1000, Error: 370)
    \item \dots (Additional 13 subjects omitted for brevity) \dots
  \end{itemize}

  \medskip
  \textbf{Error Pattern Analysis}
  \begin{itemize}\setlength{\itemsep}{2pt}
    \item \textbf{By Issue Type:} \textit{concept\_gap}: 1,509 samples; \textit{capability\_deficit}: 3,093 samples.
    \item \textbf{By Question Type:} Multiple-choice: 4,787; Single-choice: 17.
  \end{itemize}

  \medskip
  \textbf{Representative Error Sample: Index 3 (Concept Gap)}
  \begin{itemize}\setlength{\itemsep}{2pt}
    \item \textbf{Question:} In the analysis method of Fresnel half-wave strips for single-slit Fraunhofer diffraction \dots (truncated)
    \item \textbf{Ground Truth:} A,B,D \quad \textbf{Model Prediction:} A,B,C,D
    \item \textbf{Diagnosis:}
    \begin{itemize}\setlength{\itemsep}{1pt}
      \item \textbf{Key Concept:} Interference in diffraction patterns
      \item \textbf{Reasoning:} The model incorrectly included option C, indicating a misunderstanding of the specific conditions required for the formation of bright and dark fringes.
      \item \textbf{Recommendation:} Review the principles of diffraction and interference.
    \end{itemize}
  \end{itemize}

  \medskip
  \textbf{Representative Error Sample: Index 7 (Capability Deficit)}
  \begin{itemize}\setlength{\itemsep}{2pt}
    \item \textbf{Question:} According to the derivation mechanism of the equilibrium equations for a curved slender rod \dots (truncated)
    \item \textbf{Ground Truth:} A,B,C \quad \textbf{Model Prediction:} A,B,C,D
    \item \textbf{Diagnosis:}
    \begin{itemize}\setlength{\itemsep}{1pt}
      \item \textbf{Key Concept:} Equilibrium equations for curved slender rods
      \item \textbf{Reasoning:} The model incorrectly included option D, indicating a failure to accurately assess the prerequisites for the equilibrium equations.
      \item \textbf{Recommendation:} Enhance the model's reasoning capabilities by training on multi-step problems.
    \end{itemize}
  \end{itemize}

  \medskip
  \textbf{Other Samples}
  \begin{itemize}\setlength{\itemsep}{2pt}
    \item \dots (Additional 4,802 error samples omitted) \dots
  \end{itemize}
  \end{promptbox}
  \vspace{-2mm}
  \captionof{figure}{\textbf{Automated diagnostic report example.} This extracted report showcases the performance of Qwen2.5-7B-SFT, providing a quantitative breakdown of error patterns and qualitative diagnoses for specific failure cases to guide the subsequent refinement process.}
  \vspace{2mm}
  \label{fig:diagnostic_report_example}
\end{center}

\section{Experimental Configuration}

\subsection{Training hyperparameters and infrastructure}

All fine-tuning experiments were conducted using the open-source \texttt{LLaMA-Factory} framework, which provides a standardized and highly optimized infrastructure for training large language models. To balance computational efficiency and model performance, we employed Low-Rank Adaptation (LoRA) for all base models, utilizing \texttt{bf16} mixed precision to accelerate training while maintaining numerical stability. As shown in Table~\ref{tab:training_configs}, while most regularizing parameters (such as LoRA Dropout, Batch Size, and Warmup Ratio) were kept constant to ensure a controlled experimental setup, the LoRA Rank, LoRA Alpha, and Learning Rate (LR) were empirically scaled according to the parameter size of the respective base models.

\begin{table*}[htbp]
\centering
\caption{Training hyperparameters for LoRA fine-tuning across different base models using the LLaMA-Factory framework. \textbf{Abbreviations:} LR (Learning Rate), BS (Batch Size), Max Len (Maximum Sequence Length), Prec. (Mixed Precision).}
\label{tab:training_configs}
\renewcommand{\arraystretch}{1.15} % 增加行高，提升可读性
\resizebox{\textwidth}{!}{%
\begin{tabular}{@{} l ccccccccc @{}}
\toprule
\textbf{Model} & \textbf{LoRA Rank} & \textbf{LoRA Alpha} & \textbf{LoRA Drop} & \textbf{Max Len} & \textbf{LR} & \textbf{BS} & \textbf{Warmup} & \textbf{Epoch} & \textbf{Prec.} \\
\midrule
Llama-3.1-8B  & 16 & 32 & 0.05 & 2048 & 1e-4 & 1024 & 0.1 & 1.0 & bf16 \\
Qwen-2.5-3B   & 16 & 32 & 0.05 & 2048 & 2e-4 & 1024 & 0.1 & 1.0 & bf16 \\
Qwen-2.5-7B   & 16 & 32 & 0.05 & 2048 & 2e-4 & 1024 & 0.1 & 1.0 & bf16 \\
Qwen-2.5-14B  & 16 & 32 & 0.05 & 2048 & 1e-4 & 1024 & 0.1 & 1.0 & bf16 \\
Qwen-2.5-32B  & 32 & 64 & 0.05 & 2048 & 5e-5 & 1024 & 0.1 & 1.0 & bf16 \\
Qwen-3-4B     & 16 & 32 & 0.05 & 2048 & 2e-4 & 1024 & 0.1 & 1.0 & bf16 \\
Qwen-3-8B     & 16 & 32 & 0.05 & 2048 & 2e-4 & 1024 & 0.1 & 1.0 & bf16 \\
Qwen-3-14B    & 16 & 32 & 0.05 & 2048 & 1e-4 & 1024 & 0.1 & 1.0 & bf16 \\
Qwen-3-32B    & 32 & 64 & 0.05 & 2048 & 5e-5 & 1024 & 0.1 & 1.0 & bf16 \\
\bottomrule
\end{tabular}
}
\end{table*}

% \subsection{Evaluation protocols}
\subsection{Evaluation protocols}

To ensure reproducibility, all evaluations were conducted using the OpenCompass framework with the following strict configurations:

\textbf{1. OpenCompass Setup}
\begin{itemize}\setlength{\itemsep}{2pt}
    \item \textbf{Inference Engine:} Local HuggingFace deployment via OpenCompass \texttt{GenInferencer}.
    \item \textbf{Prompting:} Standardized zero-shot templates requiring the model to output only the option letters.
    \item \textbf{Decoding Strategy:} Greedy decoding (\texttt{do\_sample = False}, temperature = 0) to eliminate sampling variance.
    \item \textbf{Token Limit:} Maximum generation length capped at 15 tokens to optimize for concise option extraction.
\end{itemize}

\textbf{2. Strict Scoring Rules \& Post-processing}
\begin{itemize}\setlength{\itemsep}{2pt}
    \item \textbf{Option Extraction:} Generative outputs were processed via \texttt{parse\_multi\_choice\_answer} to filter out reasoning traces, deduplicate valid options, and format them alphabetically (e.g., "A,B,C").
    \item \textbf{Exact Match (EM):} A score of $1$ is awarded if and only if the normalized prediction perfectly matches the ground truth. 
    \item \textbf{No Partial Credit:} Any missing, extra, or incorrectly selected options result in a score of $0$.
\end{itemize}

\textbf{3. Benchmark Subset Selection (MMLU \& C-Eval)}
To validate cross-domain generalizability while minimizing computational overhead, we evaluated on representative subsets that strictly map to our discipline domains:
\begin{itemize}\setlength{\itemsep}{2pt}
    \item \textbf{MMLU (12 subsets):} \texttt{abstract\_algebra}, \texttt{college\_chemistry}, \texttt{college\_computer\_science}, \texttt{electrical\_engineering}, \texttt{high\_school\_computer\_science}, \texttt{high\_school\_european\_history}, \texttt{high\_school\_mathematics}, \texttt{high\_school\_world\_history}, \texttt{logical\_fallacies}, \texttt{machine\_learning}, \texttt{marketing}, \texttt{world\_religions}.
    \item \textbf{C-Eval (6 subsets):} \texttt{teacher\_qualification}, \texttt{business\_administration}, \texttt{middle\_school\_physics}, \texttt{computer\_architecture}, \texttt{logic}, \texttt{college\_physics}.
\end{itemize}

\end{document}